\documentclass[traditabstract]{aa}
\usepackage{amssymb}
\usepackage{graphics}
\usepackage{tabularx}
\usepackage{color}
\usepackage{natbib}
\usepackage{rotating}

\begin{document}

%%%%%%%%%%%%%%%%%%%%%%%%%%%%%%%%%%%%%%%%%%%%%%%%%%%%%%%%%%%%%%%%%%%%%%%%%%%%%%%

% Trojans: 56 (C6)
% Hilda: 9 (C6) + 3 (C7) + 6 (C8) = 18
\title{The heart of the swarm: K2 photometry and rotational characteristics
of 56 Jovian Trojan asteroids}
\titlerunning{K2 photometry of Jupiter Trojan asteroids}

\author{%
Gy. M. Szab\'o\inst{1,2} \and
A. P\'al\inst{2,3} \and
Cs. Kiss\inst{2} \and
L. L. Kiss\inst{2,4} \and
L. Moln\'ar\inst{2} \and
O. Hanyecz\inst{2,3} \and
E. Plachy\inst{2} \and
K. S\'arneczky\inst{2} \and
R. Szab\'o\inst{2}
}

\authorrunning{Gy.M. Szab\'o et al.}

\institute{%
ELTE Gothard Astrophysical Observatory,
	9704 Szombathely, Szent Imre herceg \'utja 112, Hungary \and
Konkoly Observatory, MTA Research Centre for Astronomy and Earth Sciences,
	Konkoly-Thege Mikl\'os \'ut 15-17,
	1121 Budapest, Hungary; e-mail: \texttt{apal@szofi.net} \and
Department of Astronomy, Lor\'and E\"otv\"os University,
	P\'azm\'any P\'eter s\'et\'any 1/A,
	1117 Budapest, Hungary \and
Sydney Institute for Astronomy, School of Physics A28,
	University of Sydney,
	NSW 2006, Australia
}

%%%%%%%%%%%%%%%%%%%%%%%%%%%%%%%%%%%%%%%%%%%%%%%%%%%%%%%%%%%%%%%%%%%%%%%%%%%%%%

\date{Received \today; accepted \today}

\abstract{We present fully covered phased light curves for 56 Jovian Trojan asteroids as acquired by the K2 mission of the \textit{Kepler} space telescope. This set of objects has been monitored during Campaign 6 and represents a nearly unbiased subsample of the population of small Solar System bodies. We derived precise periods and amplitudes for all Trojans, and found their distributions to be compatible with the previous statistics. We point out, however, that ground-based rotation periods are often unreliable above 20\,h, and we find an overabundance of rotation periods above 60\,h compared with other minor planet populations. From amplitude analysis we derive a rate of binarity of 20$\pm$5\%. Our spin rate distribution confirms the previously obtained spin barrier of $\sim$5\,h and the corresponding $\sim$0.5\,g\,cm$^{-3}$ cometary-like density limit, also suggesting a high internal porosity for Jovian Trojans. One of our targets, asteroid 65227 exhibits a double rotation period, which can either be due to binarity or the outcome of a recent collision.}

\keywords{Techniques: photometric}

\maketitle

%%%%%%%%%%%%%%%%%%%%%%%%%%%%%%%%%%%%%%%%%%%%%%%%%%%%%%%%%%%%%%%%%%%%%%%%%%%%%%

\section{Introduction}
\label{sec:introduction}

Trojan asteroids are located at a heliocentric distance of
$\sim5.2\,{\rm AU}$ in the L4 and L5 Lagrange points of the Sun-Jupiter
system ($1:1$ mean-motion resonance). Trojans were traditionally thought to
be formed near to their present location, but recently, different scenarios
were proposed for their origin. As indicated by the surface composition, they
may have formed in the proto-Kuiper belt, and scattered inward and
captured (the Nice model, e.g. Morbidelli et al. 2005, Tsiganis et al. 2005, Gomes et al. 2005, Levison et al. 2011) as a result of resonant interactions with the giant planets (e.g. Grav et al. 2011, Emery 2016). The ``jump'' scenario assumes the collisional scattering of Jupiter only, due to close interactions with an ice giant (Nesvorn\'y{} et al. 2013). Due to their stable resonance locking, collisional frequencies are the lowest within these groups in the inner solar system \citep{dahlgren1998,delloro2001}.

While dynamical arguments support the capture of Trojans from the
outer Solar System via collisional scattering,
there are significant differences between the Trojans and the Centaurs
or trans-Neptunian objects, indicating a possible surface modification
due to the inward migration. The photometric and spectroscopic data
revealed the presence of two main taxonomical groups in the Trojan swarms,
recognized as D and P in the Tholen (1984) classification. Both types exhibit
featureless spectra, while D type has a redder slope and includes the
majority of the Trojans. Unlike the Main Belt, the Trojan taxonomical
types are not separated in the orbital elements space, but there is an
inclination-color gradient present at least in the L4 swarm (e.g. Szab\'o et al. 2007), which represents taxonomical subgroups (Roig et al., 2008).

Bimodality of Jovian Trojan asteroids was also identified in their magnitude distribution (especially at the faint-end), in near-infrared spectra, infrared albedo, and also in colours, suggesting the existence of two distinct groups among Jovian Trojans (Wong et al. 2014, and references therein). Since this bimodality extends to several parameters, the presence of ``red'' and ``less red'' {\it populations} are suggested, with different origin and different evolution, instead of simply different taxonomical {\it types} which were mixed together (Wong et al. 2014, 2015). However, objects in these populations are still significantly bluer than the typical ``red'' objects in the Centaur and trans-Neptunian populations (Peixinho et al., 2012; Lacerda et al., 2014). A common, outer Solar system origin of Jovian Trojans and trans-Neptunian objects was recently proposed by Wong \& Brown (2016), also suggesting that the retention or loss of H$_2$S in the early Solar system was the likely reason behind the colour differences we observe today.

In addition to the photometric characteristics discussed above, light curves available for Jovian Trojans can provide shape and rotational frequency distributions, and also information on the binary fraction. These statistics can be compared with the prediction of various formation and evolution models. In this sense, binary or multiple systems especially important, as their observations provide reliable masses and densities, a key to composition and internal structure. There are several formation mechanisms proposed for multiple systems (see e.g. Merline et al., 2002; Noll et al., 2008), therefore their properties give an important clue on accretional, collisional and radiative processes as well, and may lead to identify differences between the red and the less red groups in the case of Jovian Trojans. A WISE survey found that 20\% of Jovian Trojan asteroids are either extremely elongated objects, or are binaries (Sonnett et al., 2015).

In the K2 mission of the Kepler space telescope (Howell et al. 2014) 56 Jovian Trojan asteroids have been observed in Campaign 6, and long, uninterrupted light curves have been taken that are free from aliases, giving a more comprehensive description of these populations then the sparsely sampled WISE or ground based data. In this paper we present the light curves and photometric properties of these 56 Jovian Trojan asteroids, all orbiting at the L4 Lagrange point of the Sun-Jupiter system.
In Sec.~\ref{sec:observations} we summarize the observations and the data reduction schemes used to
obtain the light curves of these Trojan asteroids.
In Sec.~\ref{sec:results} we present the statistical properties of this
sample. Our results are summarized in Sec.~\ref{sec:summary}. Tabulated data and the light curves of individual Trojans are shown in the Appendix. A similar study about Main Belt asteroids with K2 will be published in a related paper (Szab\'o et al., 2016).

\begin{figure}
%{\includegraphics[viewport=245 199 569 365,width=0.92\columnwidth]{rl.eps}}
%{\hspace{30pt}\includegraphics[width=234pt]{field.eps}}\vspace{10pt}
%\includegraphics[width=234pt]{rl1.eps}\\
%\includegraphics[width=234pt]{field3.eps}
\includegraphics[width=234pt]{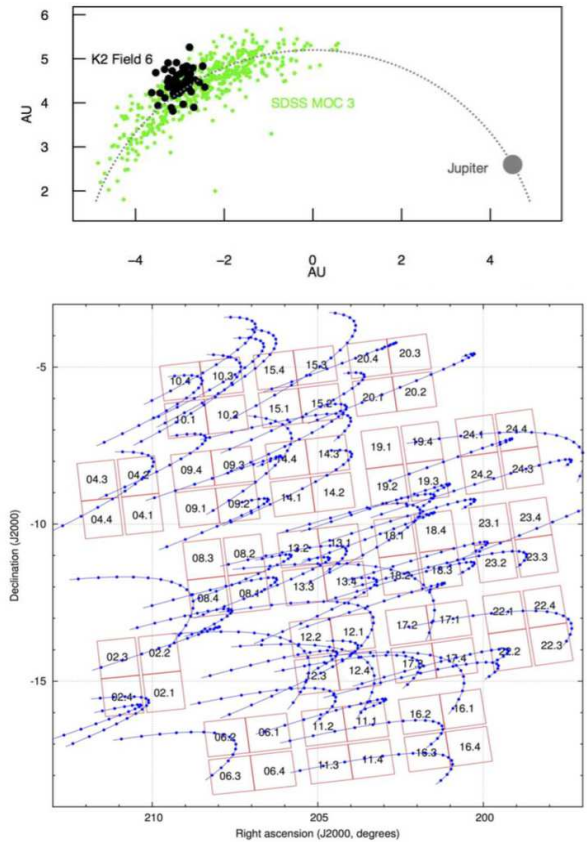}
\caption{Upper panel:
The
upper view of Field 6 Trojans (large black dots) superimposed to the SDSS MOC3 Trojans
in the L4 swarm (smaller grey dots) projected to the Jupiter's orbital plane. Earth is in origin. Note that Campaign~6 pointed exactly into the core of the L4 swarm.
Lower panel:
The field-of-view of K2 Campaign 6 superimposed with the
apparent trajectories of the 56 Jupiter Trojans discussed in this paper.
On each trajectory, the dots show the motion in 5-days long steps.}
\label{fig:field}
\end{figure}
%% %% %% %% %% %% %% %% %% %% %% %% %% %% %% %% %% %% %% %% %% %% %% %% %% %%

%% %% %% %% %% %% %% %% %% %% %% %% %% %% %% %% %% %% %% %% %% %% %% %% %% %%
\begin{figure}
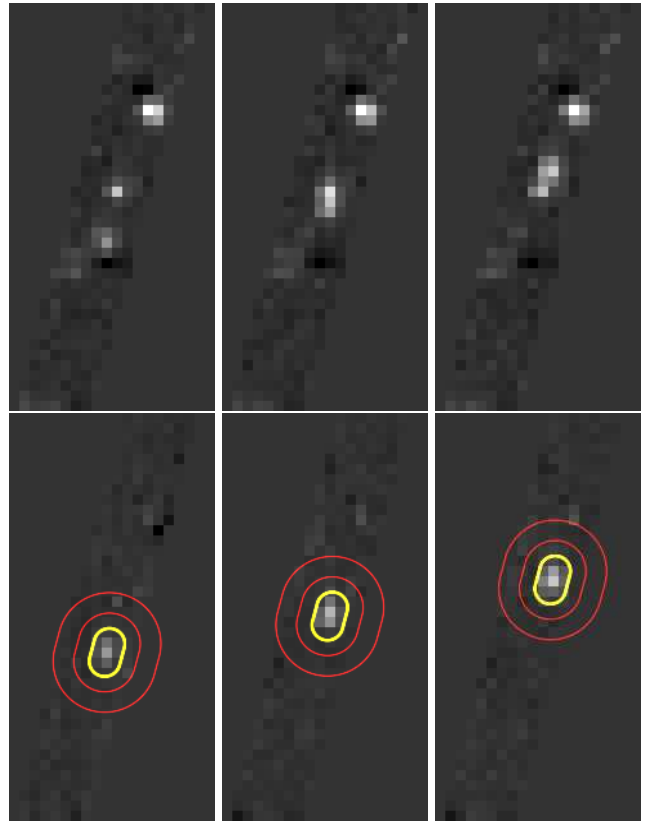

\begin{center}
\noindent
\resizebox{27mm}{!}{\includegraphics{fits-r-2633}}\hspace*{1mm}%
%\resizebox{33mm}{!}{\includegraphics{fits-r-2634}}\hspace*{1mm}%
\resizebox{27mm}{!}{\includegraphics{fits-r-2635}}\hspace*{1mm}%
%\resizebox{33mm}{!}{\includegraphics{fits-r-2636}}\hspace*{1mm}%
\resizebox{27mm}{!}{\includegraphics{fits-r-2637}}
\vspace{5mm}
\noindent
\resizebox{27mm}{!}{\includegraphics{diff-r-2633}}\hspace*{1mm}%
%\resizebox{33mm}{!}{\includegraphics{diff-r-2634}}\hspace*{1mm}%
\resizebox{27mm}{!}{\includegraphics{diff-r-2635}}\hspace*{1mm}%
%\resizebox{33mm}{!}{\includegraphics{diff-r-2636}}\hspace*{1mm}%
\resizebox{27mm}{!}{\includegraphics{diff-r-2637}}
\end{center}\vspace*{-4mm}
\caption{Typical stamps taken from the image series related to
(12974) Halitherses, which was the object with the
fastest apparent speed (exceeding 2\,pixels per long-cadence frame).
The upper row of image stamps shows the original frames between the
cadences \#2633 and\#2637 while the lower row show the registered
and differential stamps corresponding to their respective upper counterparts.
The elongated aperture (bounded by the thick yellow curve) and the
elongated annuli (bounded by the red curves) are used for the
photometry and for the estimation of the background level, respectively.
These stamps show a region of $20\times 40$ pixels, i.e. an area of
$1.33^\prime\times 2.66^\prime$ on the sky.}
\label{fig:trailstamps}
\end{figure}
%% %% %% %% %% %% %% %% %% %% %% %% %% %% %% %% %% %% %% %% %% %% %% %% %% %%

\begin{figure*}
\begin{center}
\resizebox{60mm}{!}{\includegraphics{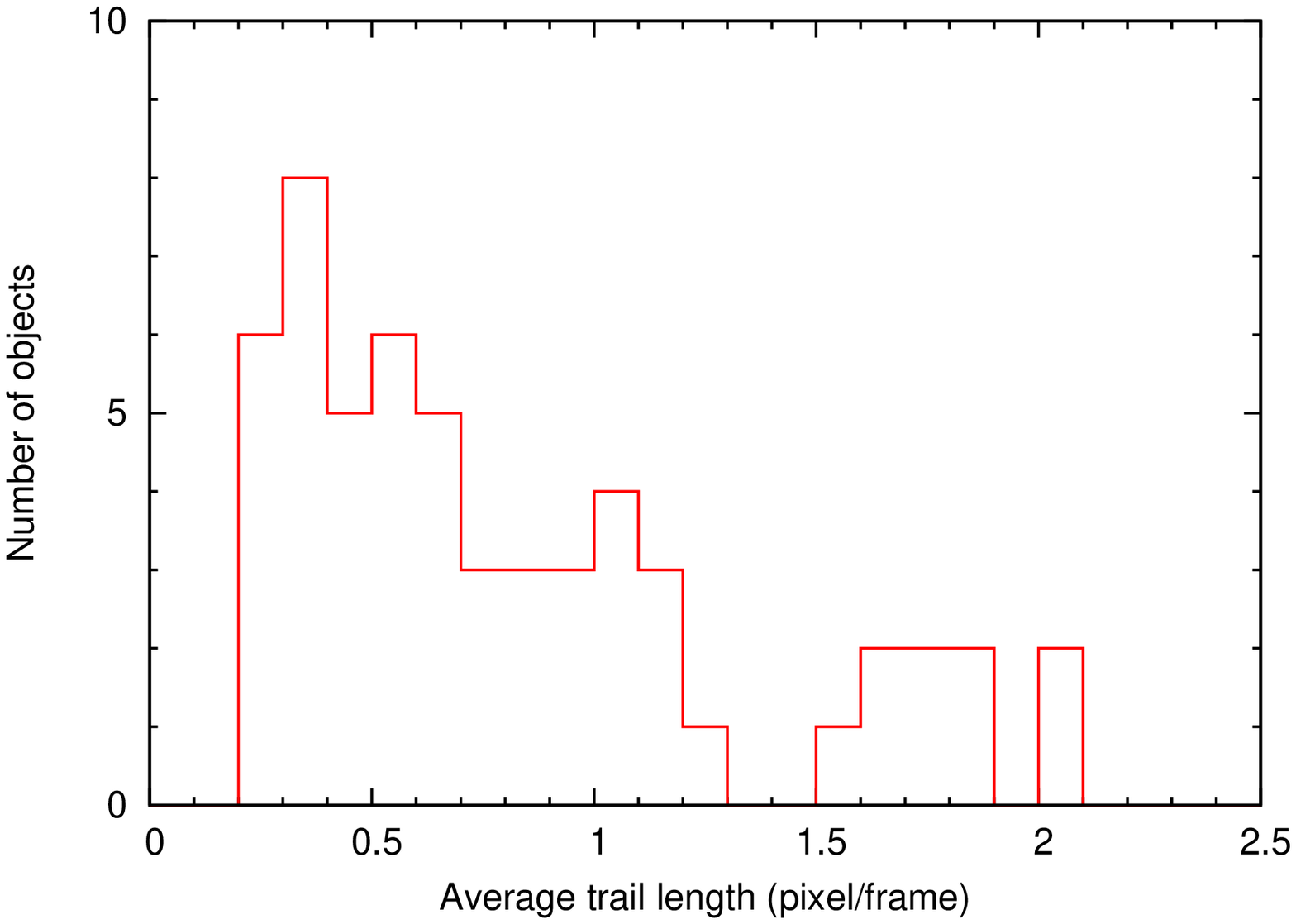}}
\resizebox{60mm}{!}{\includegraphics{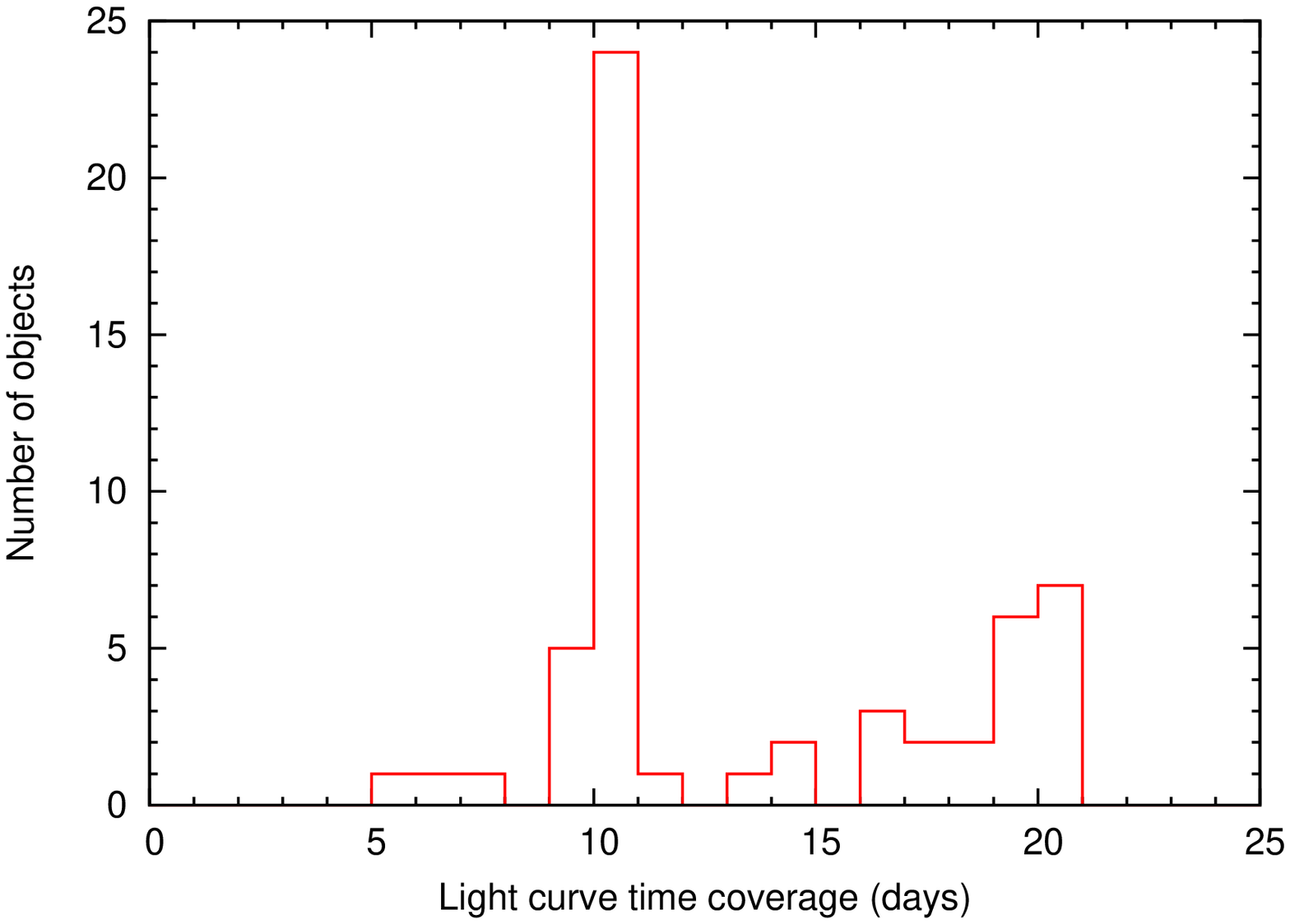}}
\resizebox{60mm}{!}{\includegraphics{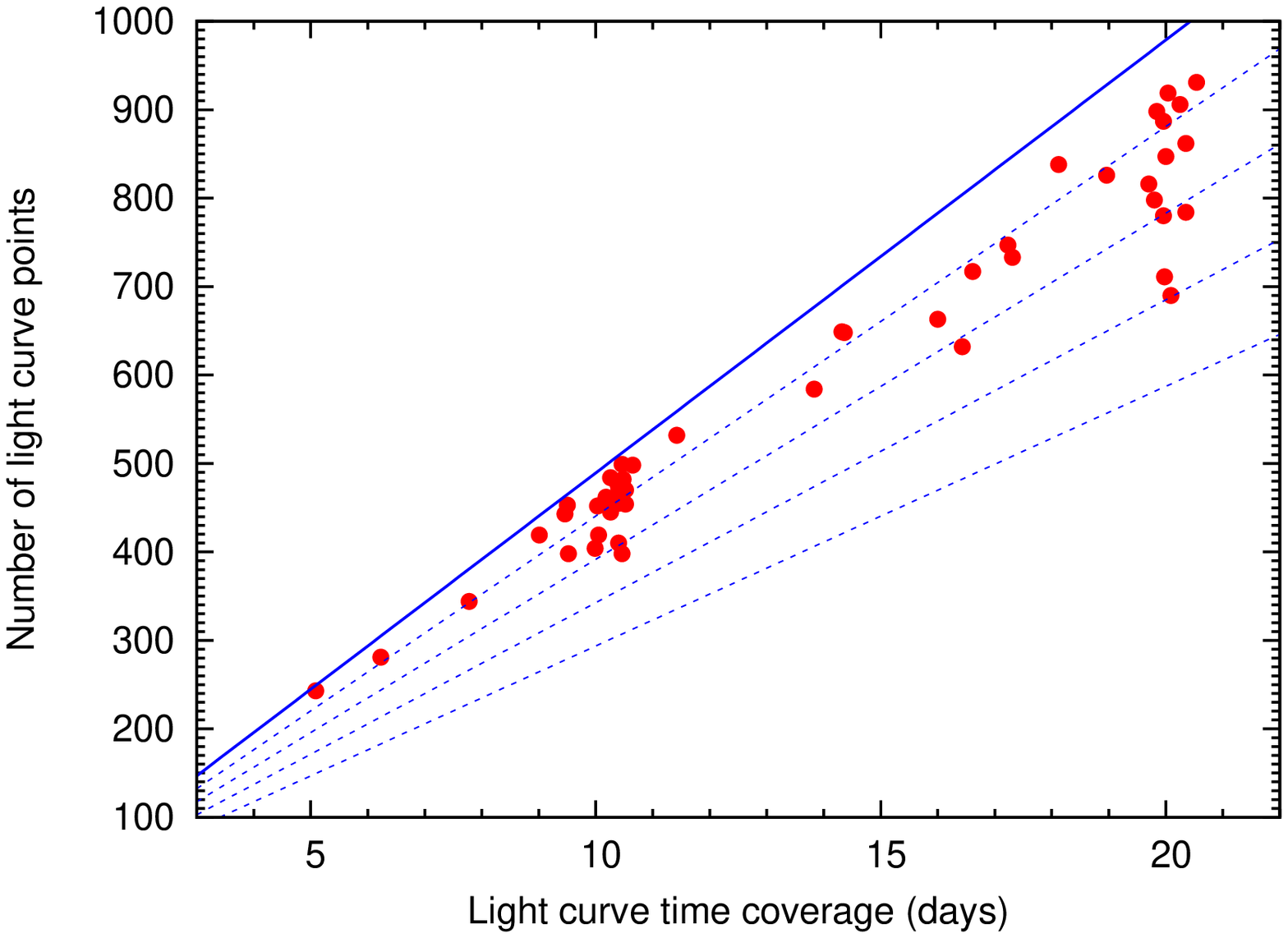}}
\end{center}\vspace*{-4mm}
\caption{Statistics related to the light curve acquisition characteristics
and the photometric data coverage. See text for further explanation.}
\label{fig:stat}
\end{figure*}
%% %% %% %% %% %% %% %% %% %% %% %% %% %% %% %% %% %% %% %% %% %% %% %% %% %%

%%%%%%%%%%%%%%%%%%%%%%%%%%%%%%%%%%%%%%%%%%%%%%%%%%%%%%%%%%%%%%%%%%%%%%%%%%%%%%

\section{Observations and data reduction}
\label{sec:observations}

\subsection{K2 observations}
\label{sec:k2observations}

 \textit{Kepler} has observed parts of the apparent trajectories of 56 Trojans
during K2 Campaign 6, between 2015 July 13.95 and 2015 September 30.87
(UTC). In total, Campaign 6 data are formed from 3686 long cadence
frames and the stamps allocated for these 56 Trojan asteroids have also
been retrieved in long cadence mode, {correpsonding to 29.41 minute sampling frequency)}. To date, all of the moving objects
observed by K2 were trans-Neptunian objects \citep{pal2015,kiss2016,pal2016}
whose apparent speed in the K2 CCD frames were so small that during
the photometric analysis, these objects can be treated as point sources
on the individual long cadence frames.
However, Trojans of Jupiter are located significantly closer to the inner
Solar System, i.e. to K2 itself. Hence, the apparent speed can
exceed $0.1^{\prime\prime}/{\rm min}$, yielding trail-like corresponding
PSFs on the long cadence frames. As we will describe later on
(Sec.~\ref{sec:datareduction}), the photometry of these trails require
non-circular apertures.
The main parameters of the 56 observed Trojans are summarized
in Table~\ref{table:keplerobs}. The apparent trajectories of these objects
w.r.t the K2 CCDs are shown in Fig.~\ref{fig:field}.

%%%%%%%%%%%%%%%%%%%%%%%%%%%%%%%%%%%%%%%%%%%%%%%%%%%%%%%%%%%%%%%%%%%%%%%%%%%%%%

\subsection{Data reduction and photometry}
\label{sec:datareduction}

In principle, the reduction of K2 long cadence data corresponding to
Jupiter Trojans follows the similar steps that have been conducted in
former analysis of moving objects observed with K2 \citep{pal2015,kiss2016}.
In order to correct for the positioning jitter of the spacecraft we retrieved nearly a dozen of additional stellar sources for each Trojan and included these in further processing \citep[see e.g. Fig.~1 of][]{pal2016}.
While all of these 56 objects have been observed in the same campaign of K2, the determination of the positioning jitter have been done separately for each object. This is essential due to the large field-of-view: namely, the same pitch, roll and yaw offsets applied during the attitude control of the space telescope yields different apparent centroid shift and field rotation on the various CCD modules.

After the derivation of the positioning jitter of the subsequent frames, we registered the frames to the same reference system in order to perform differential image analysis. A series of frames have been pre-selected to form a master median-combined image which was then subtracted from the subsequent frames. Differential aperture photometry were then performed using apertures that follow the expected shape of the apparent elongated PSFs of these objects. The apparent speed of $\lesssim 0.1^{\prime\prime}/{\rm min}$ mentioned above is equivalent to $\lesssim 2$ {\textit Kepler} CCD pixels per long cadence sampling. Therefore, it
is essential to take into account these shapes during the photometry. In Fig.~\ref{fig:trailstamps} we displayed a short series of subsequent image stamps and the related apertures corresponding to (12974) Halitherses. This object showed the the largest apparent speed, namely $1.9\dots 2.1$
pixels per long cadence frame. We note that this algorithm of
using elongated apertures for flux extraction yields the same output
in the limit of circular apertures at low apparent speed.

During the differential image procedures, the background stars cannot be
completely eliminated. This is mostly due to the residual structures
yielded both by the interpolation algorithm as well as the intrinsic
photon noise of the stars themselves. These residual structures caused
outlier points which were expunged from the final light curves.
Light curve point exclusion was also caused by data loss during the
telemetry of K2 data. On average, $\sim10\%$ of the data points were
excluded from the light curves due to the aforementioned reasons.
In Fig.~\ref{fig:stat} we plotted some corresponding statistical properties
while quantities related to the light curve statistics
are also found in Table~\ref{table:keplerobs}.

\begin{figure}
\begin{center}
%{\includegraphics[viewport=220 130 569 408,clip,width=\columnwidth]{pa.eps}}
{\includegraphics[width=\columnwidth]{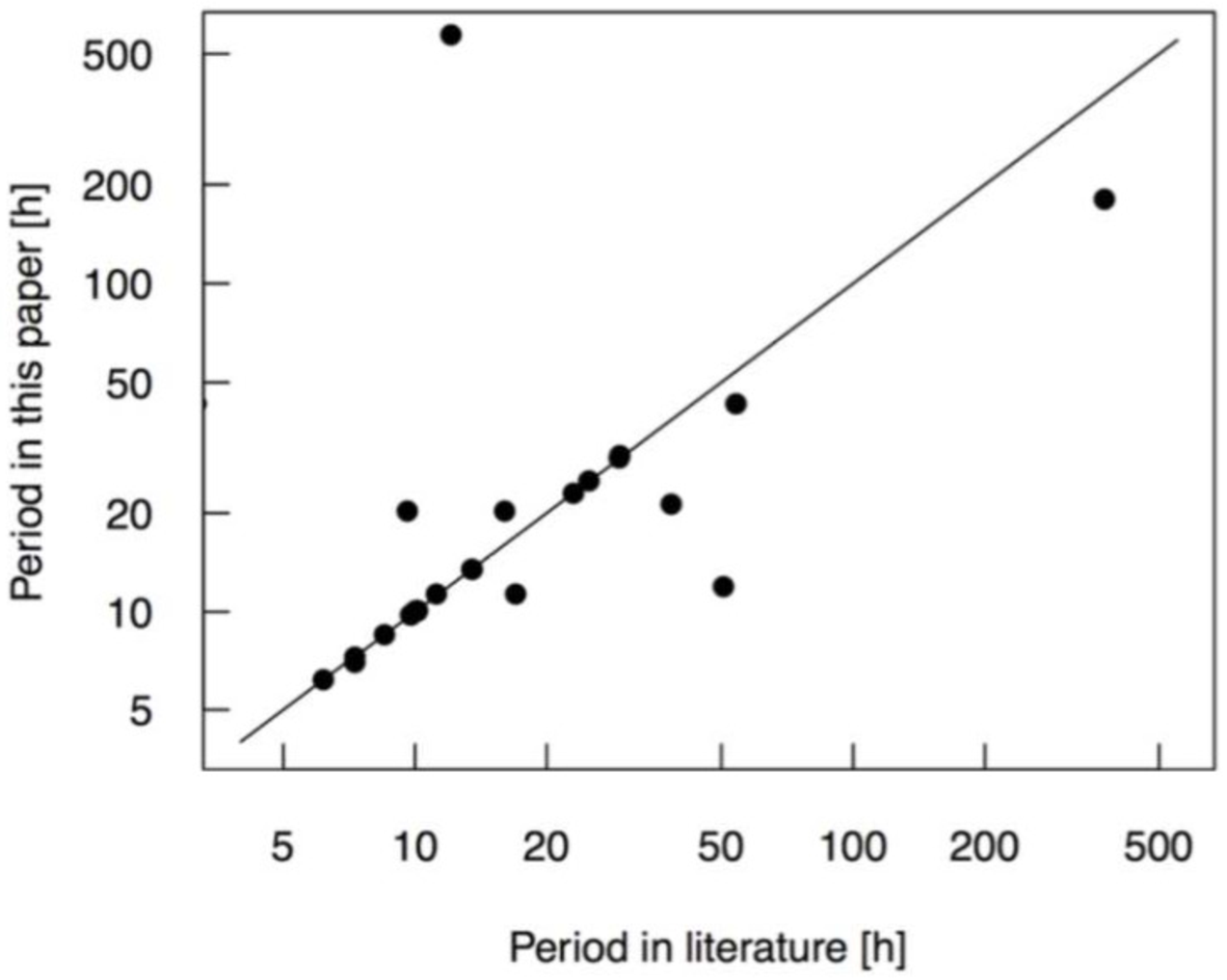}}
\end{center}%\vspace*{-4mm}
\caption{Comparison of periods in the literature and our determinations.}
\label{fig:perdet}
\end{figure}

\begin{figure}
\begin{center}
\includegraphics[clip,width=\columnwidth]{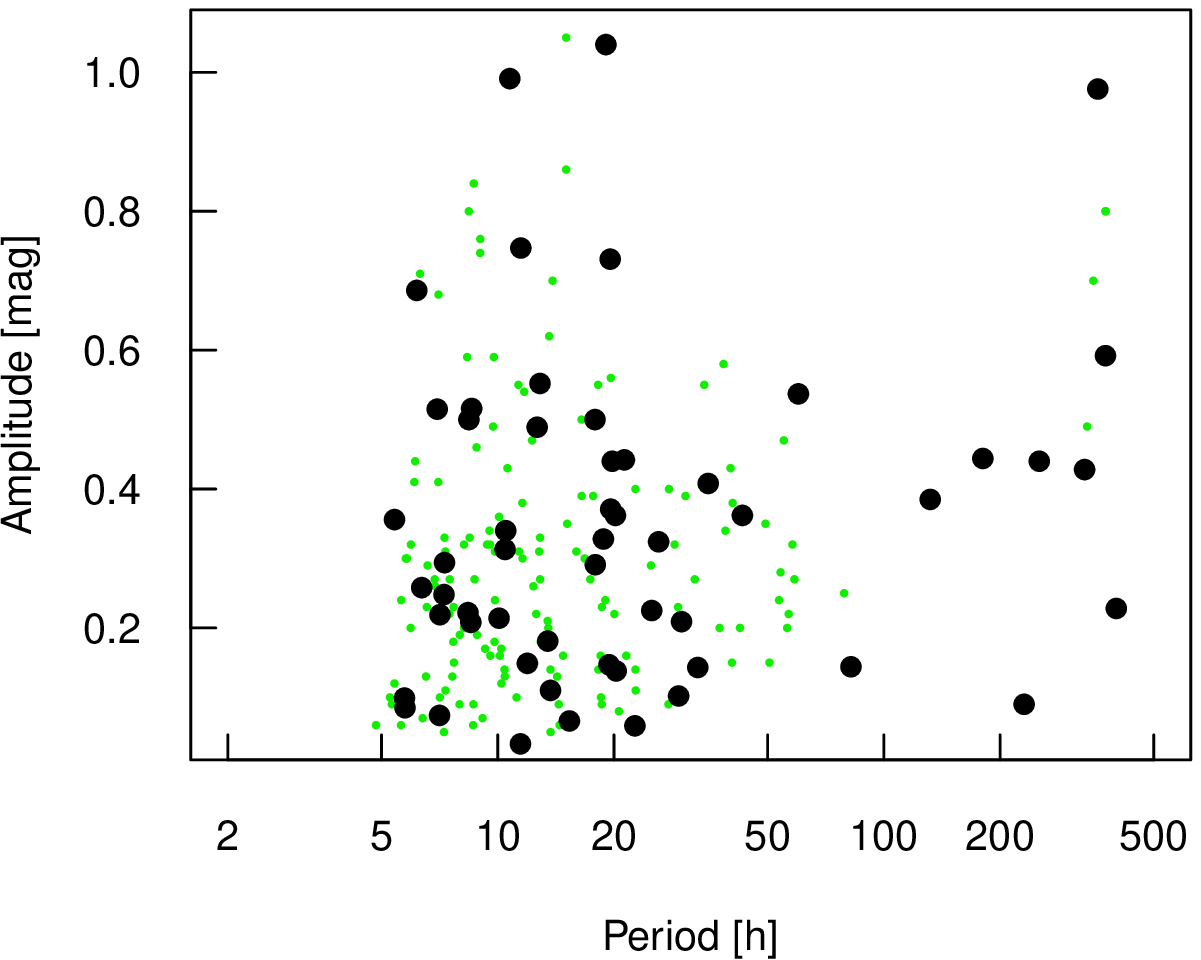}
\includegraphics[clip,width=\columnwidth,clip]{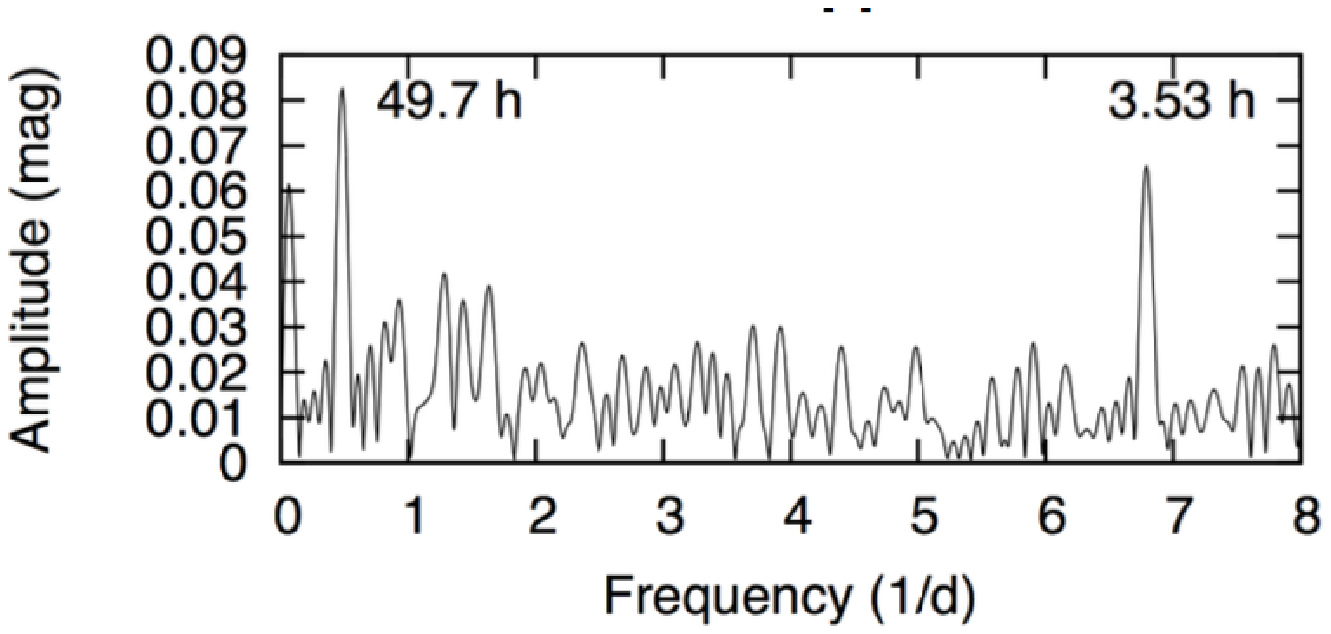}
\end{center}%\vspace*{-4mm}
\caption{Upper panel: The period--amplitude distribution of Trojan asteroids.
Our data are plotted with large black dots, data from the literature is
plotted with small green dots. See the text for references and discussion. Lower panel: The periodogram of (65227) 2002 ES$_{46}$ with multiple periods.}
\label{fig:peramp}
\end{figure}

For all of the previously discussed photometric procedures, we applied
the various tasks of the FITSH package \citep{pal2012}. We note here
that the features of aperture photometry on non-circular (in fact, arbitrary
shaped) apertures are included only in the upcoming, \texttt{0.9.2} version
of the package. This new version is currently under preparation for
public release and expected to be published on the website of the
software\footnote{http://fitsh.szofi.net/} soon.

The statistics of light curve acquisition is plotted in Fig. 3.
The left panel shows the distribution of
the apparent trail lengths of the Trojans on the long-cadence frames. It
can easily be seen that one-third of the objects have an average trail length
less than $0.5$ pixels (which yields roughly indistinguishable PSF from the
fixed sources) while the two-thirds of the objects have a trail length less
than $1$ pixel. In addition, there are only 2 objects for which the
trail length is larger than $2$ pixels on average. The middle panel
shows the distribution of the light curve coverages (i.e. the timespan
between the first and last light curve data point). Most of our objects were observed for 10--20 days, with a distinct peak around 10 days, and there is another group scattering around 20 days of coverage. The right panel shows the number of light curve points as the function  the duty cycle (percentage of the number of observed 29.4-min cadences with respect to the maximally possible during the length of observation), i.e. how many cadences have been lost due to technical problems originating mainly from the photometric pipeline (outliers, encountering too bright stars, stellar residuals, etc.) The thick blue line corresponds to the slope of $\approx 1/0.02043$\,frame/day (where $0.02043\,{\rm d}$ is the K2 long
cadence period) and the thin dashed blue lines show its 90\%, 80\%, 70\% and 60\%. The vast majority of our targets were observed with a duty cycle between 80 and 96\%, a handful of them above 70\%. The long coverage and the relatively continuous observations support the conclusion that period biases and aliases due to interruptions can only marginally affect our sample, if they bias it at all.

Of course our sample is magnitude limited, but is quite close to be complete in the observed magnitude range. In Szab\'o et al. (2015) we demonstrate that no undiscovered asteroids are found in the full-frame and part-frame K2 observations, confirming that the asteroids have been discovered down to the sensitivity limit of K2. Therefore we do not expect severe biases due to undetected small asteroids.

\subsection{Period analysis}
\label{sec:periodanalysis}

We searched for significant periodicities using the Fourier method as implemented in the {\tt Period04} program package \citep{lenz2005}, and also the Lomb-Scargle periodogram in the {\tt gatspy} Python package\footnote{https://github.com/astroML/gatspy/}. We got very similar results in several test cases, therefore we decided to stick to the Lomb-Scargle periods. We note that the errors of the individual photometric points are taken into account. Only those signals were considered that were significant on the 3$\sigma$-level compared to the background local noise periodogram. We phase-folded the light curves with the best period and its double value, then decided which gives a better fit based on a visual inspection.

In our parallel paper about Main Belt asteroid detections we show that the period determination is usually solid if the coverage exceeds 5 days and duty cycle is above 60\%{} (Szab\'o et al., 2016).
Since the conditions are well fulfilled for K2 Trojan asteroids, we could derive reliable solutions for most asteroids. In addition to the automatic processing of period determination all phased light curves were inspected visually, too. We suspect that the occurence of period biases due to unfortunate data distribution is very low. For some objects the light curve we obtained suggested a complex brightness variation. In these cases we tested manually for periods other than the primary one, but finally omitted
solutions different from the bimodal light curves. The phased light curves were also binned
by 0.02 rotational phase steps, and the full amplitude was accepted as the difference
between the minimum and maximum of this binned light curve.

\section{Results}
\label{sec:results}

\subsection{Period and amplitude distributions}

We present the determined periods and amplitudes in Table \ref{table:results}. Both parameters span over a wide range (which we discuss later in details), and indicatively, the median value of the measured periods is around 13~h, and the median amplitude is 0\fm32. The number of high amplitude asteroids is surprisingly high; all asteroids
in our sample exceed 0\fm1 amplitude, and several asteroids exceed 0\fm75 amplitude, belonging to more than $2:1$ asphericity in sky projection. These findings, as well as the overall distribution of our data points, are consistent with the results in previous publications.

Our period determinations are plotted against data from the literature
(Molnar et al. 2008, Mottola et al. 2011, French et al. 2015, Waszczak et al. 2015)
in Fig. \ref{fig:perdet}. For periods less than 10 hours our periods are in
perfect coincidence with the previous determinations, while above 15--18 h period,
we could confirm only a fraction of previous periods. Clearly, the power of the
uninterrupted K2 light curves is in its accuracy in the long period range, partly
because the measurements are free from daily aliases, and also because the stable observation
circumstances do not lead to distracting systematics that mimic rotational light variations.

In Fig. \ref{fig:peramp} we plot the amplitude--period diagram for Trojans from the
literature, (Molnar et al. 2008, Mottola et al. 2011, French et al. 2015, Waszczak et al. 2015)
compared to our results. It is known that Main Belt families are characterized by
specific brightness variation distributions (Szab\'o and Kiss, 2008). For the Trojan asteroids, we cannot confirm such parameter dependencies.

Our sample shows a significant overabundance above 60 hour periods. Possibly this is due to
the limited completeness of previous surveys in the range of very slow rotation rate.
Due to the unbiased K2 observations that last for many days in the most cases, we believe
that in this range our sample is still balanced, and reflects the common occurrence
of very slow rotators among Trojan asteroids. This result would worth comparing to a
similar sample in the Main Belt. Interestingly, this is not possible now, since the already
known distributions suffer the same incompleteness as the Trojan asteroids in the very long
period range, so we cannot compare our results to the known distributions. Even K2 asteroid
surveys are more limited in the very long period range than in the Trojan swarms, simply
because Main Belt asteroids move faster and spend much less time on silicon than Trojan asteroids
do.

\subsection{Signs of binarity}
Three light curve features are invoked as alerting signs for a binary asteroid. Leone et al. (1984) defined a light curve amplitude $>$0\fm9 at least at one viewing geometry, referring to an $a/b$ asphericity $>2.3$. This elongation cannot be explained by a rubble pile body in equilibrium, but instead, a Jacobi ellipsoid stretched along the semi major axis of two rubble piles orbiting each other. Another signal of binarity is the slow rotation, possibly reflecting a tidal synchronization with a modestly far companion. Following the recipe of Sonnett et al. to put the alerting limit at 3$\times$ the average Trojan period, here we consider a Trojan to rotate unusually slow if the period exceeds $3\times$ the average, roughly 40 hours. The third signal of possible binarity is the presence of two periods, reflecting the light variation of the main body and the companion which have not been synchronized, or the forced perturbation of the main body because of the presence of the companion.

In our sample, several asteroids exhibited the described diagnostics of binarity. Asteroids 21593, 22056, and 39289 were detected to exceed 0\fm9 amplitude, and hence, to become a good candidate for binary pile asteroids. Even, 22056 exhibited the slowest rotation rate of 358 hours (almost 15 days!), therefore we consider it to the best and very promising candidate for a binary Trojan in K2 data.

Very slow rotation only characterized a surprisingly large number of asteroids. Besides the mentioned example of 22056, we found asteroids 9807 (330\,h), 13331 (180\,h), 23958 (trend only), 24357 (131\,h), 39270 (82\,h), 63239 (60\,h), 65223 (252\,h), 65240 (230\,h), 129602 (43\,h) in the alerting range of rotational rates.

There is one asteroid with observations both in WISE data and K2, (16152) 1999~YN$_{\rm 12}$. Sonnett et al. (2015) derived a WISE magnitude range of 0\fm97$\pm$0\fm15, for this asteroid, while here we present a period of $11.47\pm0.22$ hours and 0\fm75 amplitude. Since the amplitude is based on 475 K2 photometric points, the photometric error is in the order of 0\fm01 in the amplitude. 6 years were passing between the WISE and K2 measurements, therefore the two amplitudes refer to very different aspect geometry at roughly opposing positions on the asteroid's orbit. Since both amplitudes are unusually high, this body should be really elongated and a pole position roughly perpendicular to the plane of Ecliptic. Thus, (16152) 1999~YN$_{\rm 12}$ passed the amplitude criterion of binarity, and remains a good candidate for further investigations.

Putting these detections together, we get 11 K2 Trojans with signs on binarity. This means $\approx 20 \pm 5$\%{} rate of binaries among L4 Trojans, or at least asteroids with strong signs of binarity. This determination is unbiased because the full light curve coverage and the long observation runs, and is fully compatible to the debiased rate of 14--23\%{} of Sonnett et al. from WISE data, and is consistent or slightly exceeds a previous estimate of 6--10\%{} (Mann et al., 2007) from Earth-based light curves.

\subsection{(65227) 2002 ES$_{\rm{46}}$ with double periods}
We detected multiple periodicities in the case of (65227) 2002 ES$_{\rm{46}}$, the ``one-hump'' periods are 49.7\,h and 3.53\,h, with 0\fm065 and 0\fm08 full amplitudes, respectively. The detection of these periods is secure above the the noise level of $\approx$0\fm03--0\fm04. These periods are not resonant to each other, and since the period ratio is $\approx$14.1, they seem to have independent origin in processes with very different periods. The amplitudes are tiny, indeed the smallest ones in our K2 Trojan sample (Fig. \ref{fig:peramp} lower panel).

This detection can be considered as another alerting signal for a binary asteroid. Both the periods and the amplitudes can be explained by a binary object with partial tidal locking. Assuming a two-hump light curve solution, the rotation period of one body would be 7.06\,h; while the other periodicity of 49.7 (or 99.4) hours comes from the orbital motion and possibly, the light variation of the tidally locked rotator.

This interpretation is consistent also with the small amplitudes. Since both periods come from only one body of the system, the presence of the other body significantly suppresses the measured amplitude. Assuming two equally bright components, the full light curve amplitude from each components would be 0\fm16 and 0\fm12.

Another possible interpretation would be that the object is in the state of free precession, due to a recent collision. It is known from theoretical physics that an oblate ellipsoid with principal moments of inertia $A=B$ and $C$ will be in a state of free precession if the rotation axis is inclined to the total angular momentum vector, and the ratio of $\omega$ rotation rate and the $\Omega$ precession rate will be $\Omega/\omega = (A-C)/C$, independently of the inclination angle of the rotation vector. If we assume a free precession in the case of (65227) 2002 ES${_{\rm 46}}$, an ellipsoidal shape and a homogeneous internal composition, we can derive the axis ratios to be $A/C=1.071$. Such large precession rate can therefore be observed only for an almost spherical body. This numerical result is also consistent to the observed light variation from this object on the order of a few 0\fm01.

However, we consider that the binary scenario is more likely in the case of 65227. We considered that binarity is quite common in our sample. Moreover, in the case of binarity, quite elongated objects can support the observed light variation, since the two sources mutually  suppress the amplitudes. The free precession scenario seems to be less likely since the precession rate has stringent implications to the sphericity of the rotating body, and also, the scenario assumes a recent collision in an otherwise relaxed part of the Solar System.

\subsection{Comparison of red and less red populations}
We compared the red and less red populations by means of period and amplitude distributions
of the member asteroids. For this task, we cross-correlated the K2 asteroids to the SDSS Moving Object Catalog 4 (MOC hereafter), to derive the $t$ color which separates the red and less red members the most (Szab\'o et al. 2007). In case of multiple measurements of the same asteroid, the measurements were averaged to get more precise colors.

Eleven Trojans with K2 observations were found with an entry in SDSS MOC. Four of them were identified to represent the less red population -- (8241) Agrius, 13331, 16152, 23939 -- and seven that belong to red population of Trojans -- (1749) Telamon, (5028) Halaesus, 21599, 22056, 24357, 59049, and 129602. We added the population membership as a population flag ($r$ for red, $lr$ for less red) in Table \ref{table:results}.

No specific pattern has been found for red and/or less red population members. Both groups contain normal rotators around the median value of all K2 Trojans ($\approx13$\,h period, $\approx$0\fm32 amplitude) and also, both exhibit very slow rotators, as well (e.g. 180 and 358 hours for 13331 and 22056, respectively) and elongated asteroids (e.g. 0\fm75 and 0.98 for 16152 and 22056, respectively). Because of low counts, we cannot get to more detailed conclusions, but we could confidently observe that high amplitude asteroids and slow rotators are not specific to one population only: they are quite common in both populations.

%%%%%%%%%%%%%%%%%%%%%%%%%%%%%%%%%%%%%%%%%%%%%%%%%%%%%%%%%%%%%%%%%%%%%%%
\subsection{Period distribution and the spin barrier of Jovian Trojan asteroids}

%%%%
\begin{figure}
\includegraphics[width=8.5cm]{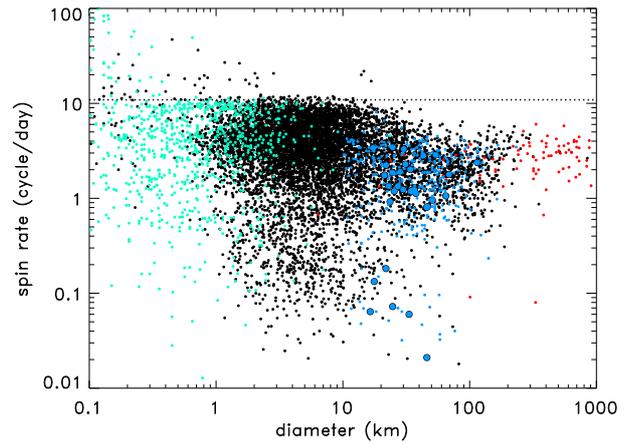}
\caption[]{Diameter vs. spin rate of asteroids. Black dots: main belt, red dots: trans-Neptunian objects, green dots: near-Earth asteroids, small blue dots: Jovian Trojans. Data is obtained from the Asteroid Light Curve Database \citep{Warner}. The Jovian Trojan asteroids with known sizes -- as determined by NEOWISE (Grav et al., 2012) -- are marked by large, filled blue circles.   \label{fig:spindiam}}
\end{figure}
%%%%%

%%%%
\begin{figure}
\includegraphics[width=8.5cm]{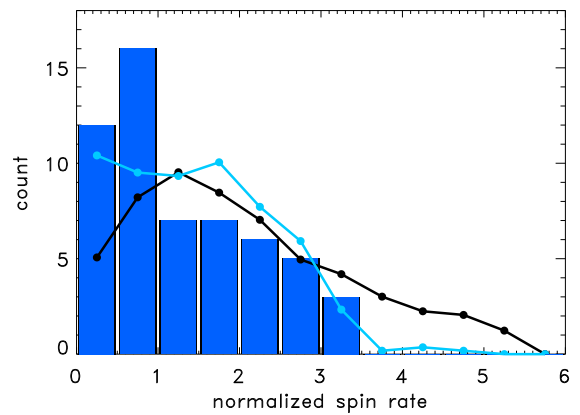}
\caption[]{Distribution of normalized spin rates in our sample (blue bars). We used $<$f$>$\,=\,1.34\,cycle\,day$^{-1}$ for normalization. Black and blue curves represent the scaled normalized spin rate distribution of main belt and Jovian Trojan asteroids, respectively, as obtained from the Asteroid Light Curve Database \citep{Warner}.   \label{fig:spindist}}
\end{figure}
%%%%%

In Fig.~\ref{fig:spindiam} we present the size vs. spin rate distribution of small bodies, including near-Earth and main belt asteroids, Jovian Trojans, Centaurs, and trans-Neptunian objects. The fast rotation of minor planets is limited by the so called spin barrier: there is a critical rotation period at which a rubble pile asteroid would fly apart due to its centripetal acceleration. For a specific body, this critical period can be estimated as $P_c \approx  3.3\cdot\sqrt{(1+A)/\rho}$, where $A$ is the light curve amplitude, $\rho$ is the density (in [g\,cm$^{-3}$]), and $P_c$ is obtained in hours \citep{Pravec+Harris}. Using this formula the knowledge of the rotation period and the light curve amplitude can provide a lower limit estimate of the body's density. This spin barrier is well established for main belt asteroids, the critical rotation period is $\sim$2.2\,h (dashed horizontal line in Fig.~\ref{fig:spindiam}), resulting in a critical density of $\sim$2.0\,g\,cm$^{-3}$. For main belt asteroids this limit is set by asteroids with diamters of 1-10\,km. However, with the Jovian Trojans, we are in the size range of $\sim$10-100\,km, i.e. we study objects that are an order of magnitude larger than the ones setting the spin barrier for main belt asteroids.

\begin{figure}
\includegraphics[width=8.5cm]{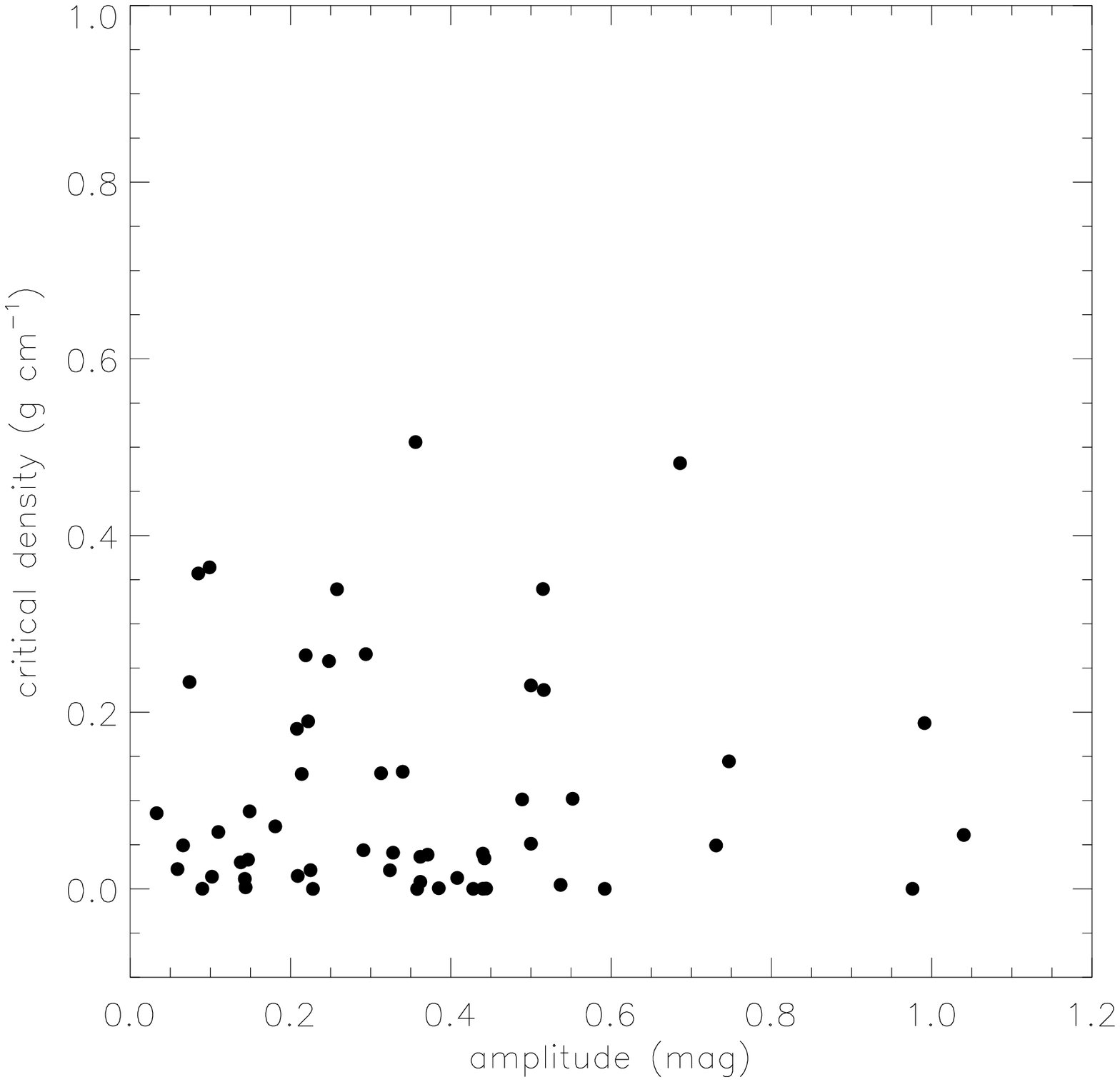}
\includegraphics[width=8.5cm]{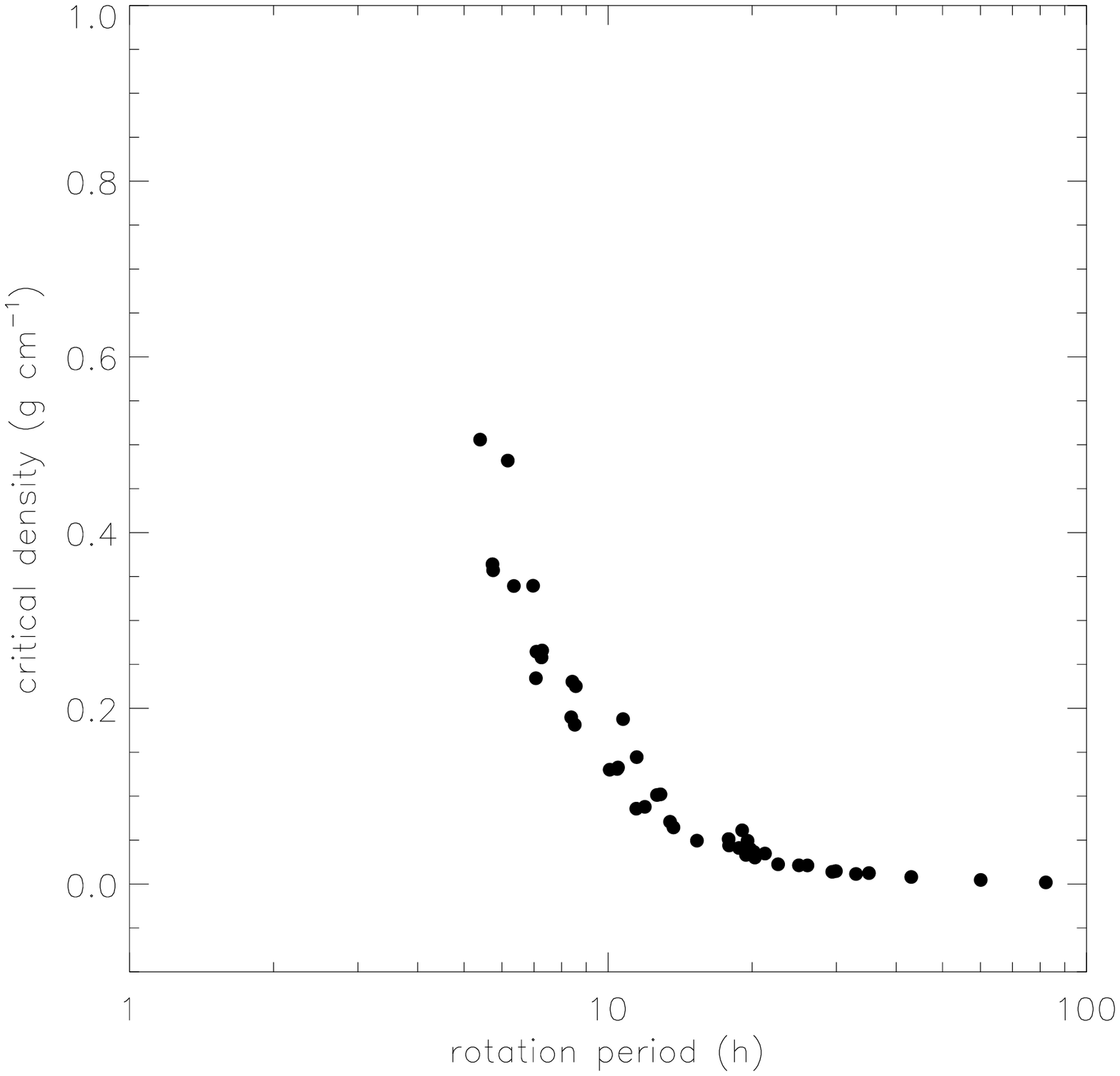}
\caption[]{Critical densities obtained for our sample, as a function of light curve amplitude (upper panel) and rotation period (lower panel).  \label{fig:spinbarrier}}
\end{figure}

Comparison of the spin rate distribution of Jovian Trojans with main belt asteroids in the Jovian Trojan size range (Fig.~\ref{fig:spindist}) shows a notable lack of Jovian Trojans in the high spin rate range. Based on the data available in their study, \citet{French} estimated densities of  $\sim$0.5\,g\,cm$^{-3}$ using the fastest rotation periods of $\sim$5\,h. This value is different from the critical density obtained for main belt asteroids ($\sim$2\,g\,cm$^{-3}$) but consistent with the density of cometary nuclei \citep{AHearn2011} as well as that of objects from the trans-Neptunian populations \citep{Brown2013,Vilenius2014}.
Applying the critical density calculation above for our sample, the highest density values we obtain are $\sim$0.5\,g\,cm$^{-3}$ (see Fig.\ref{fig:spinbarrier}), lower than that of main belt asteroids in the same size range, but the same as obtained by \citet{French}. Even in our unbiased sample, despite that rotation periods longer than $\sim$1\,h should have been identified, we were unable to find any Jovian Trojan asteroid with a high ($\sim$2\,h) spin rate. This may indicate icy compositions and porous interior for most Jovian Trojans, supporting an outer Solar system / Kuiper belt origin, as discussed in Sect.~1.

%%%%
%%%%%

Jovian Trojans show in general an excess in the number of slow rotators when compared with main belt asteroids (blue and black curves in Fig.~\ref{fig:spindist}). In our sample this is even more expressed, as there is a significant number of asteroids with periods $P>50$\,h, $\sim$20\% of the objects observed by K2 in this work. Very long rotation period may be an indication of binarity, as it is e.g. the case for Jovian Trojan (617) Patroclus where the $>$100\,h rotation period is explained by tidal breaking \citep{Mueller2010}. Slow rotation of small ($<$\,30\,km) and low density ($<$\,1\,g\,cm$^{-3}$) objects may have also been set by the YORP effect \citep{French}. Among our slow rotators there are objects that fall into this susceptible size range according to the sizes derived by \citet{Grav} based on NEOWISE observations.

%%%%%%%%%%%%%%%%%%%%%%%%%%%%%%%%%%%%%%%%%%%%%%%%%%%%%%%%%%%%%%%%%%%%%%%%%
\subsection{Light curve amplitude distribution}

\begin{figure}
\includegraphics[width=8.5cm]{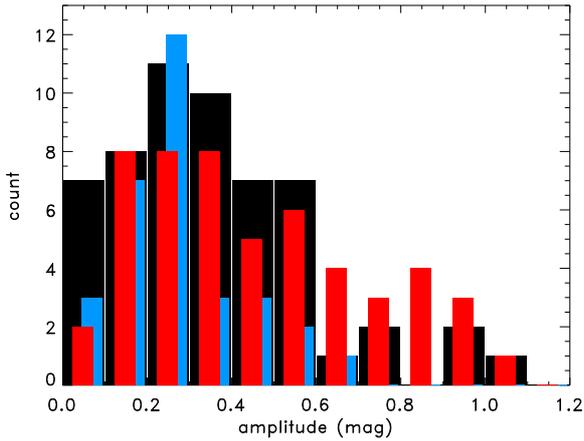}
\caption[]{Light curve amplitude distribution of Jovian Trojan asteroids. Black bars represent our original, uncorrected sample. The red and blue bars correspond the bias-corrected amplitudes of our sample and that in \citet{Binzel}.  \label{fig:amplhist}}
\end{figure}

In Fig.~\ref{fig:amplhist} we plotted the amplitude distribution of our Trojan asteroid sample (black bars) and compared it with the amplitude distribution obtained by \citet{Binzel} for a somewhat smaller sample. The magnitude limit of the detectability of periodic light curve variation for our sample is estimated to be $\Delta m_{min}$\,$\approx$\,0\fm02 that corresponds to a $a/b$ axis ratio of $(a/b)\,>10^{0.4\cdot \Delta m_{min}}\,=\,1.02$, if only shape effects are taken into account. We note again that light curve variations were detected for all asteroids in our sample, therefore our sample can be considered to be an unbiased sample, rather than the sample by \citet{Binzel}, where the targets were mostly Jovian Trojan asteroids for which the existence of detectable brightness variations were previously known. A selection bias for the \citet{Binzel} sample can also be inferred from the fact that a Kolmogorov-Smirnov test that compares the two cases gives a probability of only $\sim$7\% that the two samples are drawn from the same distribution.

Our light curves are certainly affected by the spin axis orientations. To test the impact of the geometry on the amplitude statistics, we have corrected the original amplitudes following the method given by \citet{}{Binzel}, i.e. choosing the largest value if multiple amplitudes are available in the literature \citet[using data from the Asteroid Light Curve Database][]{Warner}, and applying a correction assuming $\vartheta$\,=\,60\degr aspect angle in the case of single amplitudes.

%%%%
%%%%%

The higher frequency of large amplitudes in our sample with respect to main belt asteroids is even more pronounced than in the sample of \citet{Binzel} (see fig. 20 in their paper). \citet{Binzel} interpreted this as a higher number of elongated objects in the Trojan population, however, it is still unknown how the collisional history or other evolutionary effects can explain this deviation.

%%%%%%%%%%%%%%%%%%%%%%%%%%%%%
\section{Summary}
\label{sec:summary}

In this paper we investigated the Trojan L4 asteroids detections by K2 and got the following conclusions:

\begin{itemize}
\item{}The K2 sample shows a significant fraction ($\approx 20\%{}$) of very slow ($P>50$~h) rotation periods. The K2 sample is still unbiased in this period range, therefore this observation reflects the actual occurrence of very slow rotators.
\item{}The K2 sample shows a significant overabundance of large amplitude asteroids. 3 of 56 asteroids exceeded the 0\fm9 amplitude, and 5 ($\approx 10\%{}$) Trojans exceed 0\fm714, belonging to $a/c>2$ asphericity.
\item{}In the case of $(65227)~2002~ES_{\rm 46}$ we detected double periodicity, 49.7 and 3.53 hours and 0\fm3 and 0\fm4 amplitude, respectively. This is by far the largest asteroid known with double period.
\item{}The excess of large amplitude asteroids, the very slow rotators among Trojans, and the presence of double periods can all be explained by a high rate of binary asteroids in the L4 cloud. We estimated the occurrence of binarity between 20-25\%{}, in agreement with previous estimates.
\item{}Red and less red populations were found to be identical for light variation properties. The similarity applies for both amplitude and period distributions, and also, the presence of very slow rotators and high amplitude asteroids.
\item{}We detected a notable lack of fast rotators among Trojans. We interpreted it as the effect of a density barrier, and estimated the upper limit of the density of 0.5~g/cm$^3$ in the K2 sample, in agreement with previous estimates.
\item{}We derived the amplitude distribution of K2 Trojan asteroids and debiased assuming random spin orientation. Both the observed and unbiased distribution differ from the results of Binzel \& Sauter (2011) since K2 observed significantly more asteroids in the high amplitude wing of the distribution. How the collisional history or other evolutionary effects led to the excess of elongated bodies in the Trojan cloud still needs further investigations.
\end{itemize}

%%%%%%%%%%%%%%%%%%%%%%%%%%%%%%%%%%%%%%%%%%%%%%%%%%%%%%%%%%%%%%%%%%%%%%%%%%%%%%

\begin{acknowledgements}
Funding for the \textit{Kepler} and K2 missions is provided by the NASA Science Mission directorate. The authors acknowledge the Kepler team for the extra efforts to allocate special pixel masks to track moving targets. This work has been supported by the Lend\"ulet Programme of the Hungarian Academy of Sciences (LP2012-31, LP2014-17), by the OTKA K-109276 and K-104607, the GINOP-2.3.2-15-2016-00003 grant and NKFIH K-115709 and PD-116175 grants of the Hungarian National Research, Development and Innovation Office, a T\'eT-14FR-1-2015-0012 grant, and by the City of Szombathely under
agreement No.\ 67/177-21/2016. The research leading to these results has received
funding from the European Community’s Seventh Framework
Programme (FP7/2007-2013) under grant agreement
No.\ 312844 (SPACEINN),
the ESA PECS Contract Nos.\ 4000110889/14/NL/NDe and
4000109997/13/NL/KML, the
European Union’s Horizon 2020 Research and Innovation Programme, Grant Agreement no 687378.
L.M.\ was supported by the J\'anos
Bolyai Research Scholarship of the Hungarian Academy of
Sciences.

All of the data
presented in this paper were obtained from the Mikulski
Archive for Space Telescopes (MAST). STScI is operated
by the Association of Universities for Research in Astronomy,
Inc., under NASA contract NAS5-26555. Support for
MAST for non-HST data is provided by the NASA Office of
Space Science via grant NNX13AC07G and by other grants
and contracts.

\end{acknowledgements}

%%%%%%%%%%%%%%%%%%%%%%%%%%%%%%%%%%%%%%%%%%%%%%%%%%%%%%%%%%%%%%%%%%%%%%%%%%%%%%

{}

\newpage

%%%%%%%%%%%%%%%%%%%%%%%%%%%%%%%%%%%%%%%%%%%%%%%%%%%%%%%%%%%%%%%%%%%%%%%%%%%%%%%%%%555
\begin{appendix}
\section{Figures and tables}
%% %% %% %% %% %% %% %% %% %% %% %% %% %% %% %% %% %% %% %% %% %% %% %% %% %%
\begin{table}[h]
\caption{Summary of the observational characteristics of the 56 Trojans
observed by K2 during Campaign 6. The columns are the following:
a) MPC designation of the asteroid;
b) the average apparent trail length (in pixels) of the object on the K2 CCD frames;
c) the variations of the apparent trail length throughout
the observations;
d) the number of light curve points used in the further analysis; and
e) the total time coverage of the light curves in days. }
\label{table:keplerobs}
\begin{center}\begin{tabular}{lrrrr}
\hline
$^a$Object & $^b$Trail & $^c$Trail & $^d$LC & $^e$Time  \\
           & length    &     var.  & points &cov.\\
(MPC number) & (pixel) & (pixel) & & (days) \\
\hline
(1143) Odysseus                          & 2.05 & 0.11 &  375 &    9.52 \\
(1749) Telamon                           & 0.36 & 0.11 &  629 &   20.09 \\
(3801) Thrasymedes                       & 1.74 & 0.11 &  367 &   10.40 \\
(4035) 1986\,WD                          & 0.84 & 0.10 &  442 &   10.42 \\
(4057) Demophon                          & 0.26 & 0.14 &  871 &   20.25 \\
(4138) Kalchas                           & 0.66 & 0.15 &  489 &   10.46 \\
(5028) Halaesus                          & 0.84 & 0.09 &  466 &   10.48 \\
(5123) 1989\,BL                          & 1.11 & 0.13 &  461 &   10.46 \\
(5244) Amphilochos                       & 0.32 & 0.10 &  808 &   18.12 \\
(5436) Eumelos                           & 0.79 & 0.12 &  444 &   10.40 \\
(5652) Amphimachus                       & 0.31 & 0.12 &  747 &   19.96 \\
(8241) Agrius                            & 1.84 & 0.13 &  461 &   10.38 \\
(9807) 1997\,SJ\ensuremath{_{4}}         & 0.59 & 0.13 &  383 &   10.46 \\
(10989) Dolios                           & 0.25 & 0.14 &  860 &   19.84 \\
(11251) Icarion                          & 0.37 & 0.10 &  620 &   14.32 \\
(12238) Actor                            & 1.17 & 0.10 &  488 &   10.65 \\
(12974) Halitherses                      & 2.02 & 0.11 &  456 &   10.40 \\
(13184) Augeias                          & 1.01 & 0.14 &  435 &   10.36 \\
(13185) Agasthenes                       & 0.76 & 0.12 &  406 &    9.01 \\
(13331) 1998\,SU\ensuremath{_{52}}       & 1.13 & 0.14 &  432 &   10.40 \\
(13366) 1998\,US\ensuremath{_{24}}       & 1.55 & 0.12 &  387 &   10.05 \\
(13372) 1998\,VU\ensuremath{_{6}}        & 1.71 & 0.11 &  430 &    9.46 \\
(13379) 1998\,WX\ensuremath{_{9}}        & 0.67 & 0.13 &  449 &   10.52 \\
(14690) 2000\,AR\ensuremath{_{25}}       & 0.27 & 0.14 &  894 & 1 20.54 \\
(14791) Atreus                           & 0.22 & 0.14 &  514 &   11.42 \\
(15529) 2000\,AA\ensuremath{_{80}}       & 1.86 & 0.13 &  432 &   10.20 \\
(16152) 1999\,YN\ensuremath{_{12}}       & 1.00 & 0.13 &  475 &   10.26 \\
(21593) 1998\,VL\ensuremath{_{27}}       & 0.30 & 0.04 &  235 &    5.09 \\
(21599) 1998\,WA\ensuremath{_{15}}       & 0.42 & 0.08 &  888 &   20.04 \\
(22056) 2000\,AU\ensuremath{_{31}}       & 0.30 & 0.11 &  833 &   20.35 \\
(23939) 1998\,TV\ensuremath{_{33}}       & 1.65 & 0.12 &  453 &   10.42 \\
(23947) 1998\,UH\ensuremath{_{16}}       & 1.61 & 0.14 &  437 &   10.16 \\
(23958) 1998\,VD\ensuremath{_{30}}       & 0.90 & 0.04 &  767 &   19.80 \\
(24357) 2000\,AC\ensuremath{_{115}}      & 0.54 & 0.06 &  812 &   20.00 \\
(24534) 2001\,CX\ensuremath{_{27}}       & 1.04 & 0.13 &  446 &    9.50 \\
(24537) 2001\,CB\ensuremath{_{35}}       & 0.34 & 0.06 &  641 &   16.00 \\
(35363) 1997\,TV\ensuremath{_{28}}       & 0.63 & 0.09 &  335 &    7.78 \\
(39270) 2001\,AH\ensuremath{_{11}}       & 0.42 & 0.09 &  781 &   19.70 \\
(38574) 1999\,WS\ensuremath{_{4}}        & 0.53 & 0.06 &  682 &   16.61 \\
(39286) 2001\,CX\ensuremath{_{6}}        & 0.69 & 0.08 &  273 &    6.23 \\
(39289) 2001\,CT\ensuremath{_{28}}       & 0.25 & 0.12 &  707 &   17.31 \\
(57041) 2001\,EN\ensuremath{_{12}}       & 0.58 & 0.06 &  682 &   19.98 \\
(58480) 1996\,RJ\ensuremath{_{33}}       & 0.75 & 0.14 &  451 &   10.44 \\
(59049) 1998\,TC\ensuremath{_{31}}       & 0.90 & 0.04 &  851 &   19.96 \\
(63239) 2001\,BD\ensuremath{_{25}}       & 0.49 & 0.08 &  558 &   13.83 \\
(65210) Stichius                         & 0.42 & 0.09 &  719 &   17.23 \\
(65223) 2002\,EU\ensuremath{_{34}}       & 0.29 & 0.12 &  620 &   14.36 \\
(65227) 2002\,ES\ensuremath{_{46}}       & 0.37 & 0.15 &  440 &   10.18 \\
(65240) 2002\,EU\ensuremath{_{106}}      & 0.53 & 0.06 &  596 &   16.43 \\
(65257) 2002\,FU\ensuremath{_{36}}       & 0.92 & 0.13 &  427 &   10.26 \\
(83984) 2002\,GL\ensuremath{_{77}}       & 0.68 & 0.12 &  387 &    9.99 \\
(88227) 2001\,BU\ensuremath{_{42}}       & 0.59 & 0.06 &  744 &   20.35 \\
(88241) 2001\,CD\ensuremath{_{23}}       & 0.40 & 0.09 &  793 &   18.96 \\
(129602) 1997\,WA\ensuremath{_{12}}      & 1.04 & 0.11 &  452 &   10.40 \\
(228102) 2008\,SY\ensuremath{_{172}}     & 0.82 & 0.05 &  435 &   10.52 \\
\hline
\end{tabular}\end{center}\vspace*{-3mm}
\end{table}
%% %% %% %% %% %% %% %% %% %% %% %% %% %% %% %% %% %% %% %% %% %% %% %% %% %%

\begin{table}[h]
\caption{ }
\label{table:results}
\begin{center}\begin{tabular}{rrrrc}
\hline
$^a$Object & Period & Period & Amplitude & Population \\
           &        & error  &  \\
Number     & (h)  & (h)    & (mag) \\
\hline
1143 &  10.079 & 0.194 & 0.214 \\
1749 &  22.662 & 0.193 & 0.059 & r\\
3801 &  20.270 & 0.672 & 0.138 \\
4035 &  13.475 & 0.156 & 0.181 \\
4057 &  29.925 & 0.765 & 0.209 \\
4138 &  29.411 & 2.001 & 0.102 \\
5028 &  25.052 & 1.091 & 0.225 & r\\
5123 &  19.800 & 0.140 & 0.44 \\
5244 &  19.566 & 0.088 & 0.731 \\
5436 &  21.276 & 0.315 & 0.442 \\
5652 &   8.374 & 0.112 & 0.222 \\
8241 &  17.902 & 0.230 & 0.291 & lr \\
9807 & 331.034 &117.56 & 0.428 \\
10989 & 26.101 & 0.286 & 0.324 \\
11251 & 10.448 & 0.068 & 0.313 \\
12238 &  7.281 & 0.041 & 0.294 \\
12974 &  6.971 & 0.030 & 0.515 \\
13184 & 11.934 & 0.119 & 0.149 \\
13185 & 11.453 & 0.124 & 0.033 \\
13331 &180.451 &31.938 & 0.444 & lr\\
13366 &400     &105.26 & 0.228 \\
13372 & 20.176 & 0.327 & 0.362 \\
13379 & 13.698 & 0.138 & 0.11 \\
14690 &  8.519 & 0.030 & 0.208 \\
14791 & 19.615 & 0.259 & 0.371 \\
15529 &375     &91.019 & 0.592 \\
16152 & 11.477 & 0.223 & 0.747  & lr \\
21593 & 10.747 & 0.196 & 0.991 \\
21599 & 12.651 & 0.121 & 0.489 & r\\
22056 &358.208 &16.791 & 0.976 & r\\
23939 & 12.868 & 0.104 & 0.552  & lr\\
23947 & 15.335 & 0.506 & 0.066 \\
23958 &1142.85 &154.44 & 0.358 \\
24357 &131.868 & 0     & 0.385 & r\\
24534 & 19.417 & 0.368 & 0.147 \\
24537 & 10.489 & 0.064 & 0.34 \\
35363 & 18.779 & 0.374 & 0.328 \\
38574 &  7.085 & 0.084 & 0.219 \\
39270 & 82.191 & 4.450 & 0.144 \\
39286 &  8.421 & 0.065 & 0.5 \\
39289 & 19.062 & 0.160 & 1.04 \\
57041 &  8.562 & 0.027 & 0.516 \\
58480 & 32.967 & 1.815 & 0.143 \\
59049 &  6.172 & 0.012 & 0.686 & r\\
63239 & 60.075 & 1.940 & 0.537 \\
65210 & 35.087 & 0.626 & 0.408 \\
65223 &252.631 &67.368 & 0.44 \\
65227 &  7.066 & 0.025 & 0.074 \\
65240 &230.769 &54.945 & 0.09 \\
65257 & 17.863 & 0.235 & 0.5 \\
83984 &  5.752 & 0.037 & 0.085 \\
88227 &  6.355 & 0.018 & 0.258 \\
88241 &  5.734 & 0.013 & 0.099 \\
353363 & 5.403 & 0.020 & 0.356 \\
129602 &43.010 & 1.188 & 0.362 & r\\
228102 & 7.259 & 0.039 & 0.248 \\
\hline
\end{tabular}\end{center}\vspace*{-3mm}
\end{table}
%% %% %% %% %% %% %% %% %% %% %% %% %% %% %% %% %% %% %% %% %% %% %% %% %% %%

%% %% %% %% %% %% %% %% %% %% %% %% %% %% %% %% %% %% %% %% %% %% %% %% %% %%

%\newpage

%% %% %% %% %% %% %% %% %% %% %% %% %% %% %% %% %% %% %% %% %% %% %% %% %% %%

%\newpage

\begin{figure*}[h]
\begin{center}
\begin{tabular}{ccc}
1143 & 1749 & 3801\\
\includegraphics[viewport=5 40 510 355,clip,height=3.7cm]{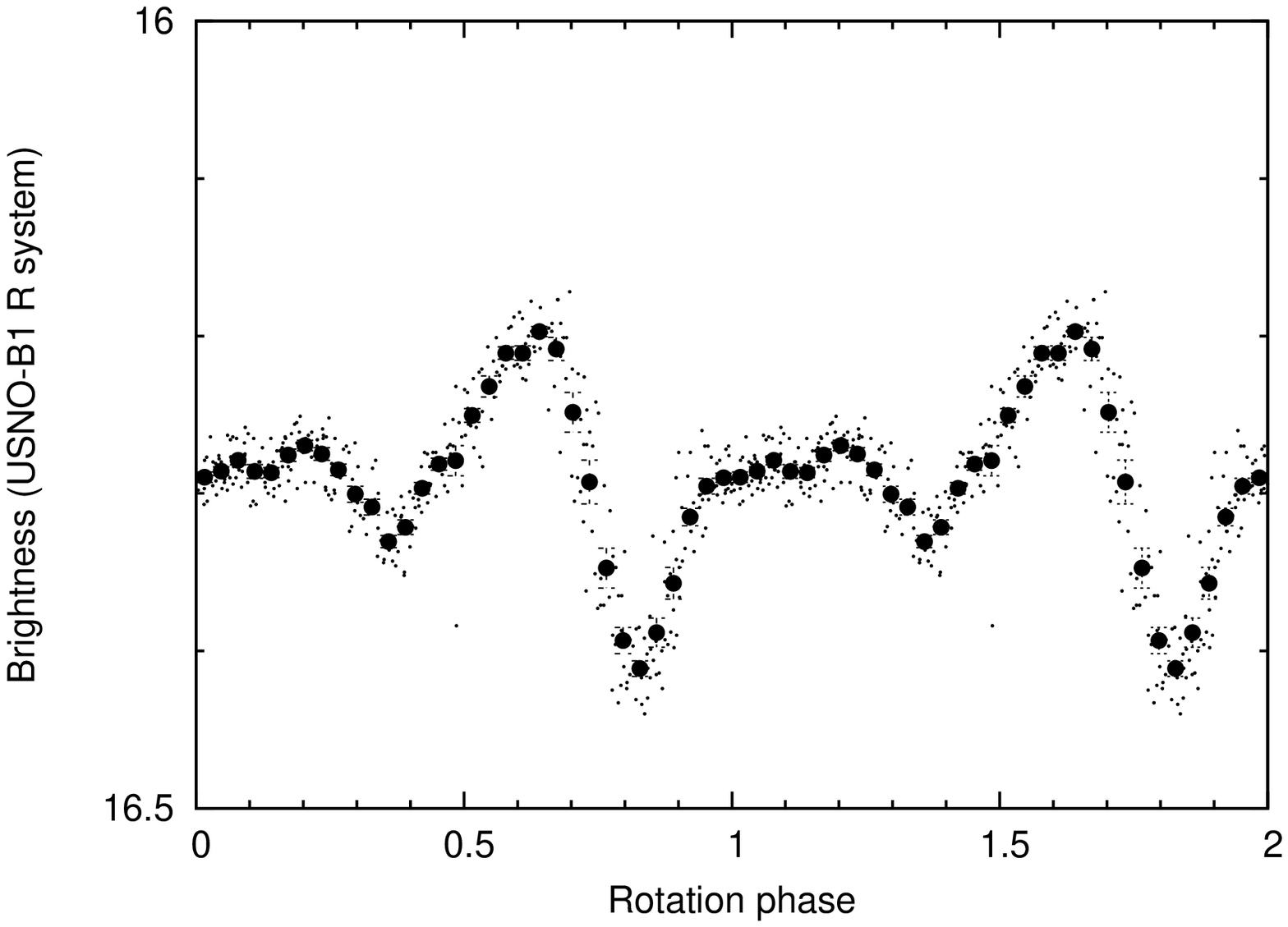}&
\includegraphics[viewport=5 40 510 355,clip,height=3.7cm]{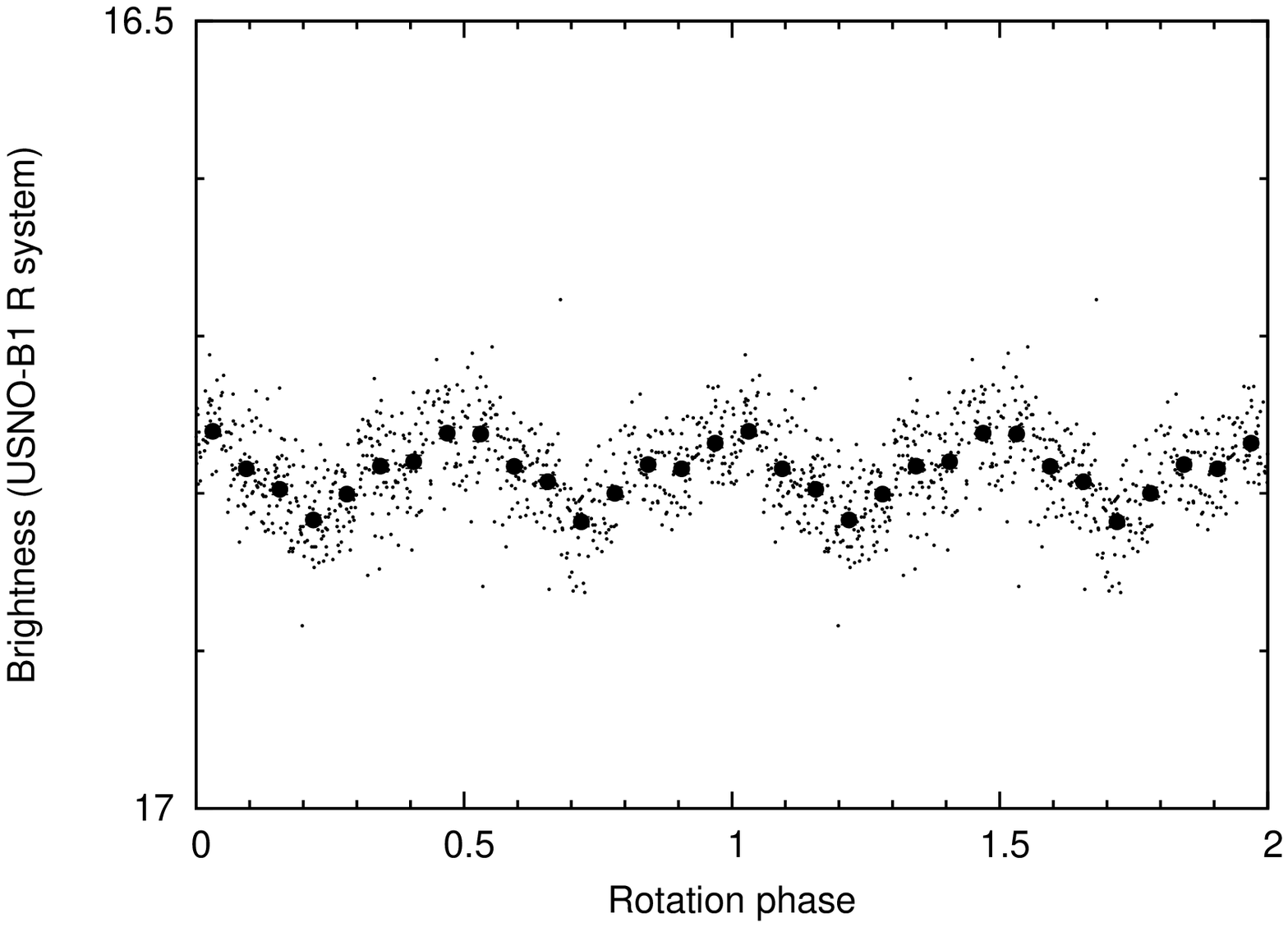}&
\includegraphics[viewport=5 40 510 355,clip,height=3.7cm]{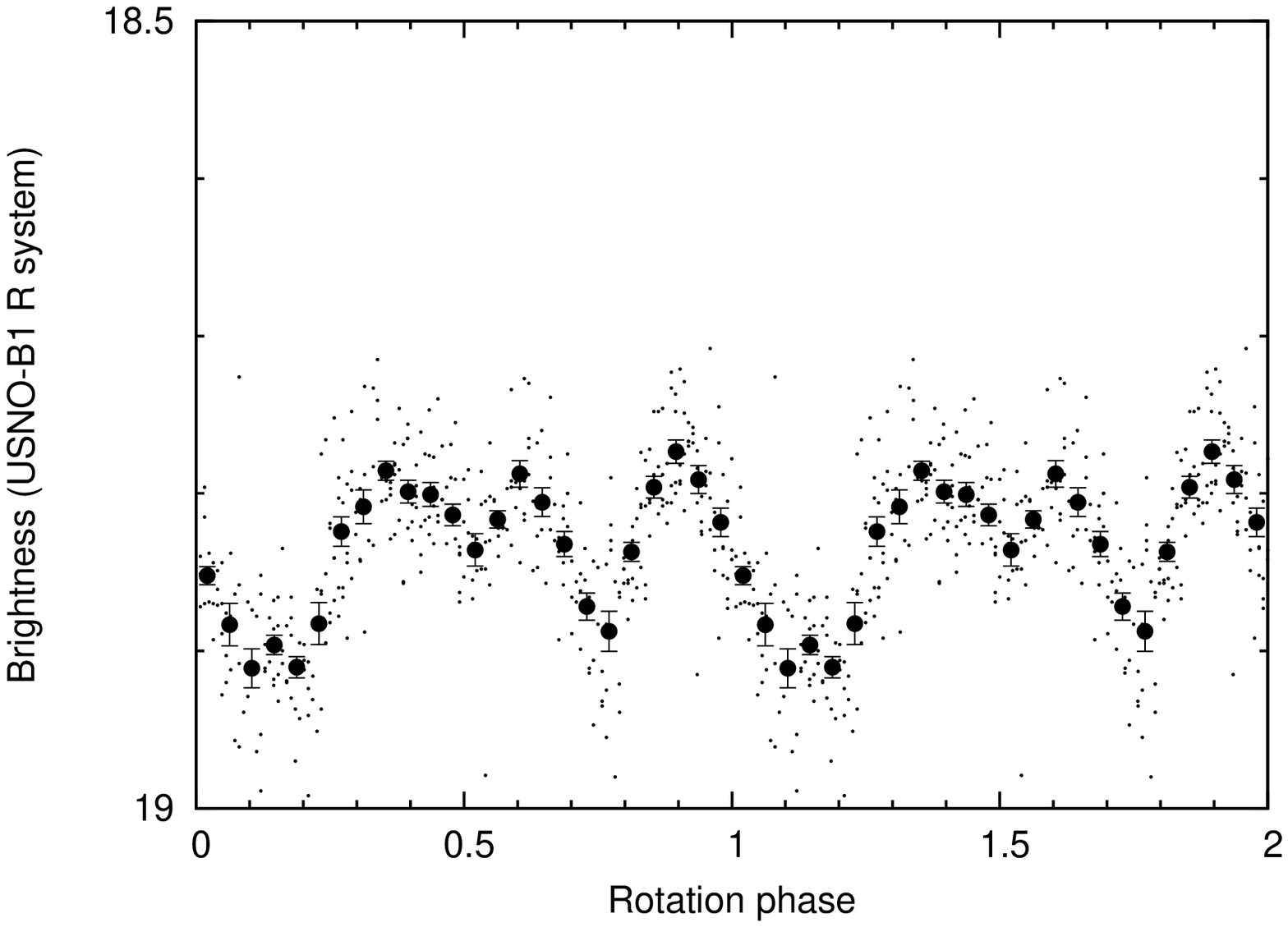}\\
4035 & 4057 & 4138\\
\includegraphics[viewport=5 40 510 355,clip,height=3.7cm]{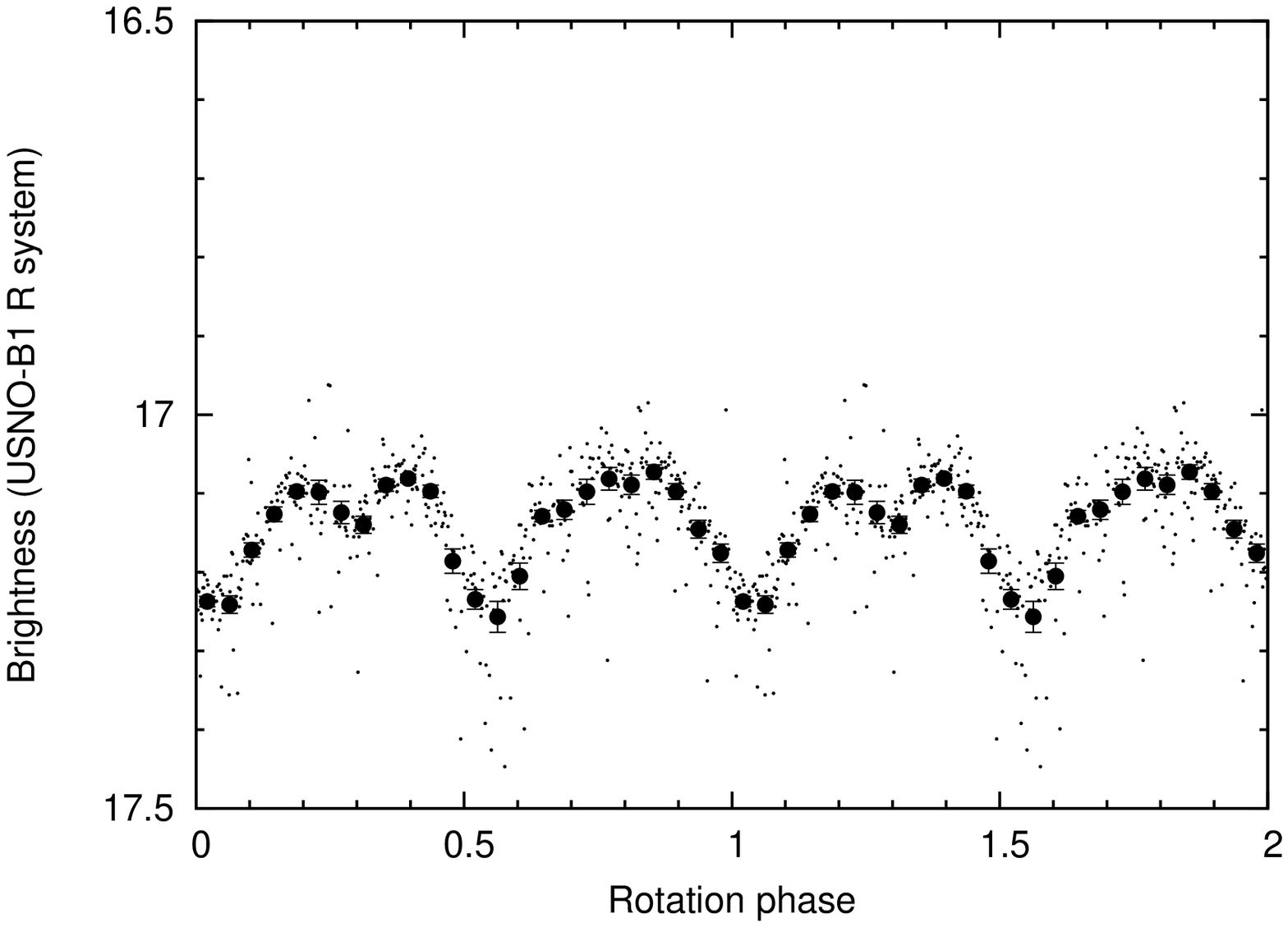}&
\includegraphics[viewport=5 40 510 355,clip,height=3.7cm]{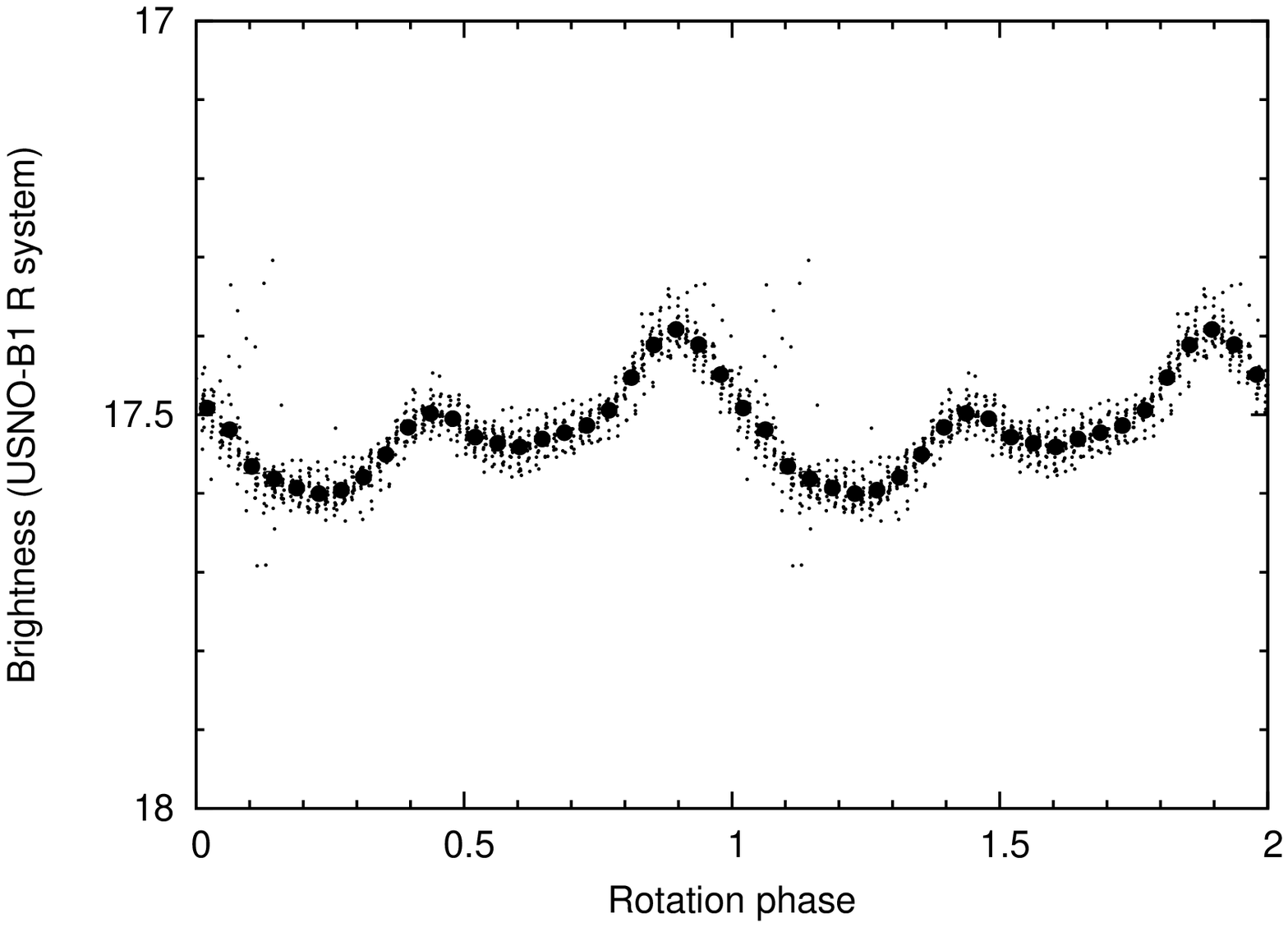}&
\includegraphics[viewport=5 40 510 355,clip,height=3.7cm]{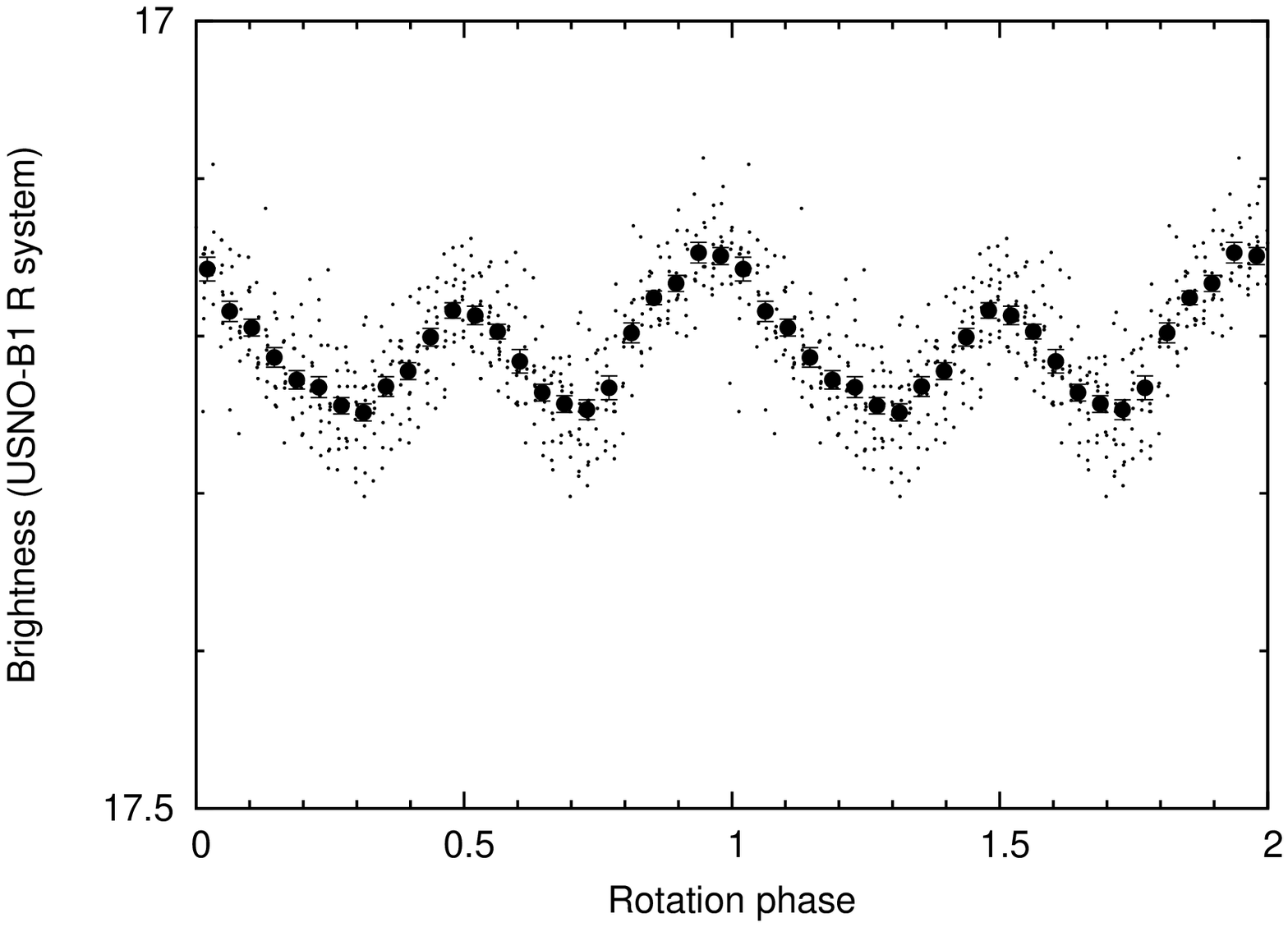}\\
5028 & 5123 & 5244\\
\includegraphics[viewport=5 40 510 355,clip,height=3.7cm]{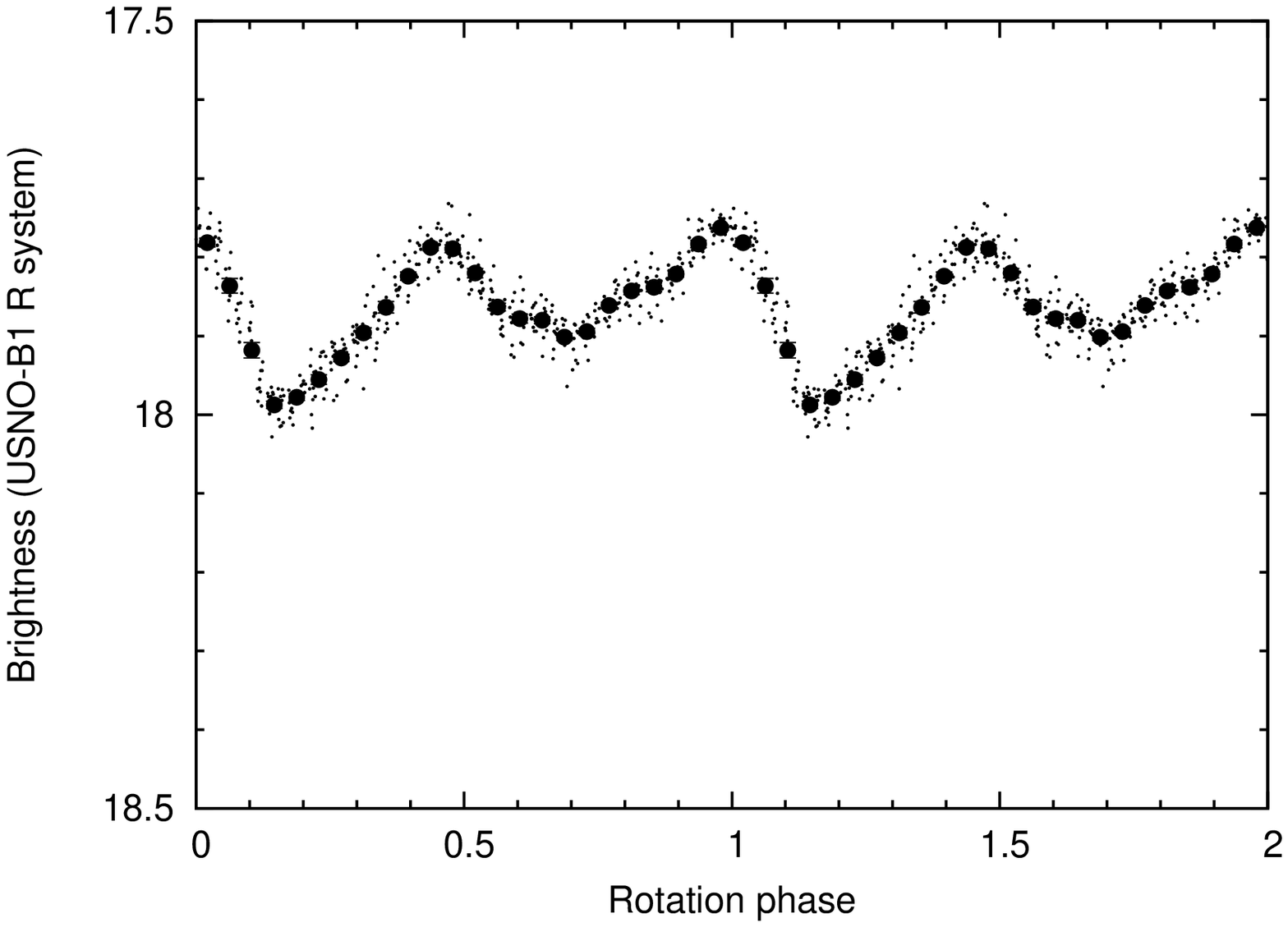}&
\includegraphics[viewport=5 40 510 355,clip,height=3.7cm]{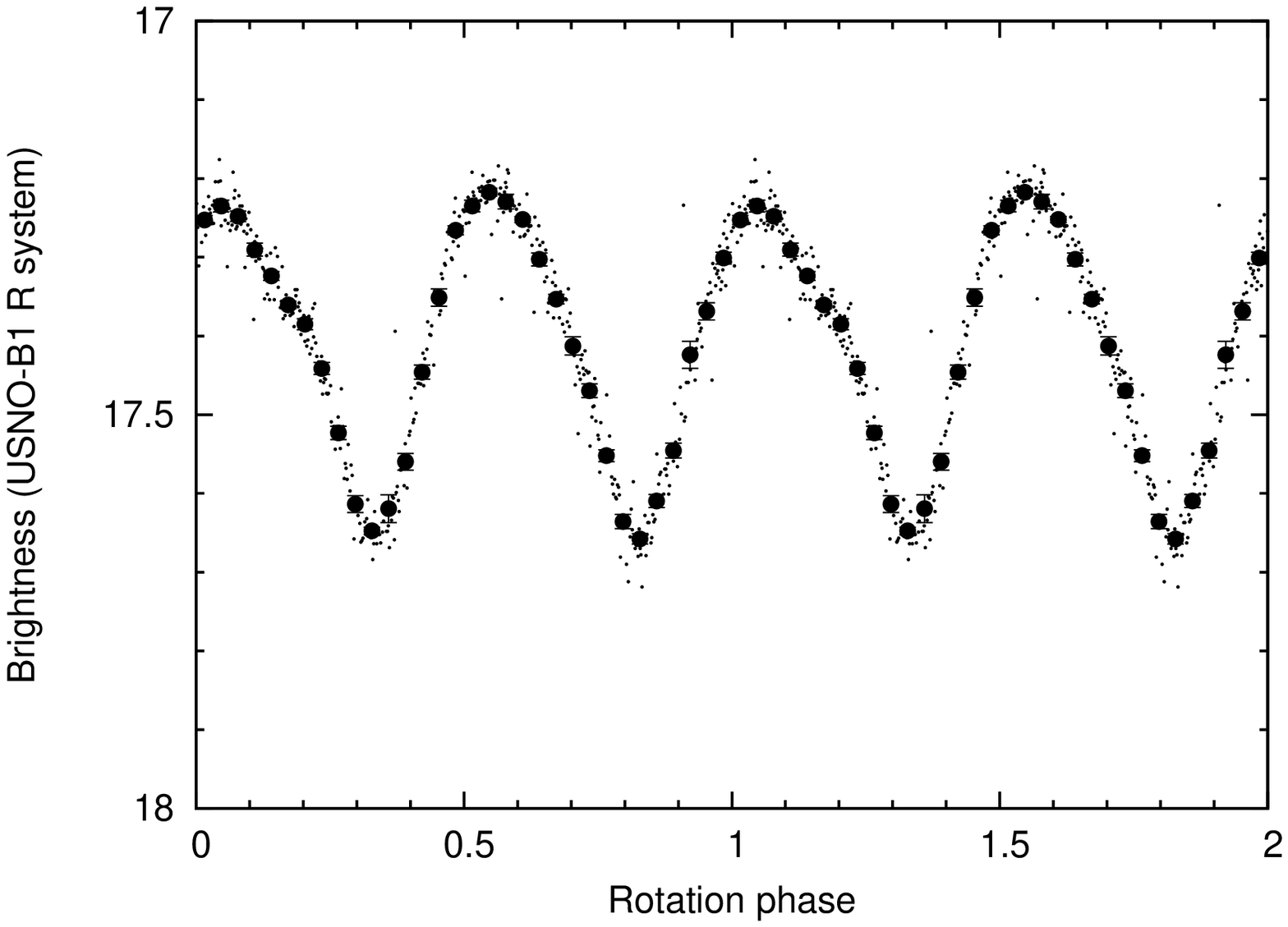}&
\includegraphics[viewport=5 40 510 355,clip,height=3.7cm]{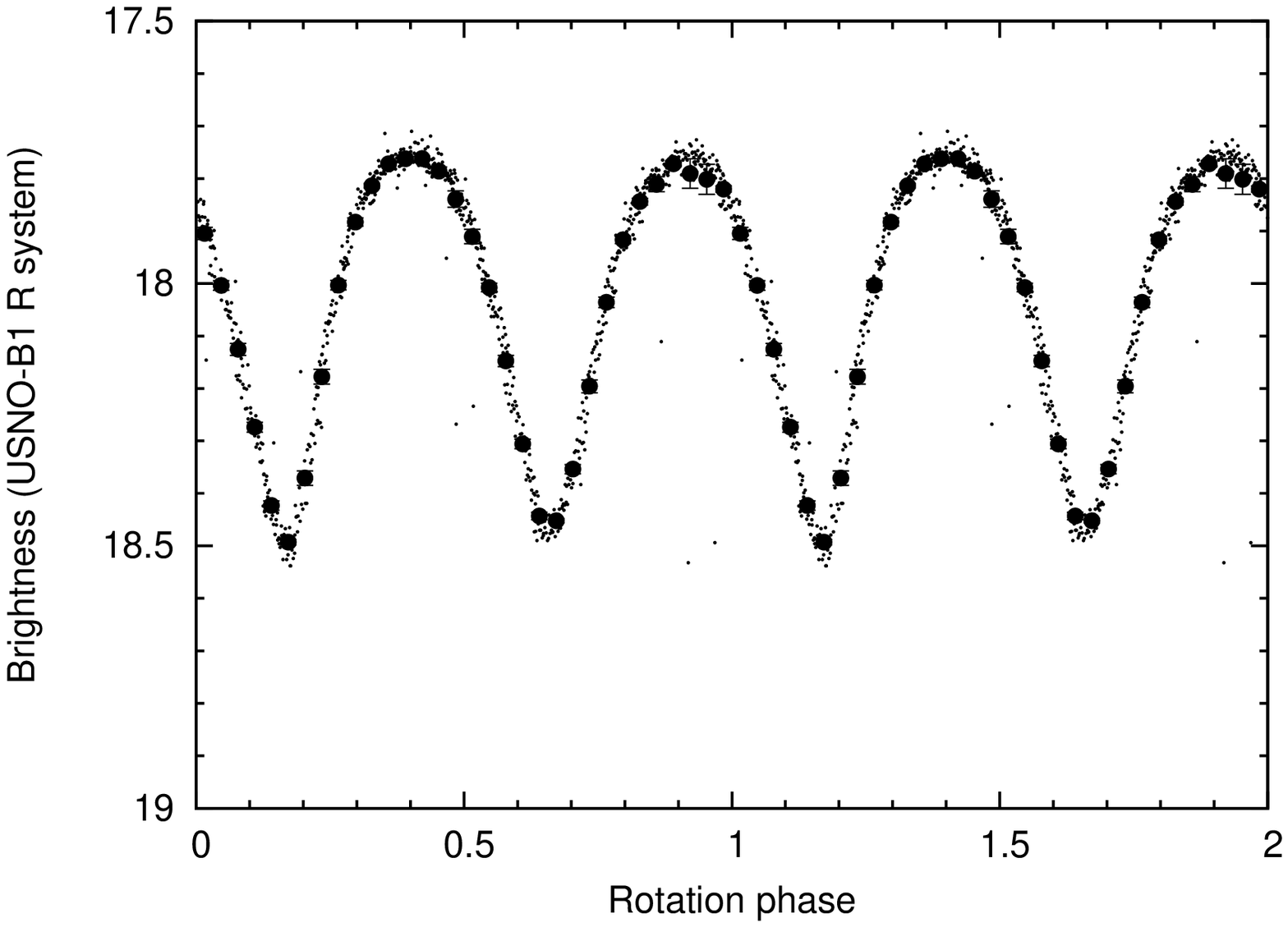}\\
5436 & 5652 & 8241\\
\includegraphics[viewport=5 40 510 355,clip,height=3.7cm]{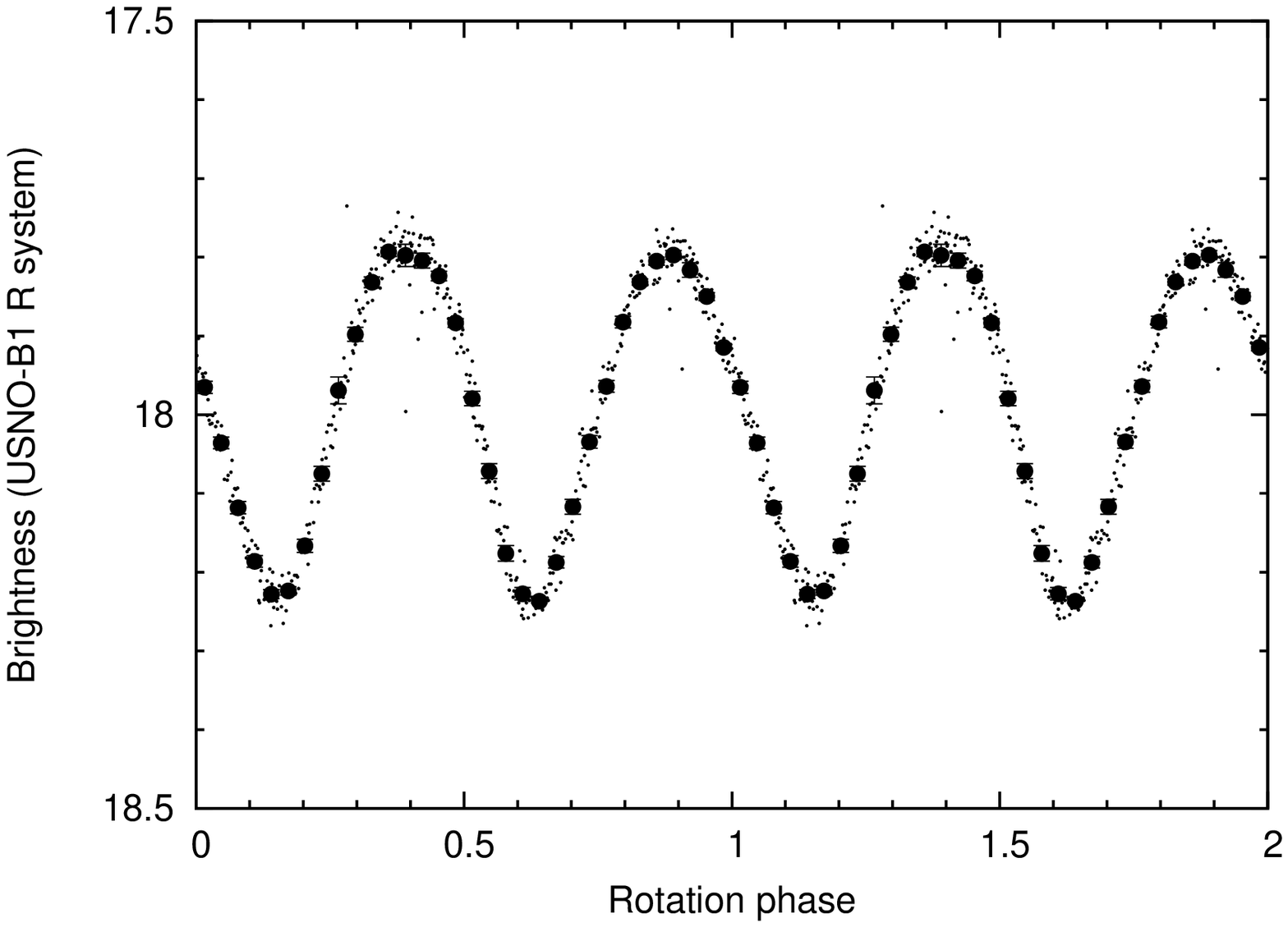}&
\includegraphics[viewport=5 40 510 355,clip,height=3.7cm]{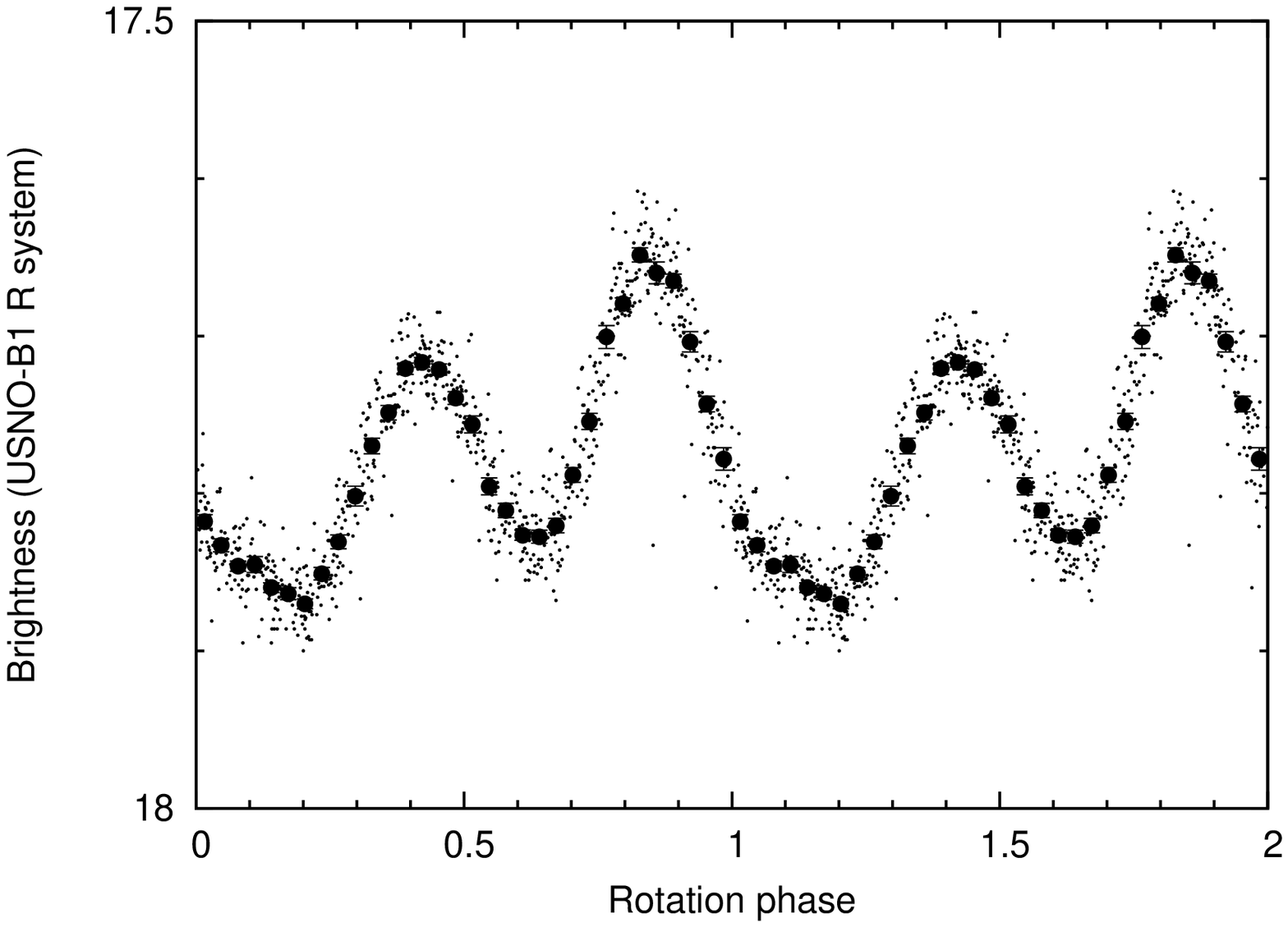}&
\includegraphics[viewport=5 40 510 355,clip,height=3.7cm]{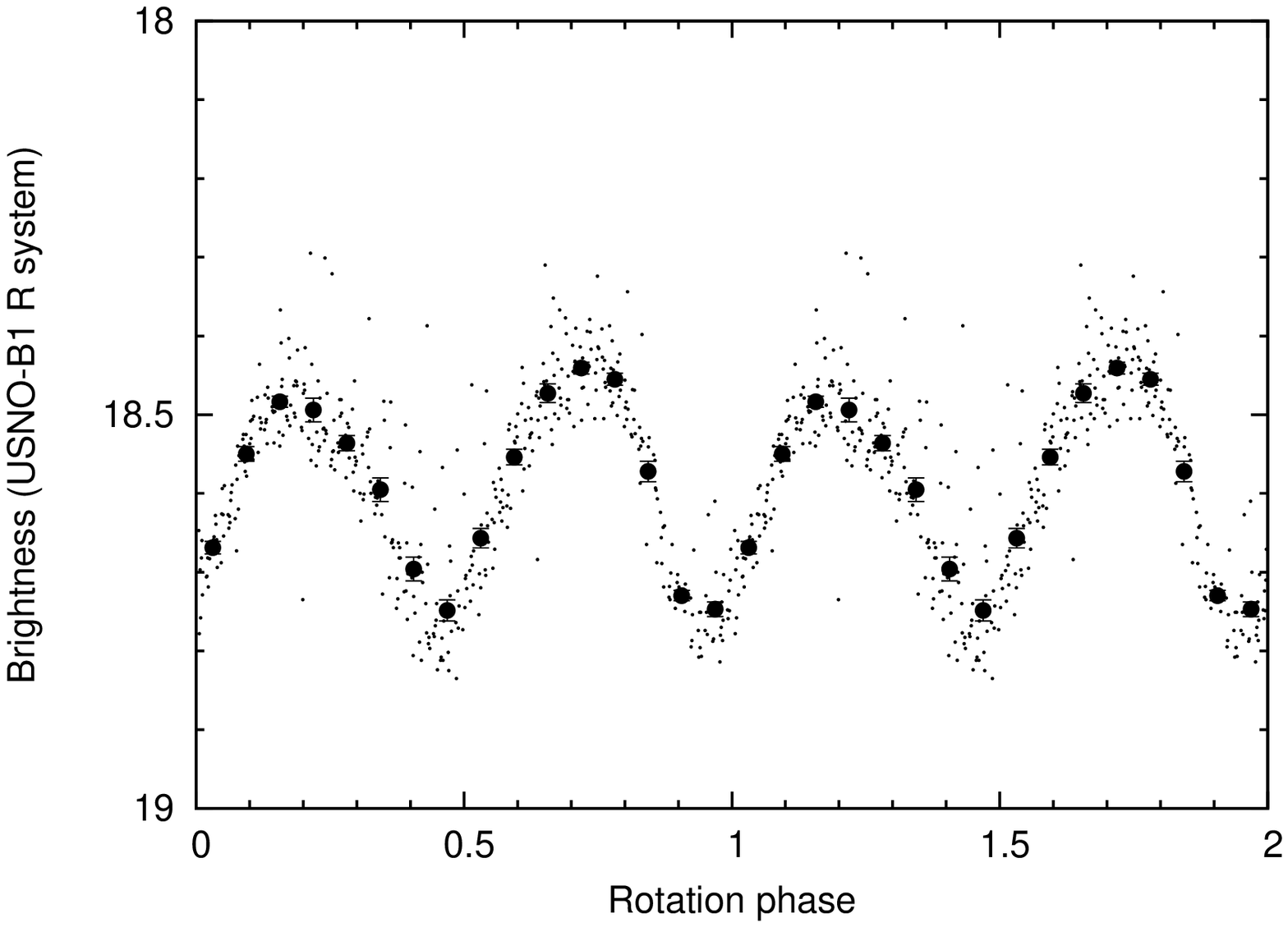}\\
9807 & 10989 & 11251\\
\includegraphics[viewport=5 40 510 355,clip,height=3.7cm]{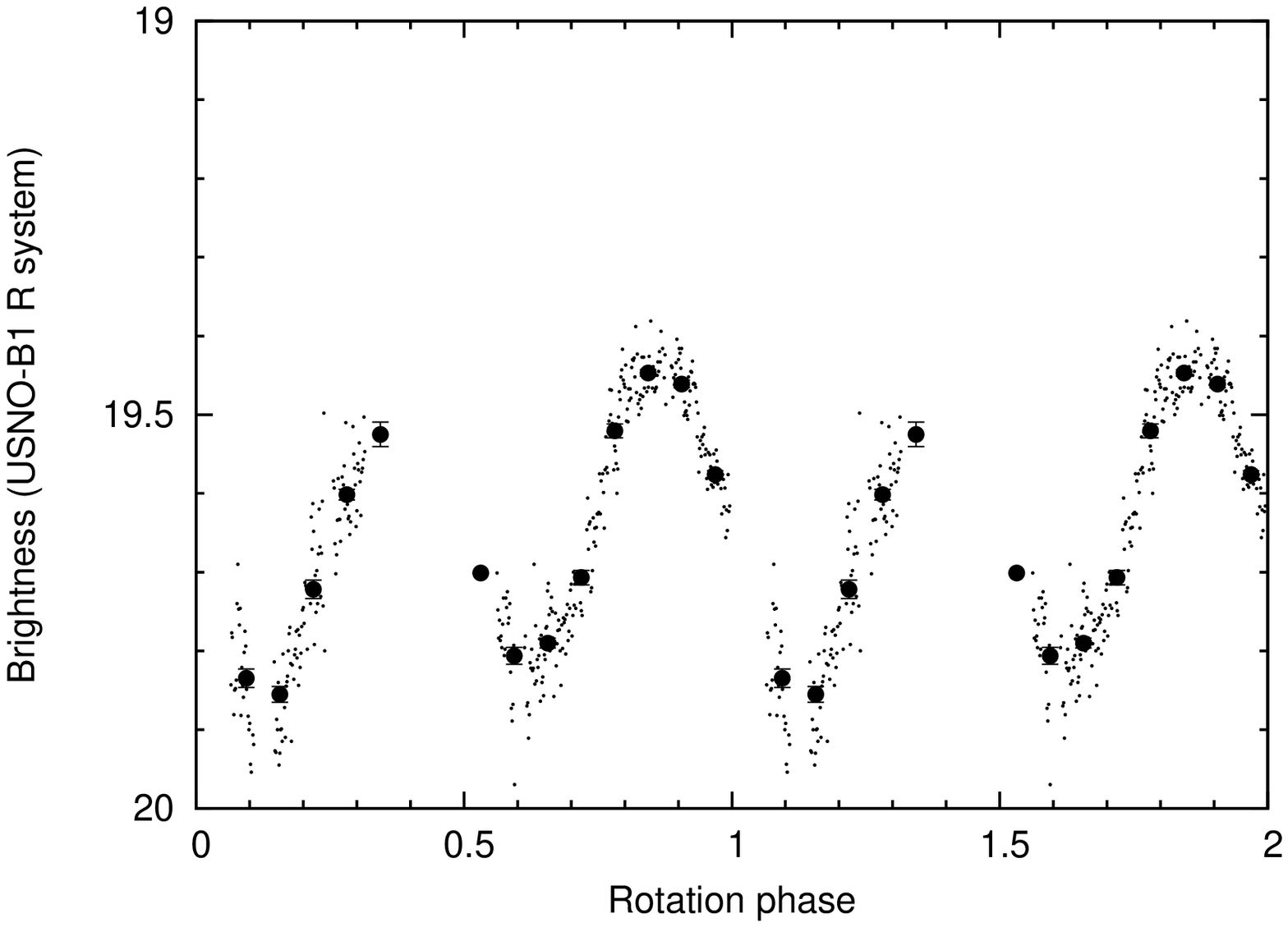}&
\includegraphics[viewport=5 40 510 355,clip,height=3.7cm]{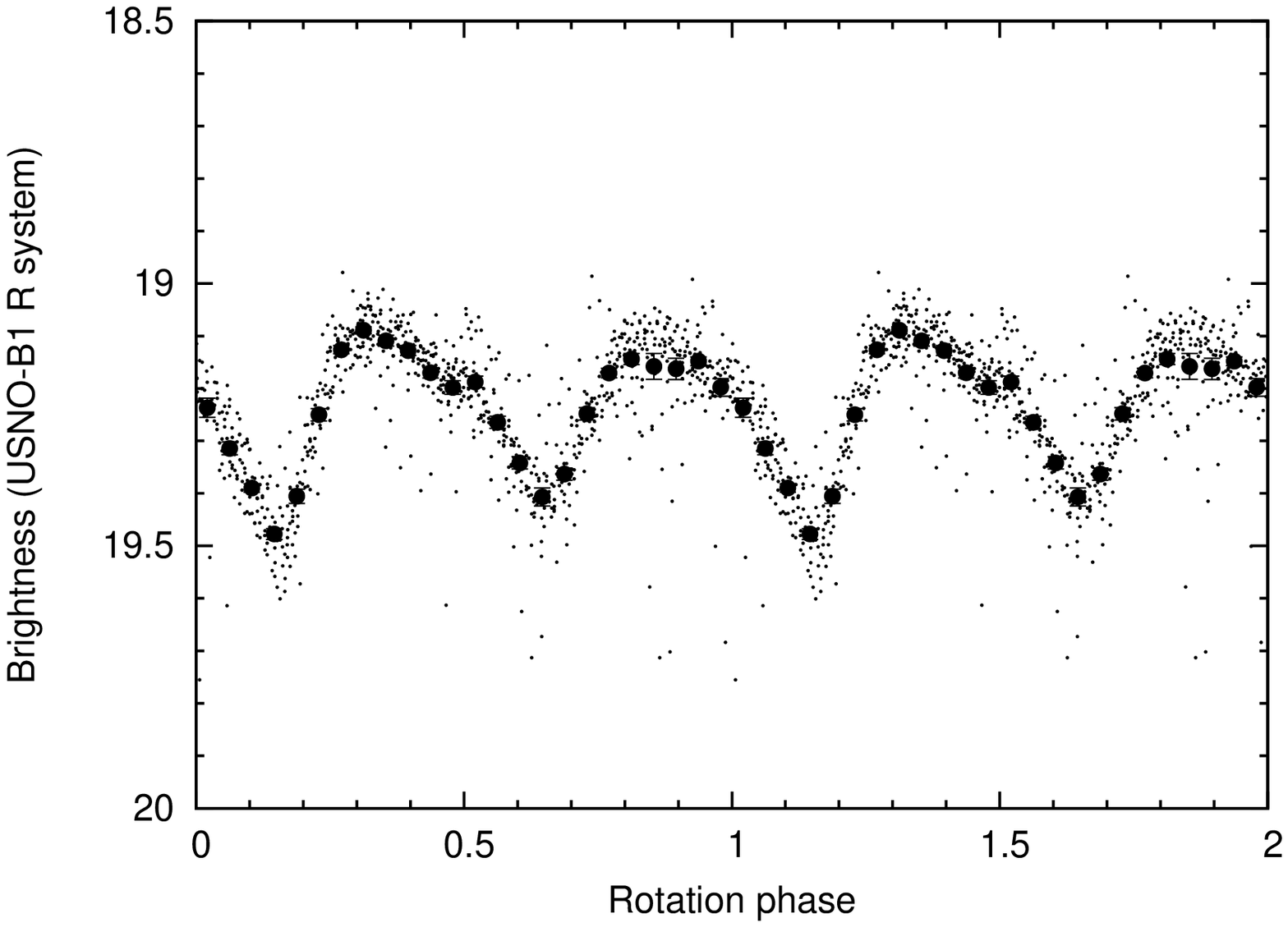}&
\includegraphics[viewport=5 40 510 355,clip,height=3.7cm]{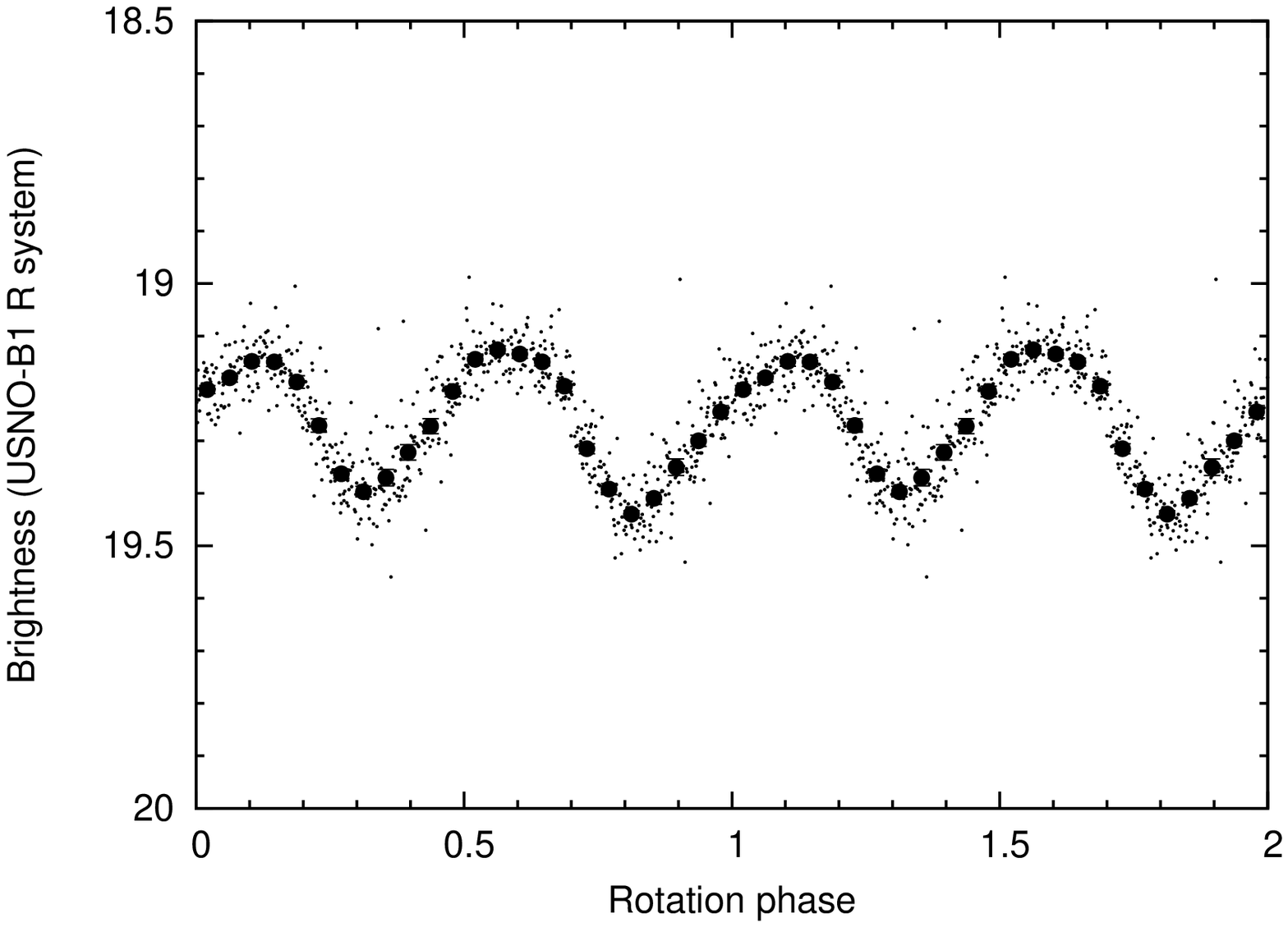}\\
\end{tabular}
\end{center}
\caption{The phased light curves of the Trojan asteroids in K2 Field 6.}
\end{figure*}

\begin{figure*}[h]
\begin{center}
\begin{tabular}{ccc}
12238 & 12974 & 13184 \\
\includegraphics[viewport=5 40 510 355,clip,height=3.7cm]{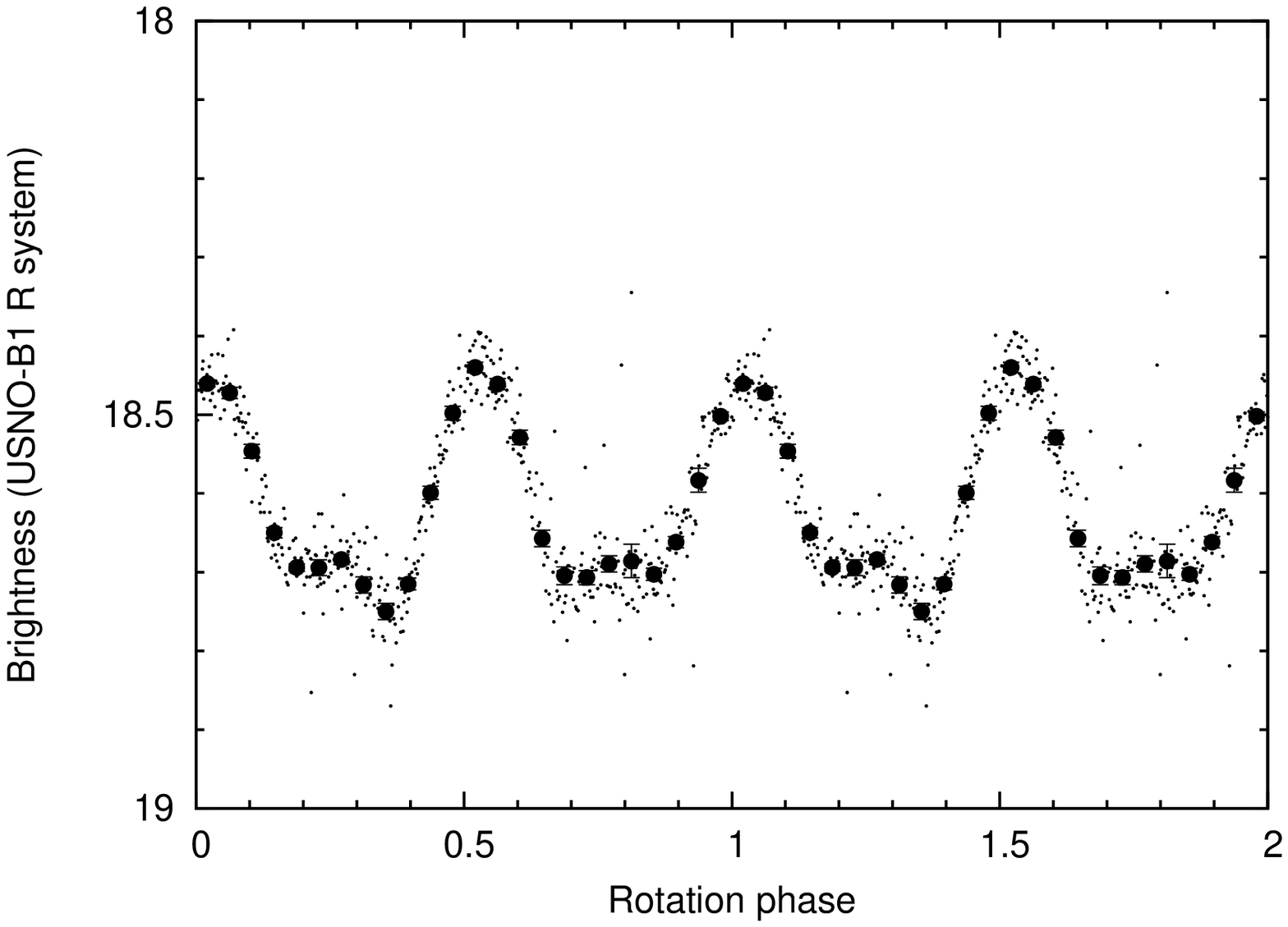}&
\includegraphics[viewport=5 40 510 355,clip,height=3.7cm]{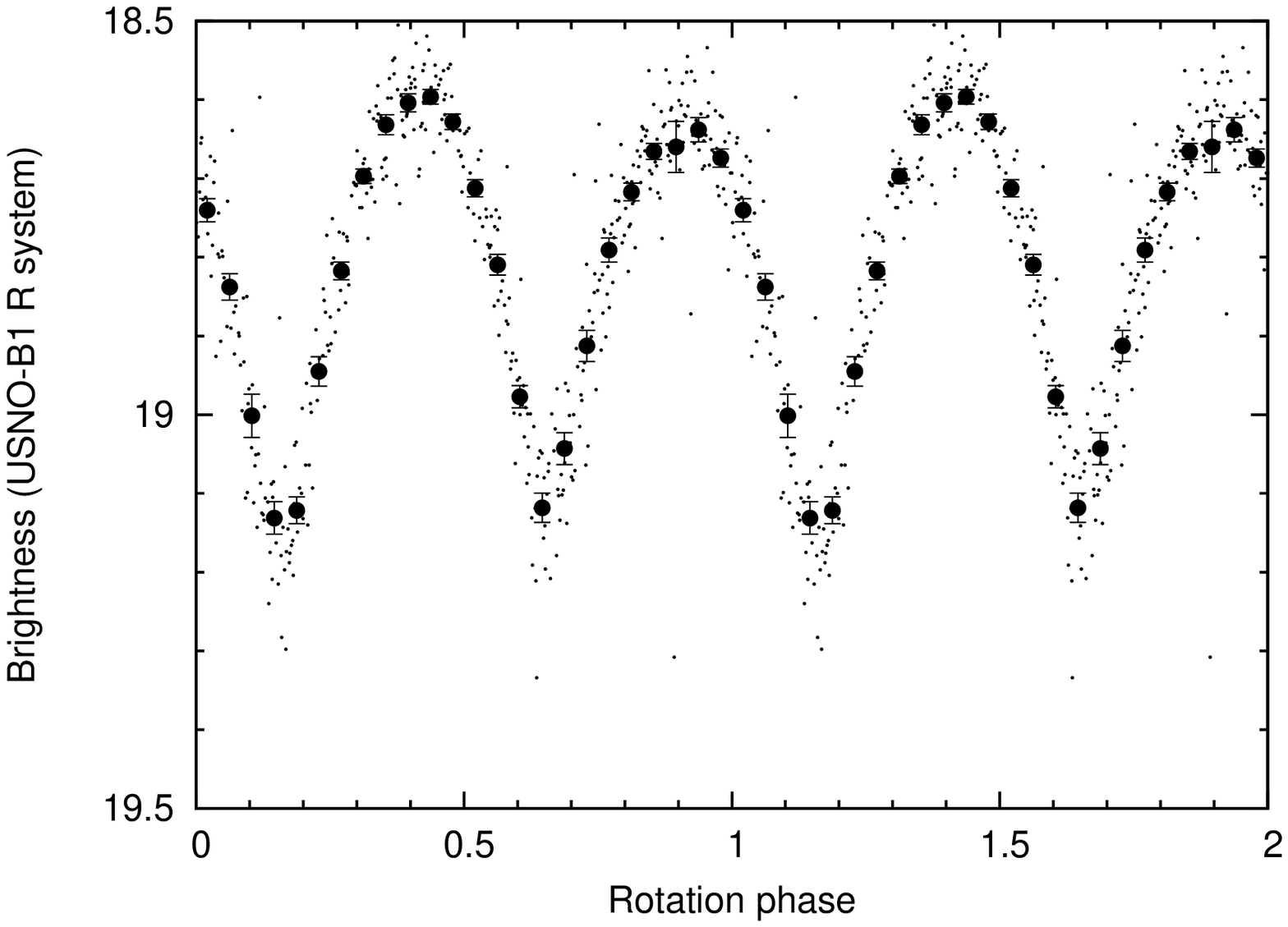}&
\includegraphics[viewport=5 40 510 355,clip,height=3.7cm]{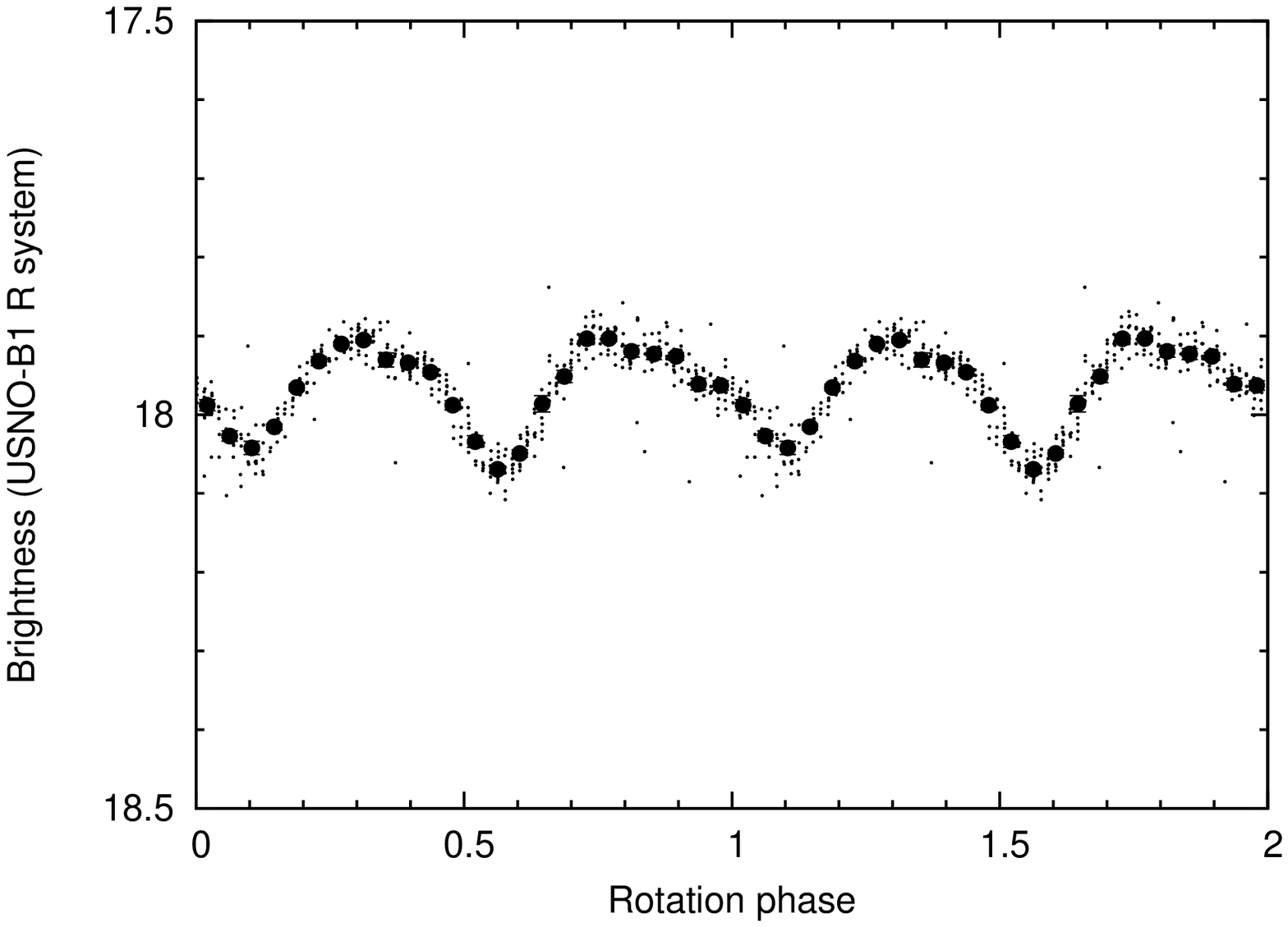}\\
13185 & 13331 & 13366\\
\includegraphics[viewport=5 40 510 355,clip,height=3.7cm]{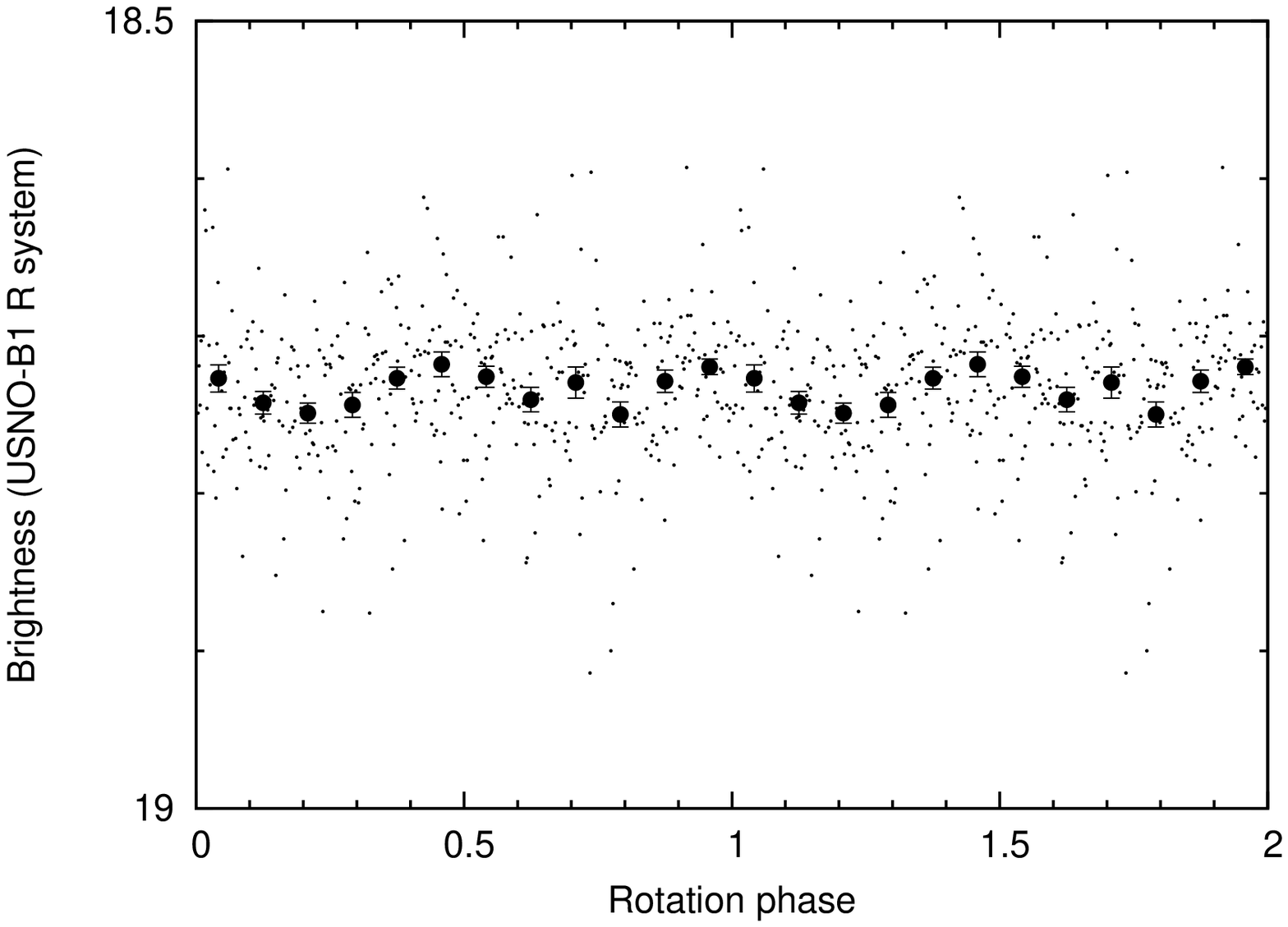}&
\includegraphics[viewport=5 40 510 355,clip,height=3.7cm]{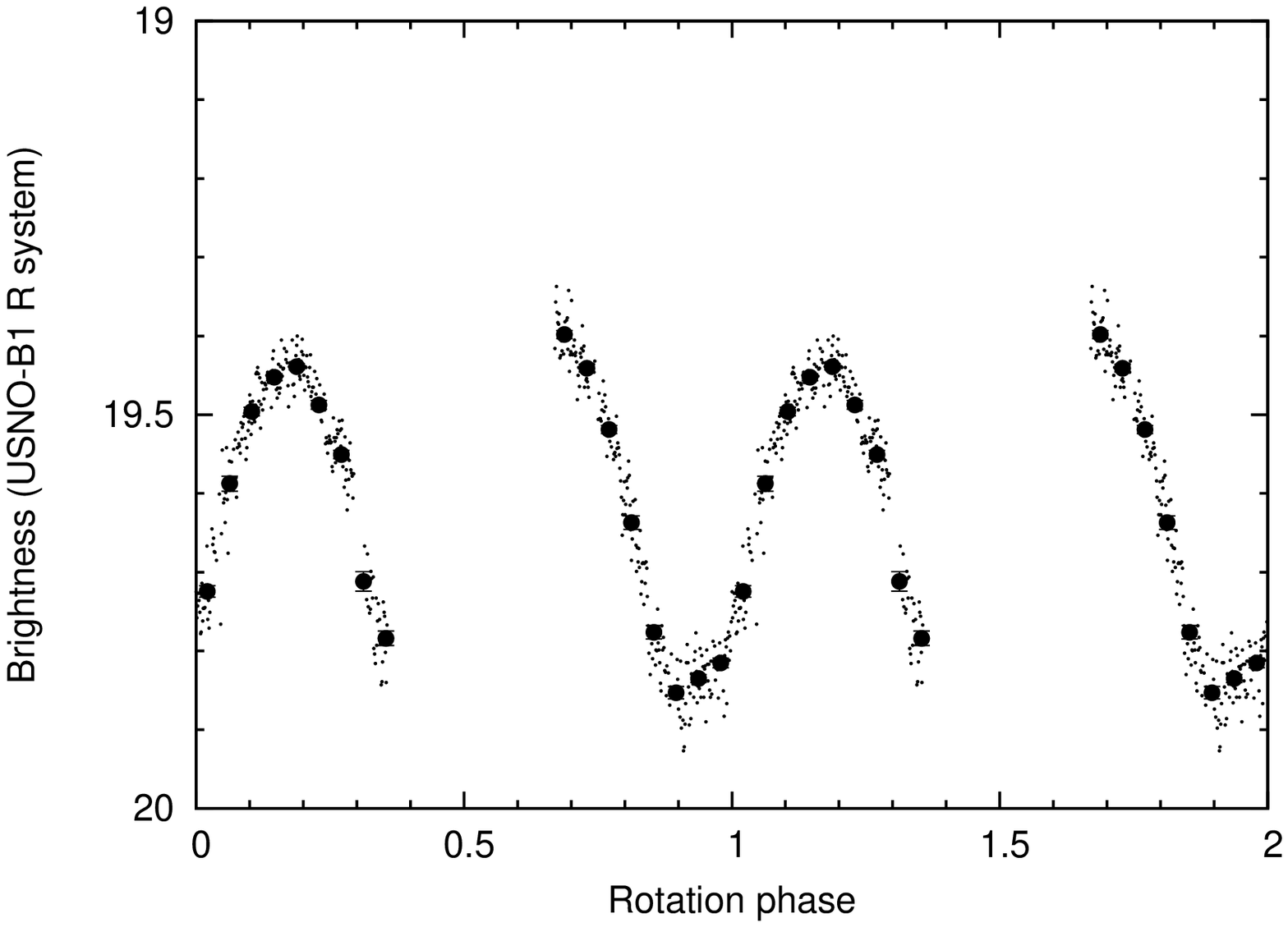}&
\includegraphics[viewport=5 40 510 355,clip,height=3.7cm]{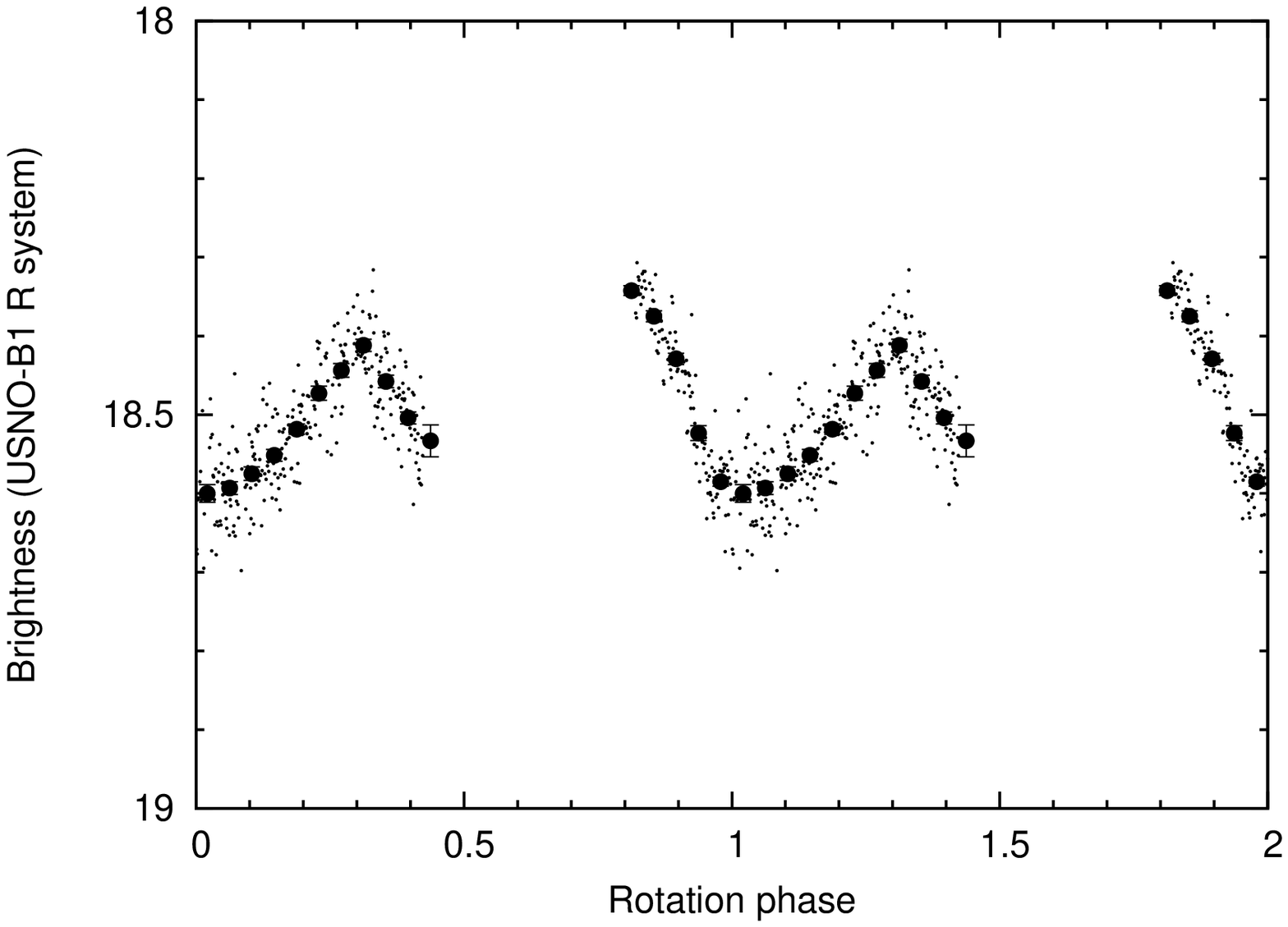}\\
13372 & 13379 & 14690\\
\includegraphics[viewport=5 40 510 355,clip,height=3.7cm]{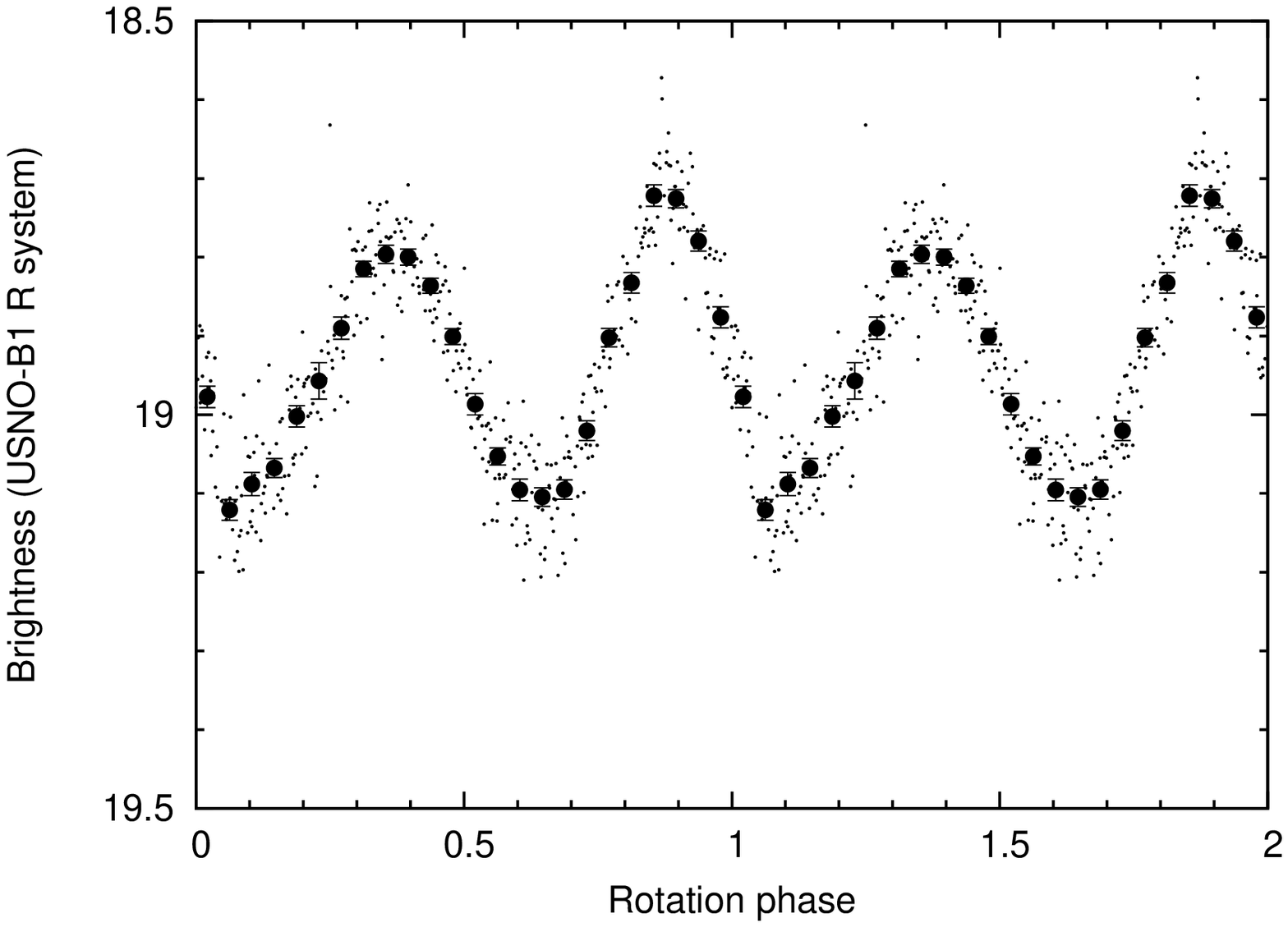}&
\includegraphics[viewport=5 40 510 355,clip,height=3.7cm]{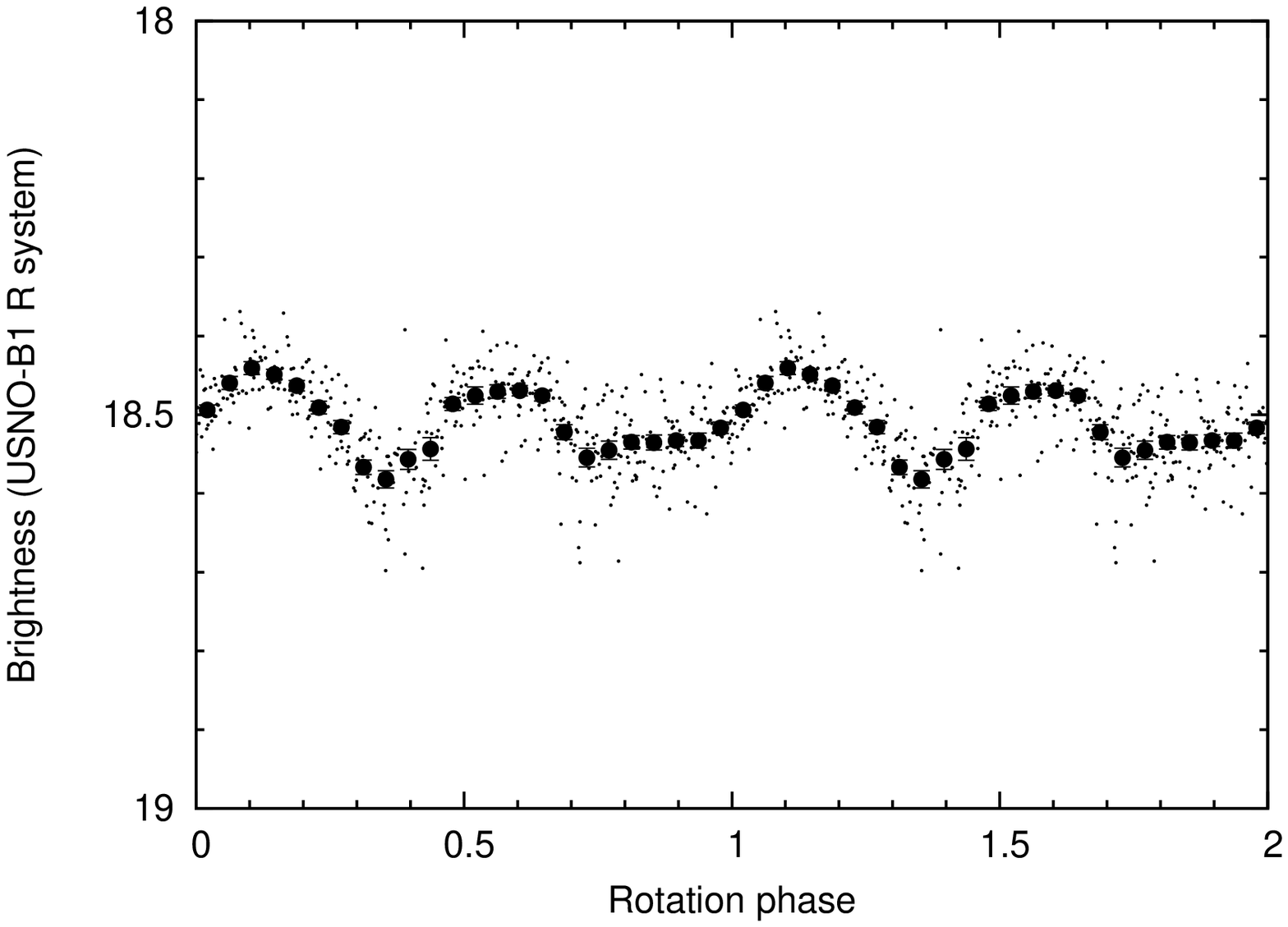}&
\includegraphics[viewport=5 40 510 355,clip,height=3.7cm]{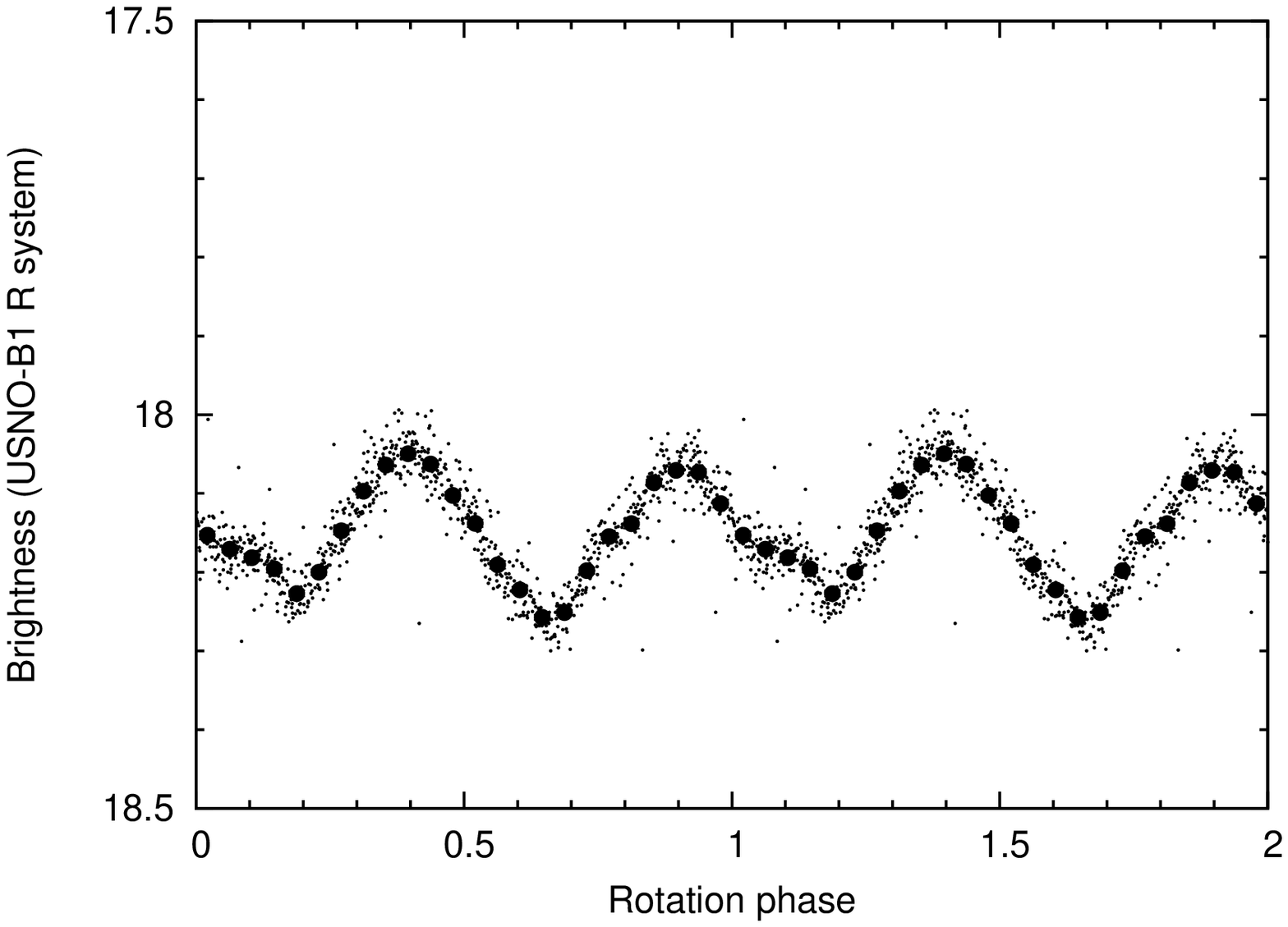}\\
14791 & 15529 & 16152\\
\includegraphics[viewport=5 40 510 355,clip,height=3.7cm]{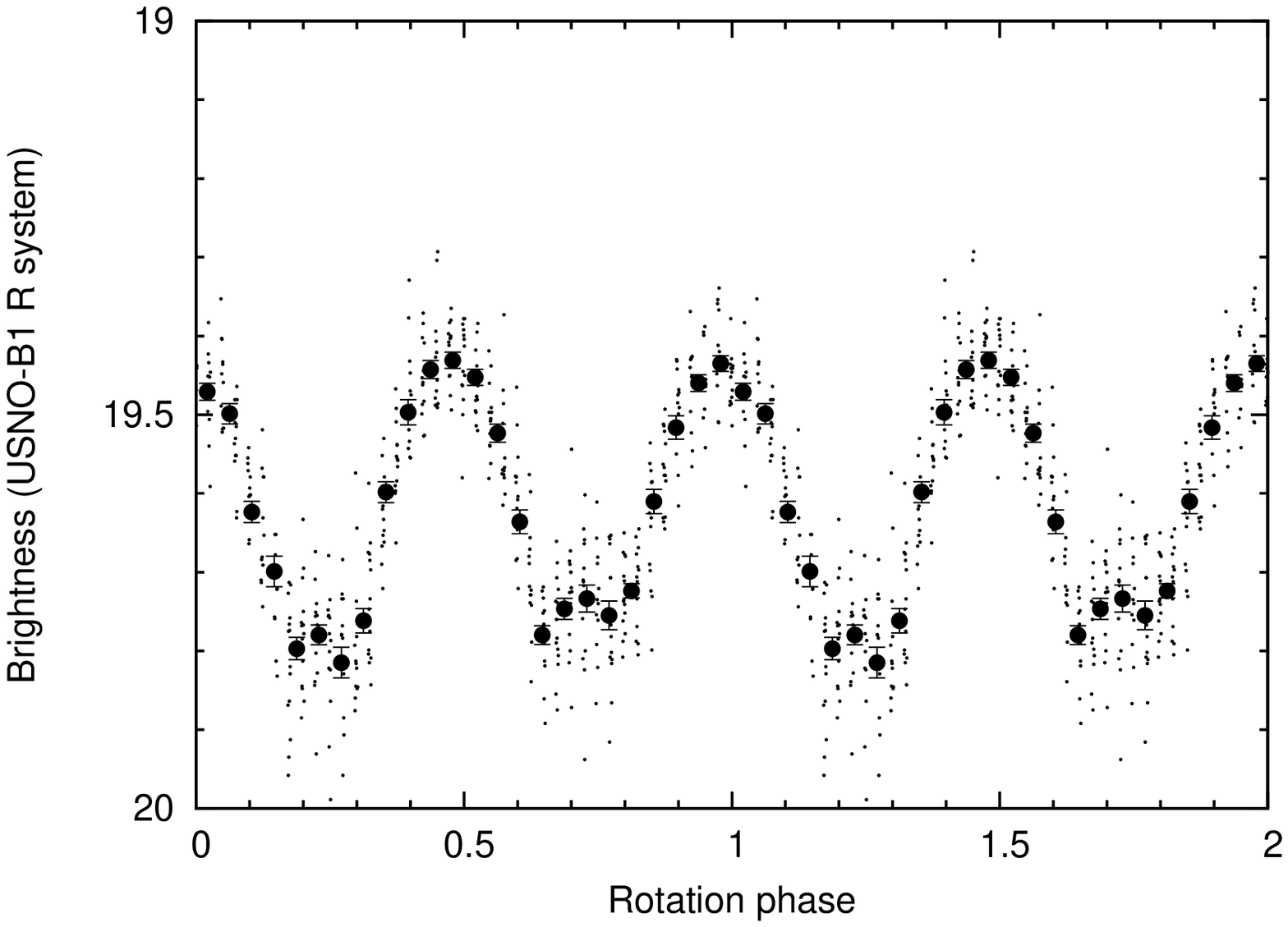}&
\includegraphics[viewport=5 40 510 355,clip,height=3.7cm]{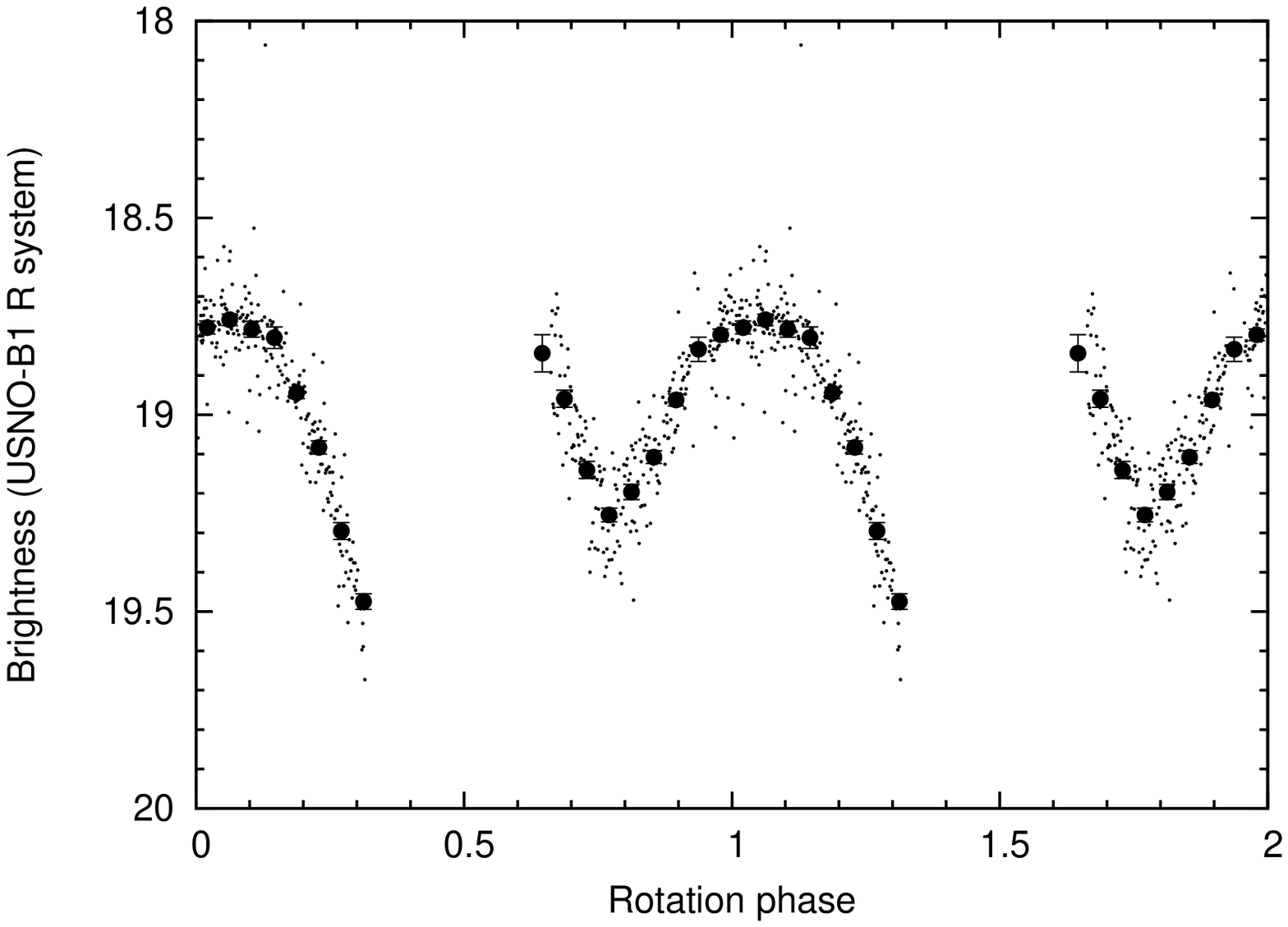}&
\includegraphics[viewport=5 40 510 355,clip,height=3.7cm]{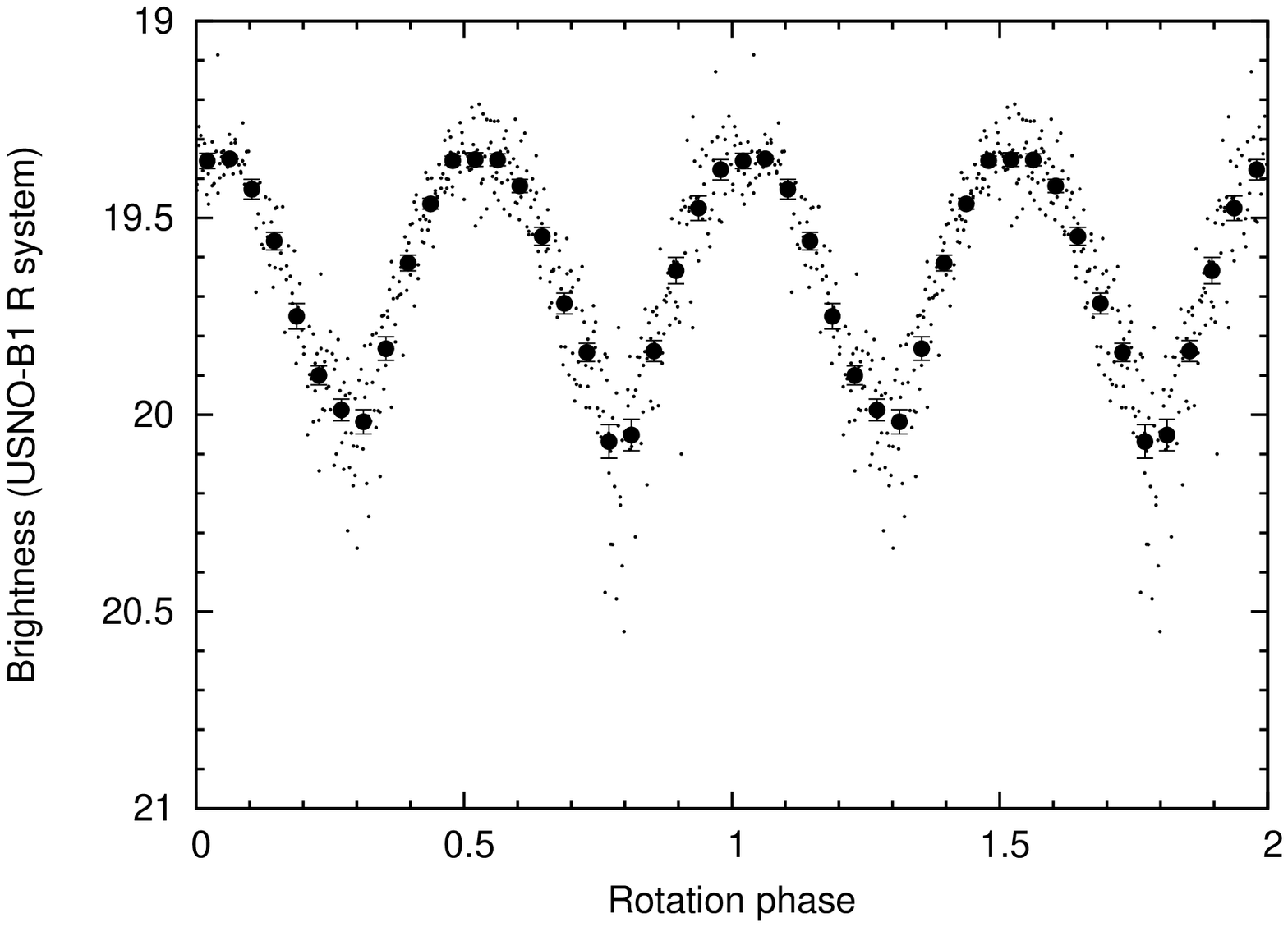}\\
21593 & 21599 & 22056\\
\includegraphics[viewport=5 40 510 355,clip,height=3.7cm]{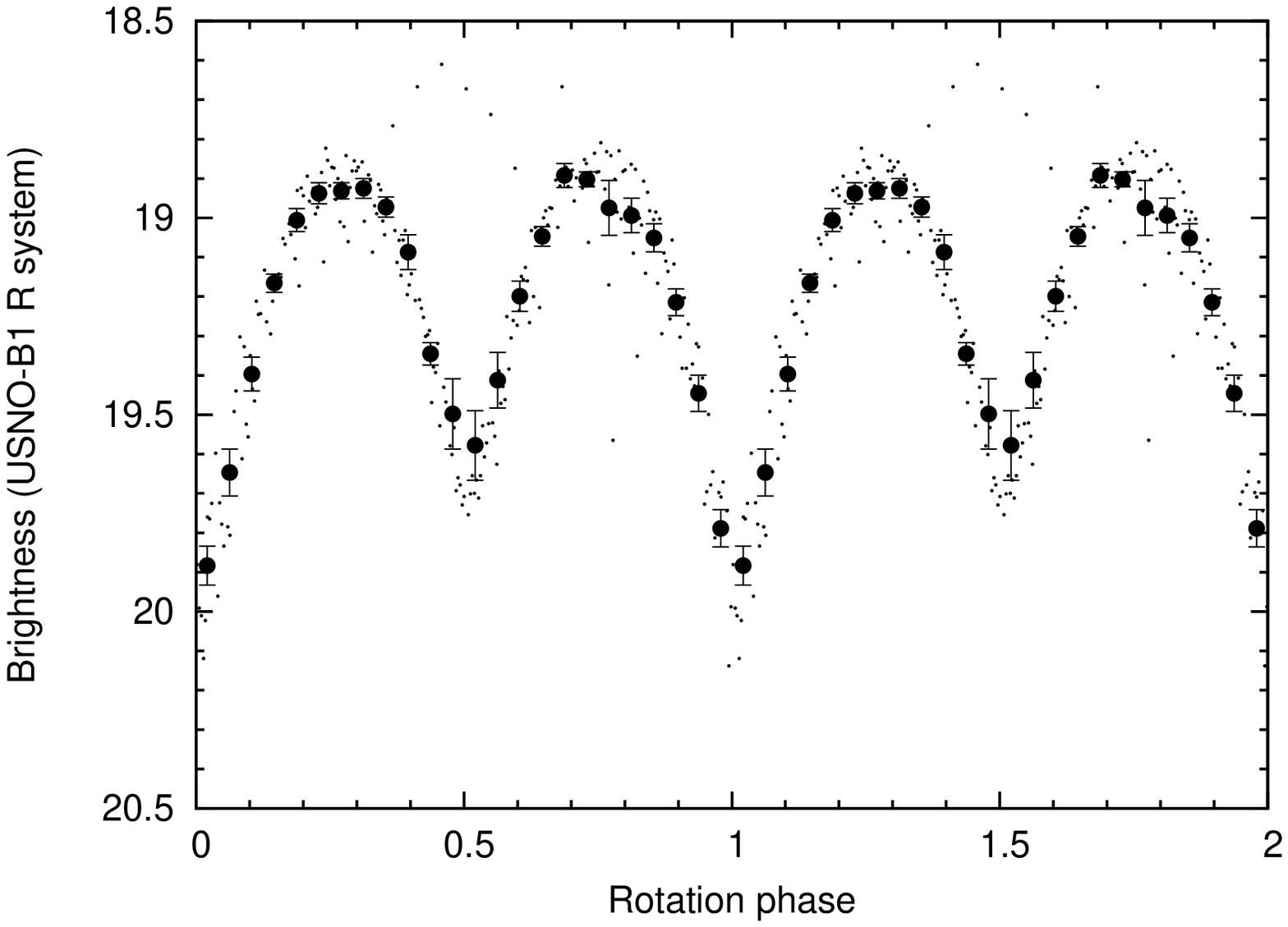}&
\includegraphics[viewport=5 40 510 355,clip,height=3.7cm]{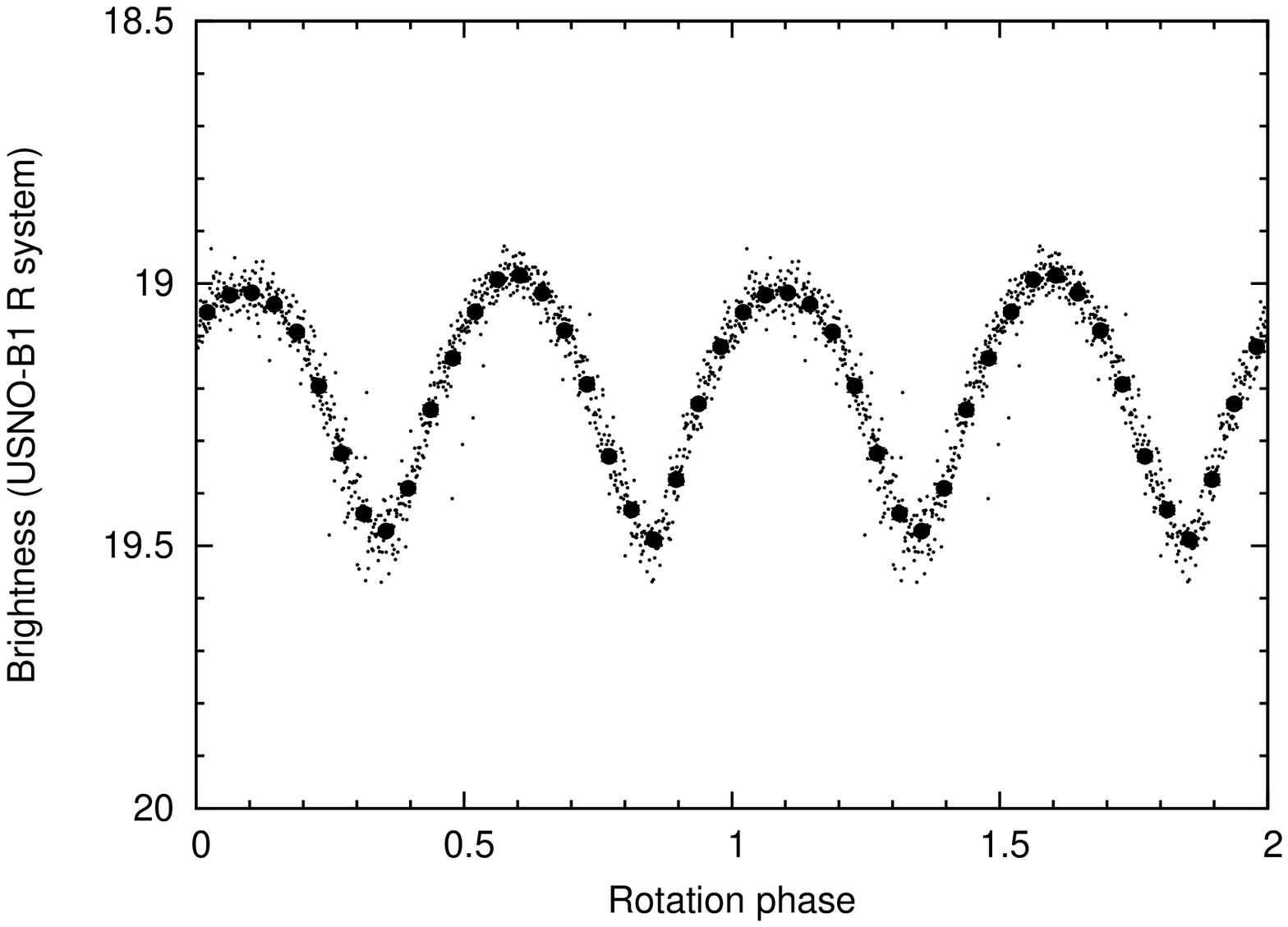}&
\includegraphics[viewport=5 40 510 355,clip,height=3.7cm]{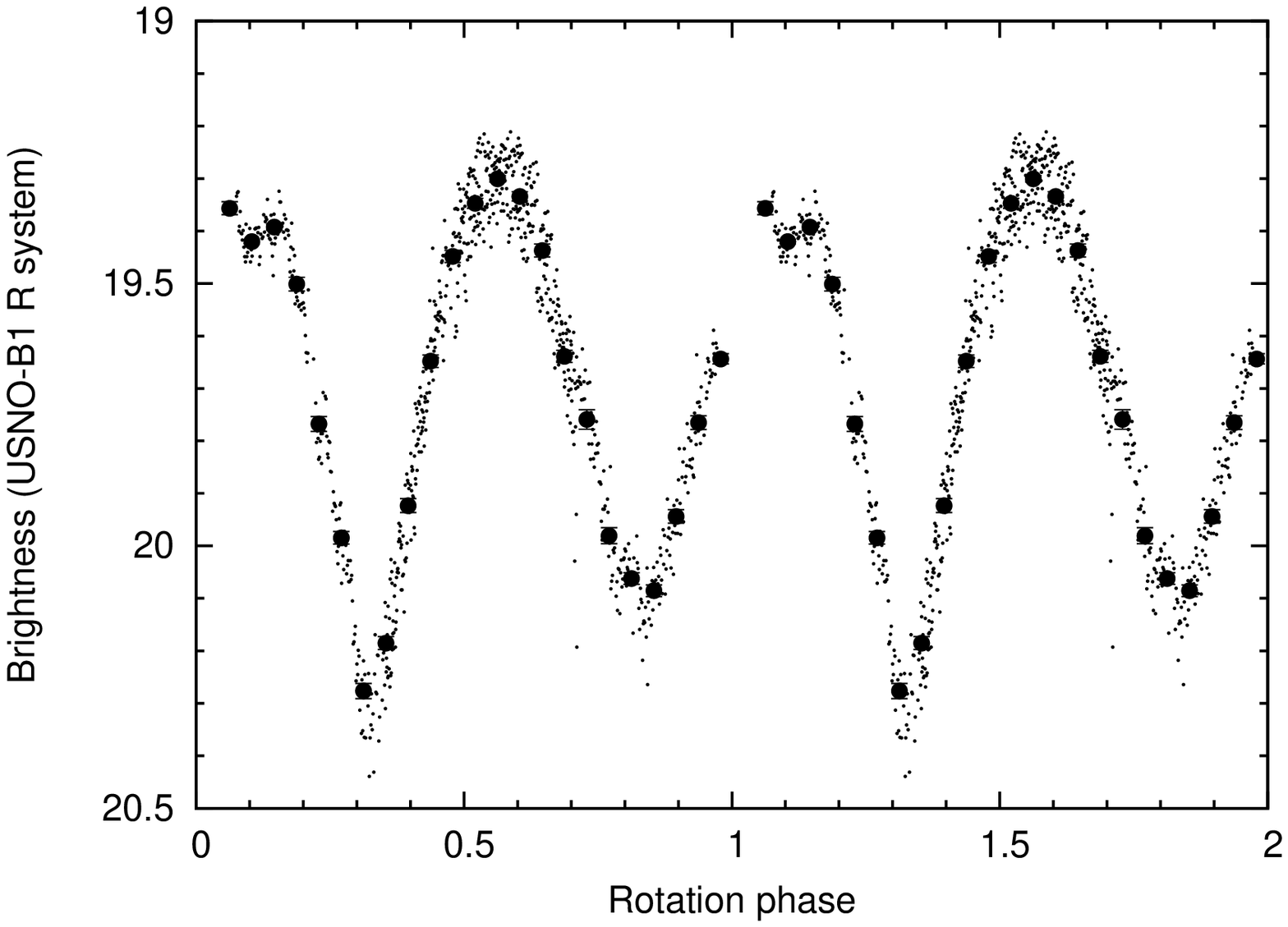}\\
\end{tabular}
\end{center}
\caption{The phased light curves of the Trojan asteroids in K2 Field 6.}
\end{figure*}

\begin{figure*}[h]
\begin{center}
\begin{tabular}{ccc}
23939 & 23947 & 23958\\
\includegraphics[viewport=5 40 510 355,clip,height=3.7cm]{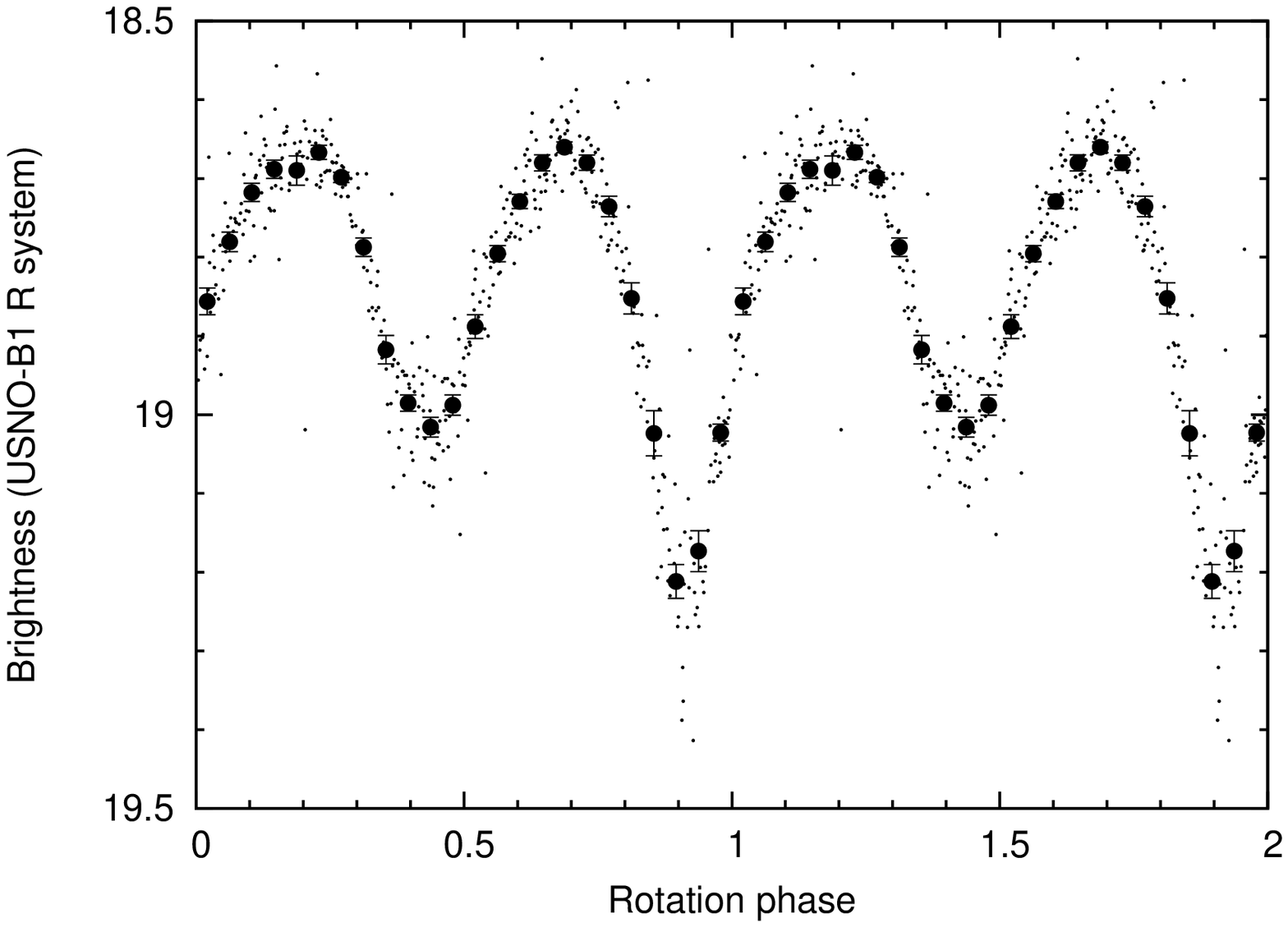}&
\includegraphics[viewport=5 40 510 355,clip,height=3.7cm]{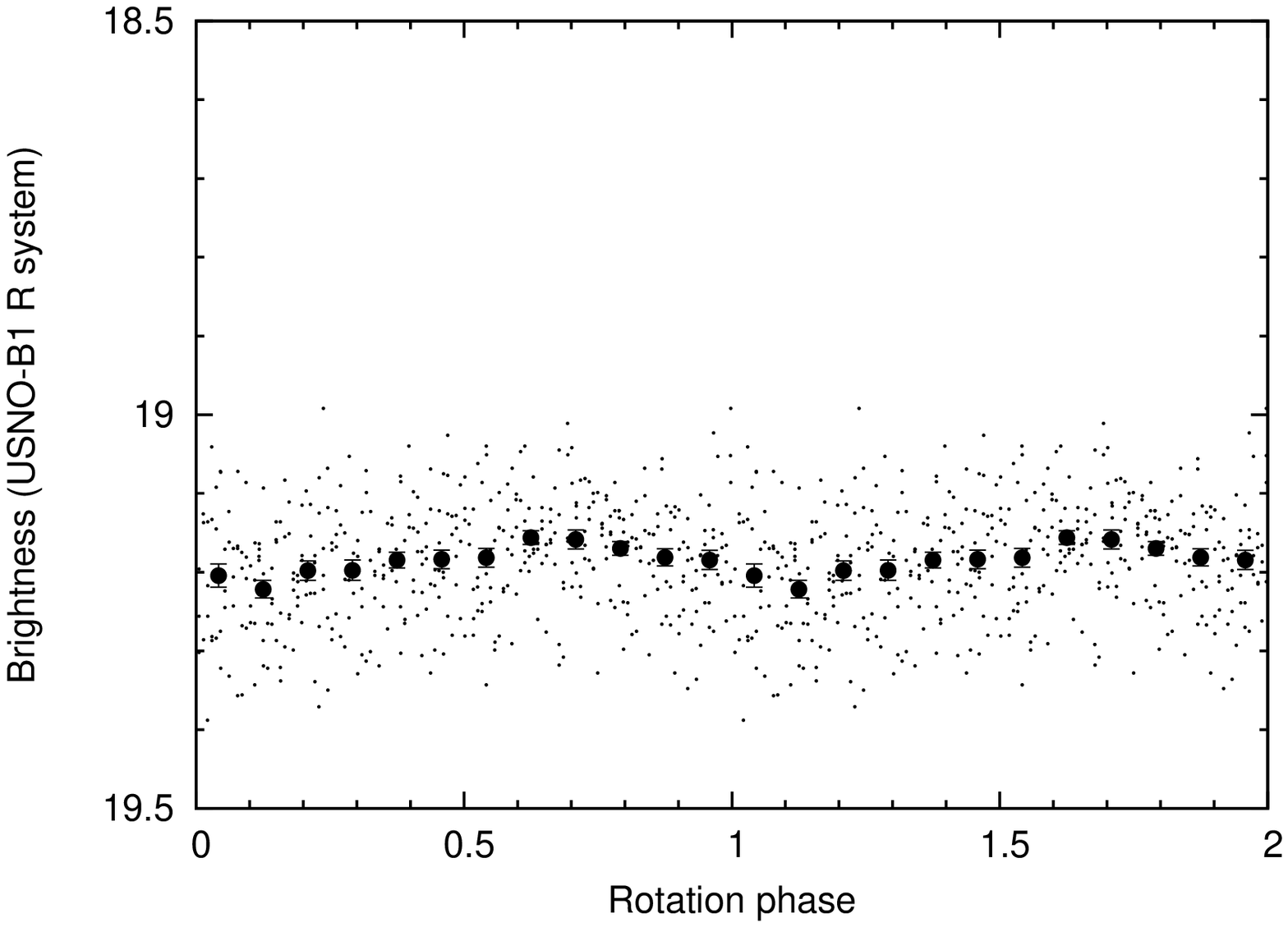}&
\includegraphics[viewport=5 40 510 355,clip,height=3.7cm]{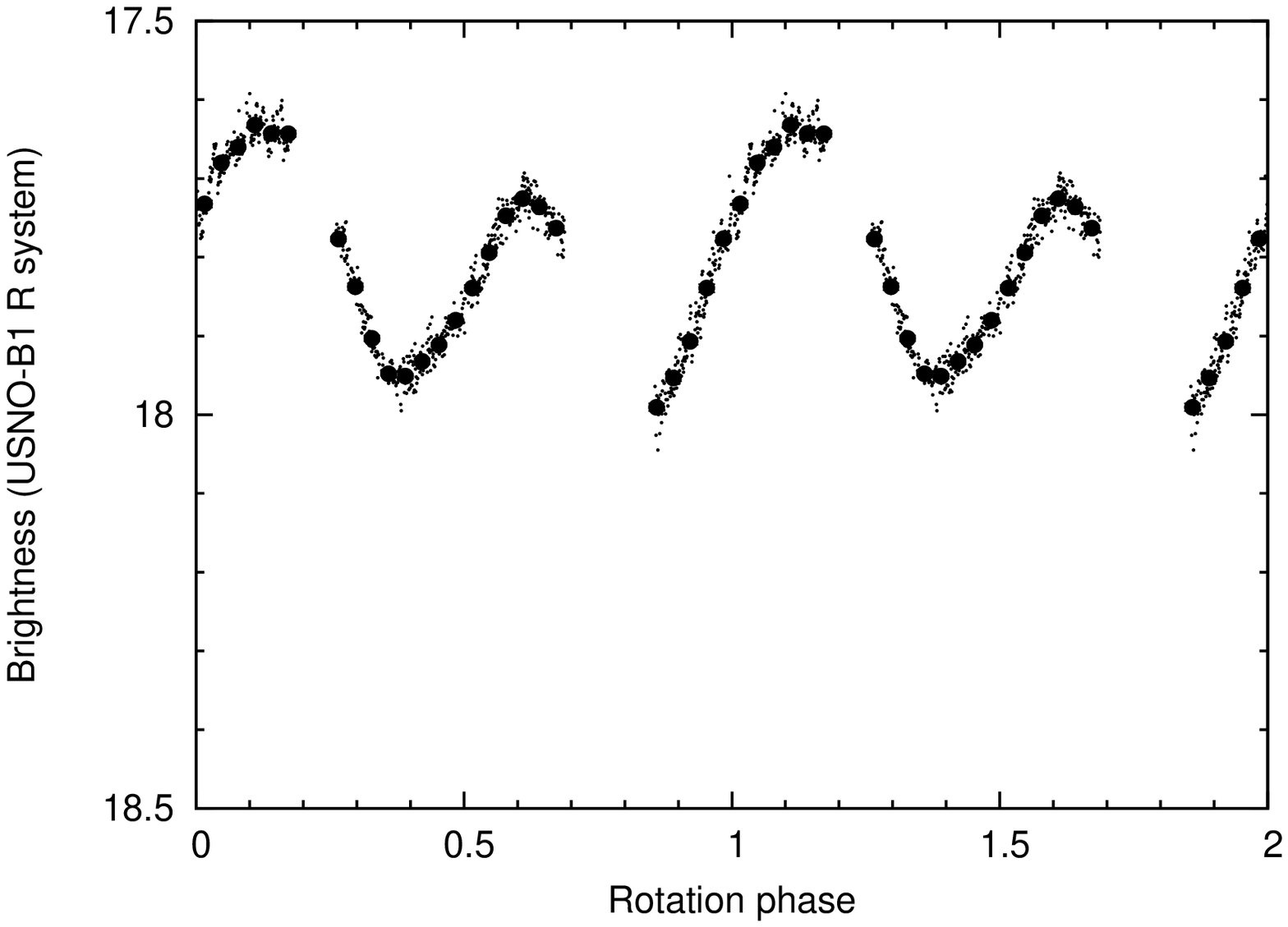}\\
24357 & 24534 & 24537\\
\includegraphics[viewport=5 40 510 355,clip,height=3.7cm]{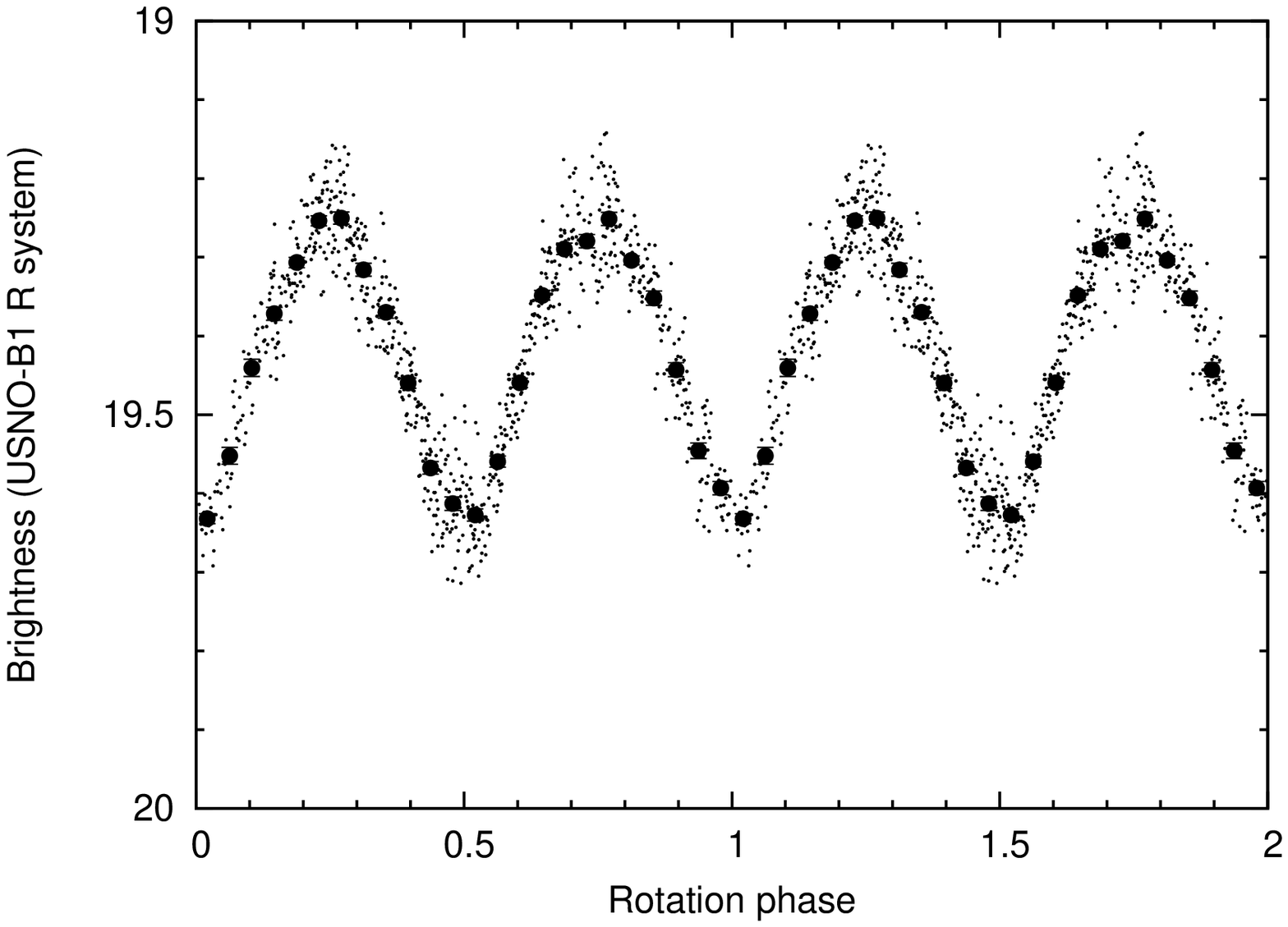}&
\includegraphics[viewport=5 40 510 355,clip,height=3.7cm]{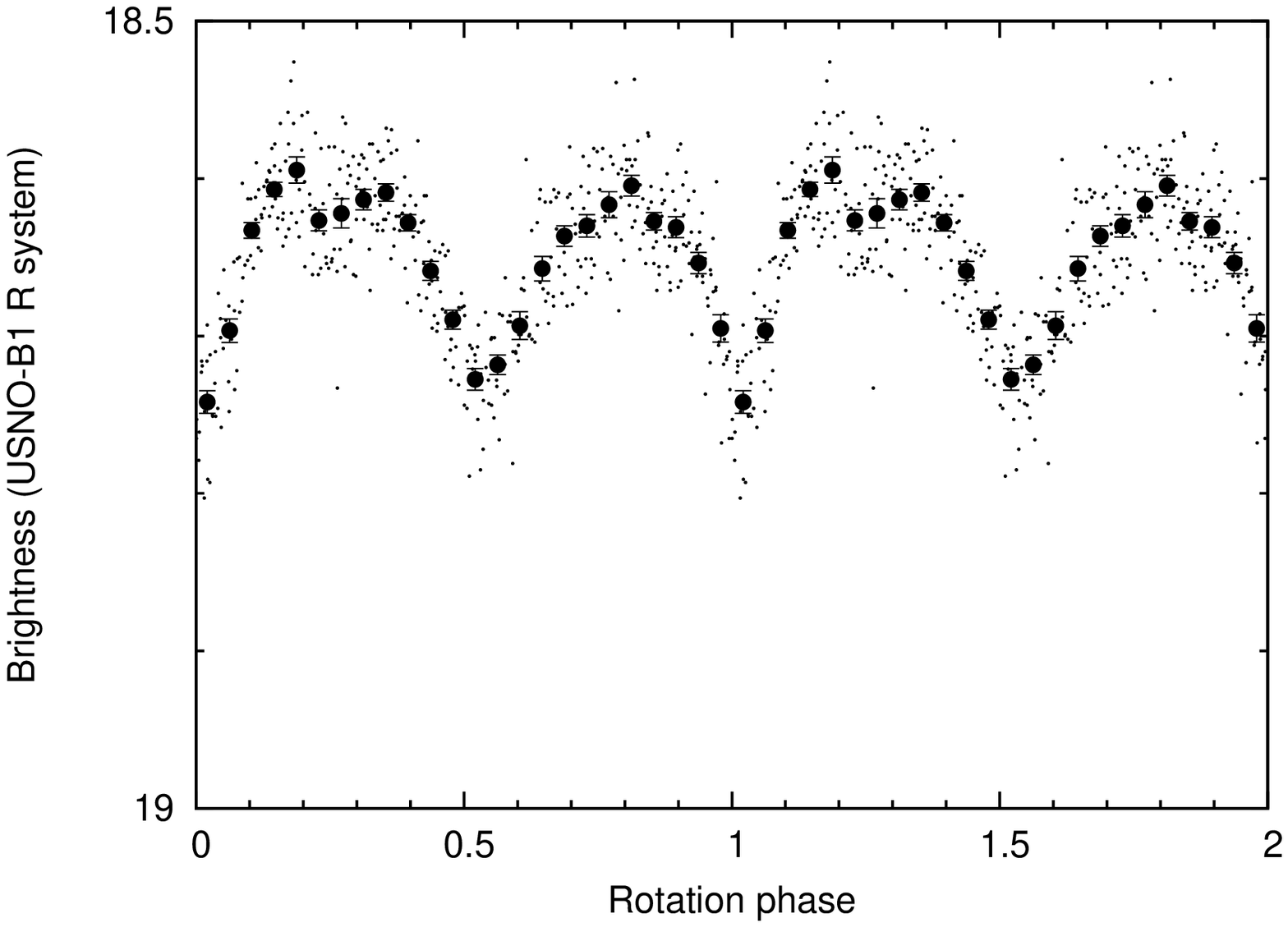}&
\includegraphics[viewport=5 40 510 355,clip,height=3.7cm]{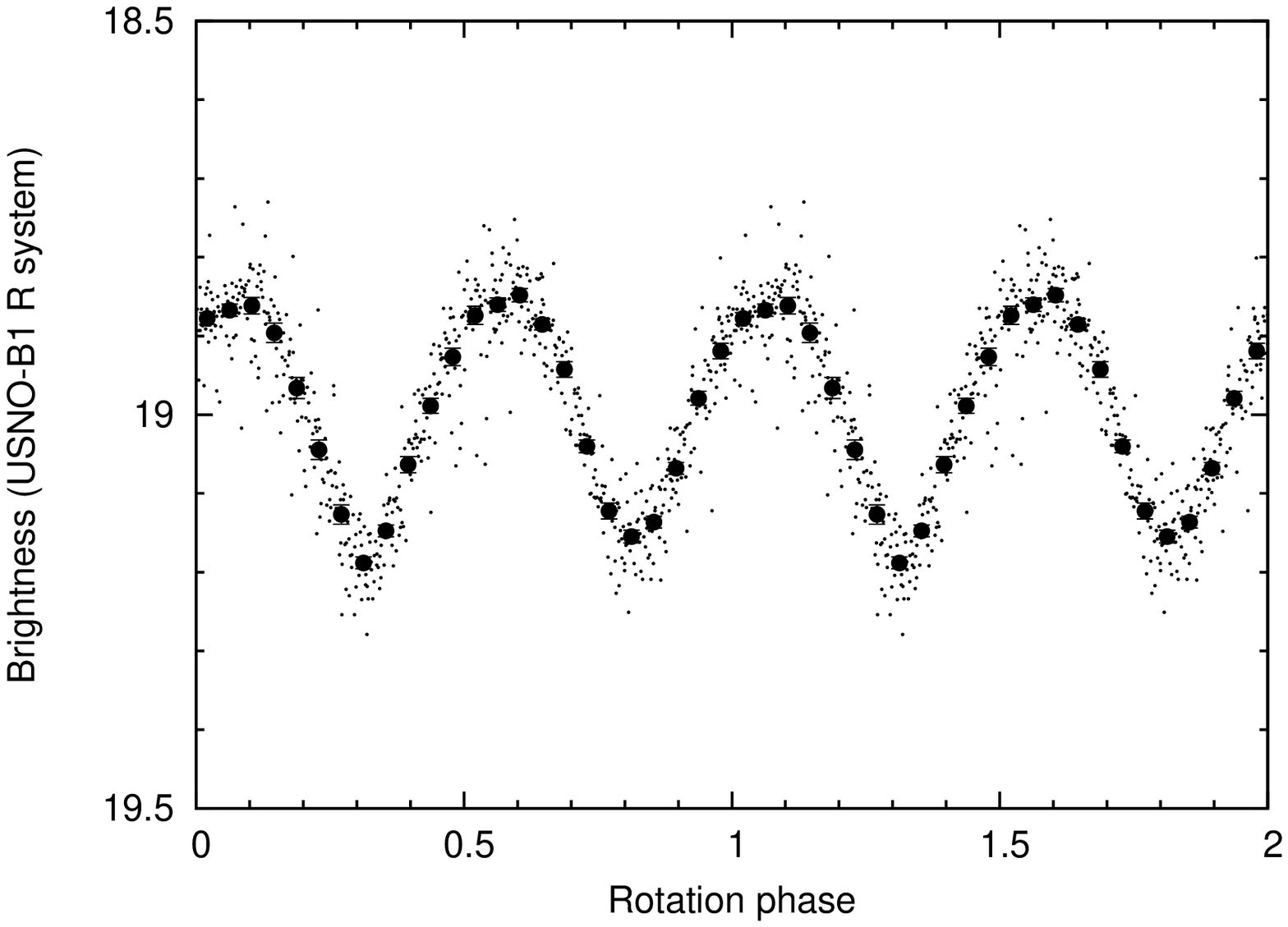}\\
35363 & 38574 & 39270\\
\includegraphics[viewport=5 40 510 355,clip,height=3.7cm]{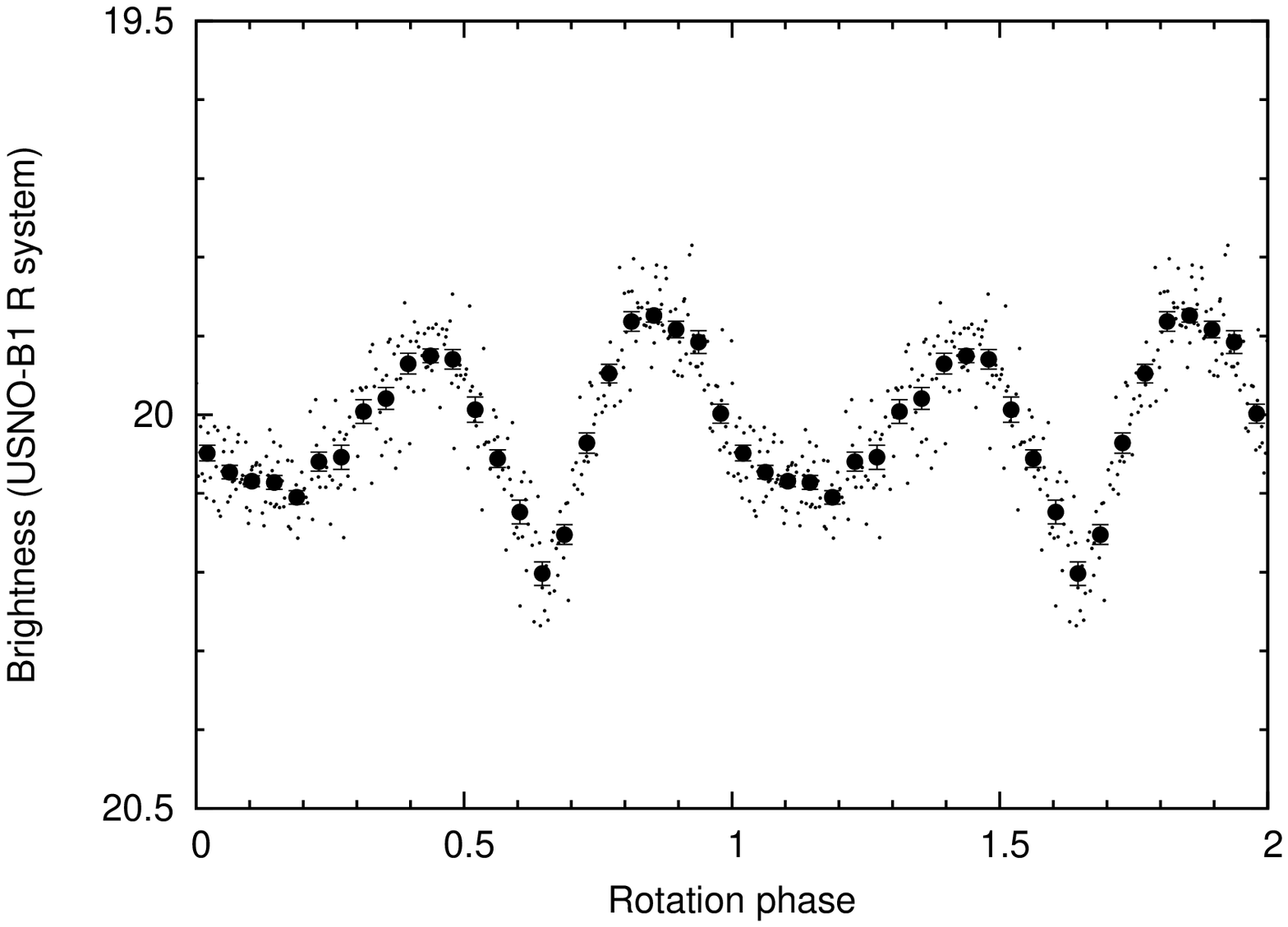}&
\includegraphics[viewport=5 40 510 355,clip,height=3.7cm]{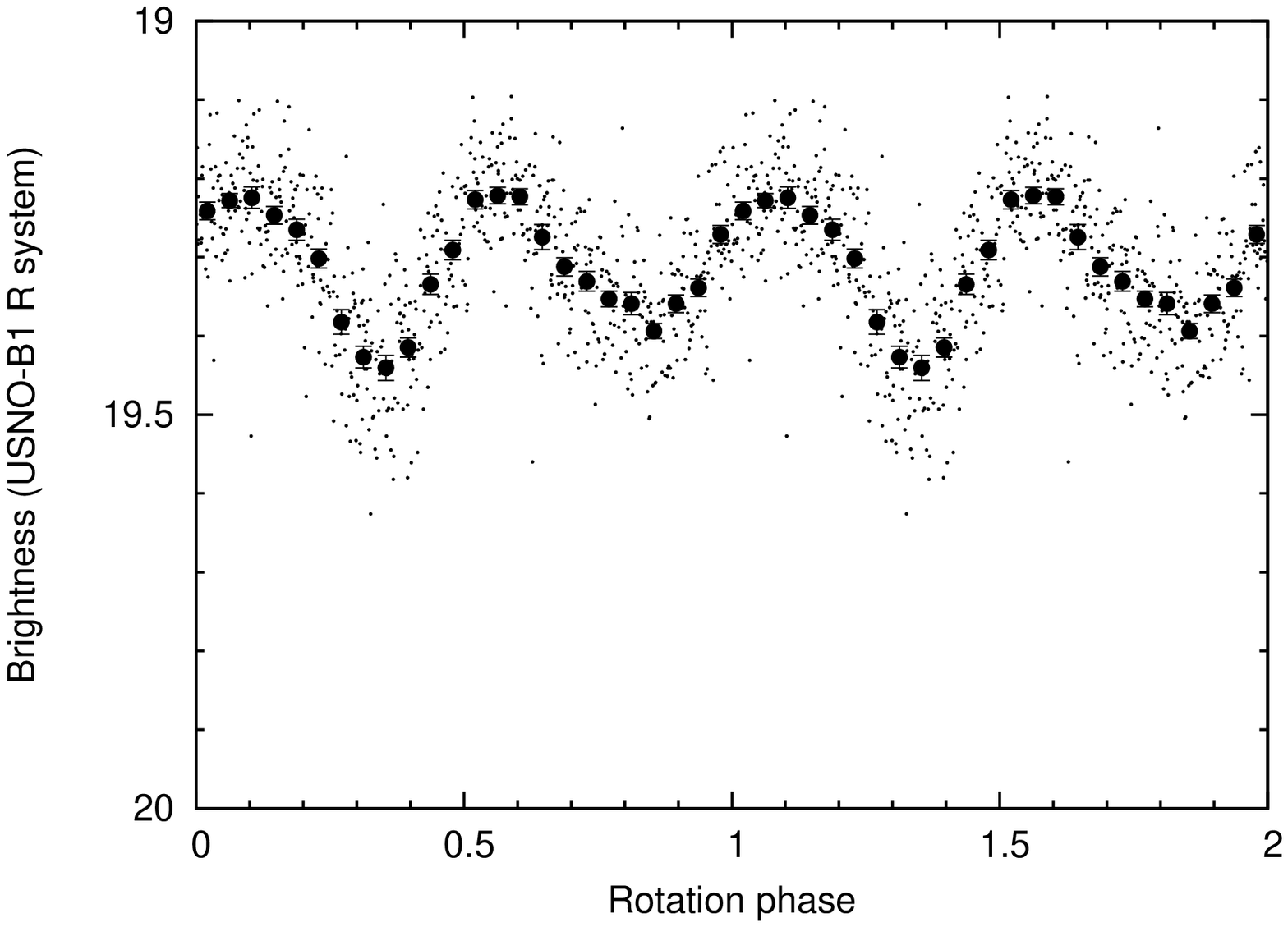}&
\includegraphics[viewport=5 40 510 355,clip,height=3.7cm]{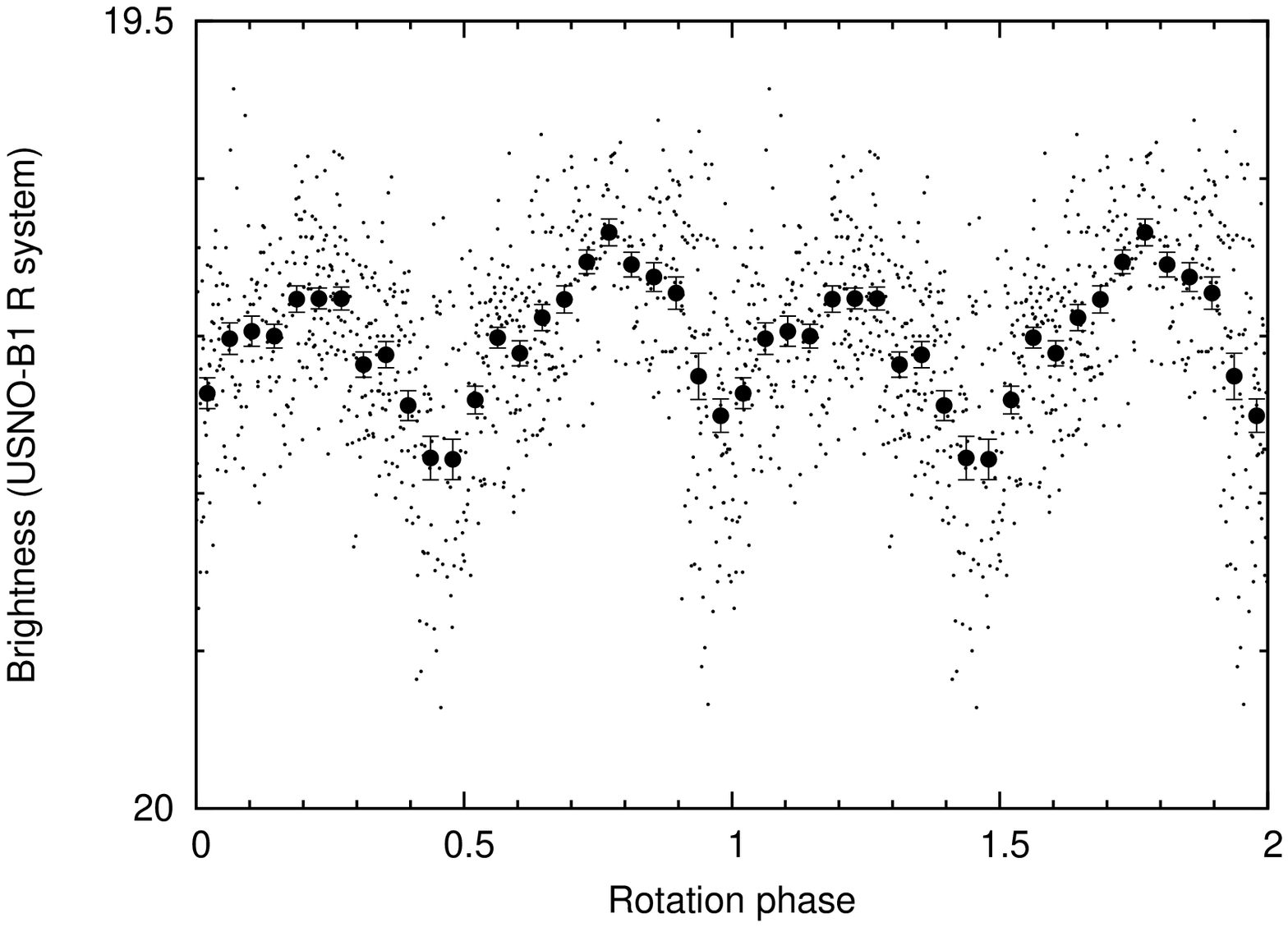}\\
39286 & 39289 & 57041\\
\includegraphics[viewport=5 40 510 355,clip,height=3.7cm]{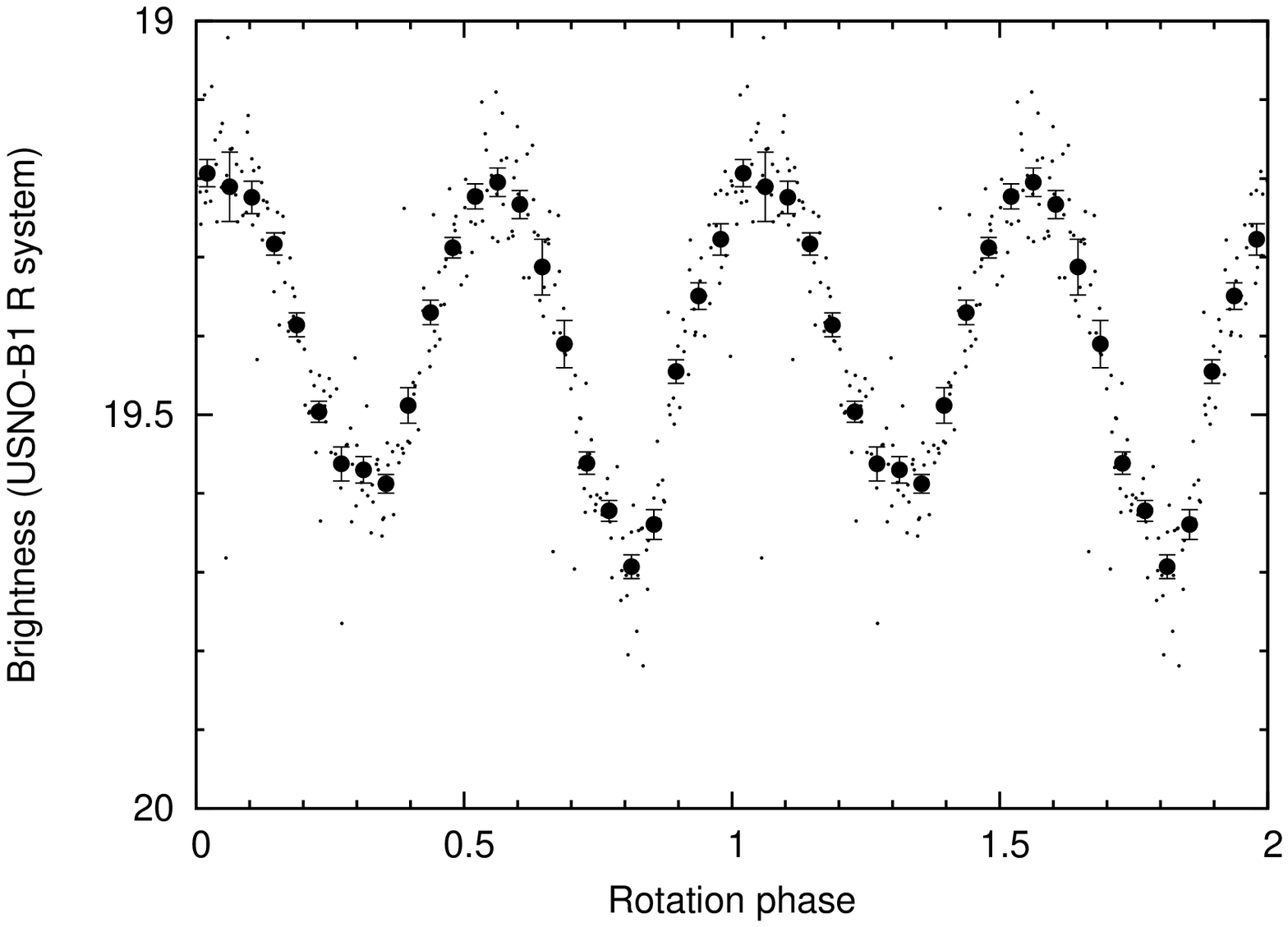}&
\includegraphics[viewport=5 40 510 355,clip,height=3.7cm]{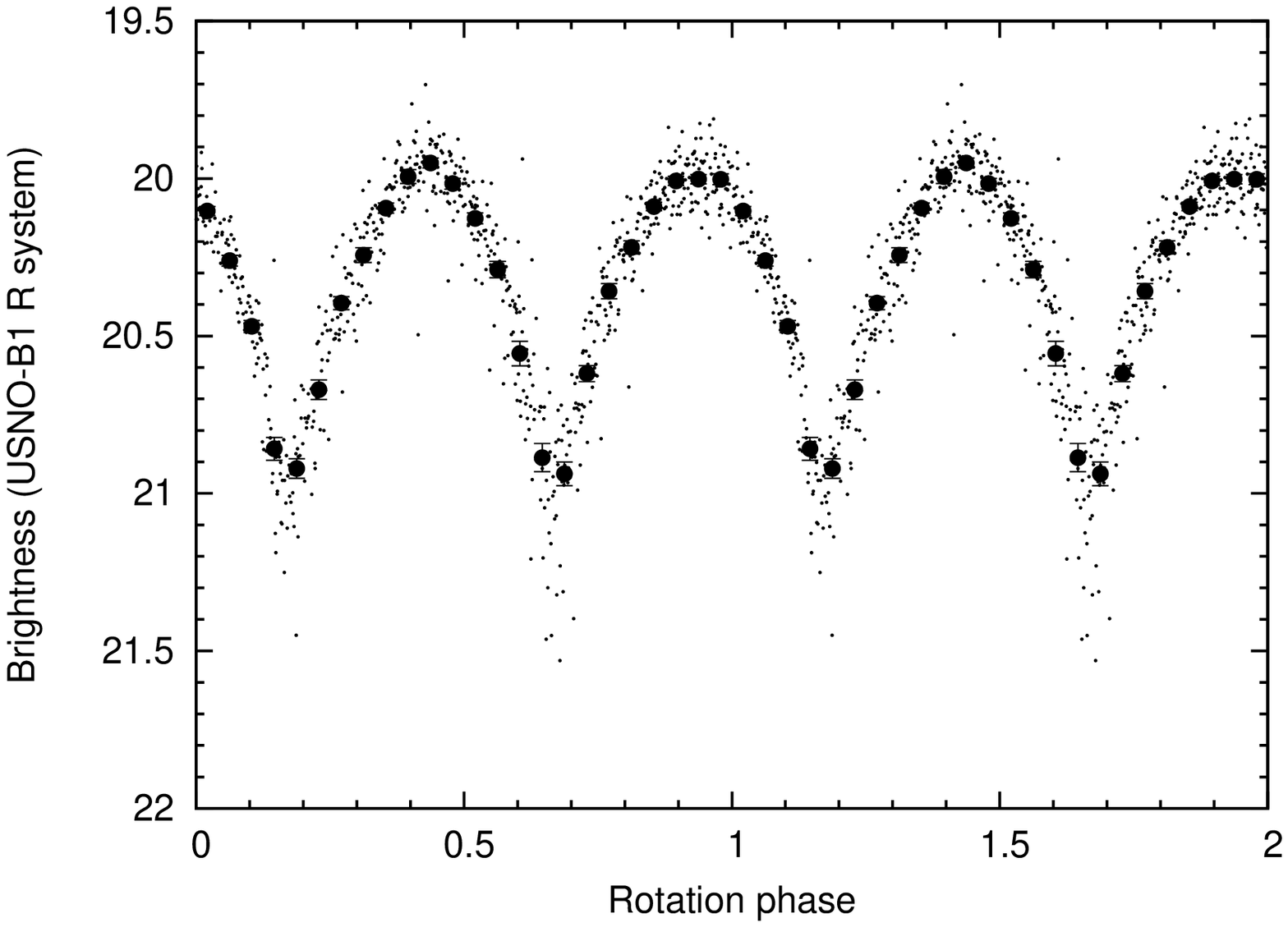}&
\includegraphics[viewport=5 40 510 355,clip,height=3.7cm]{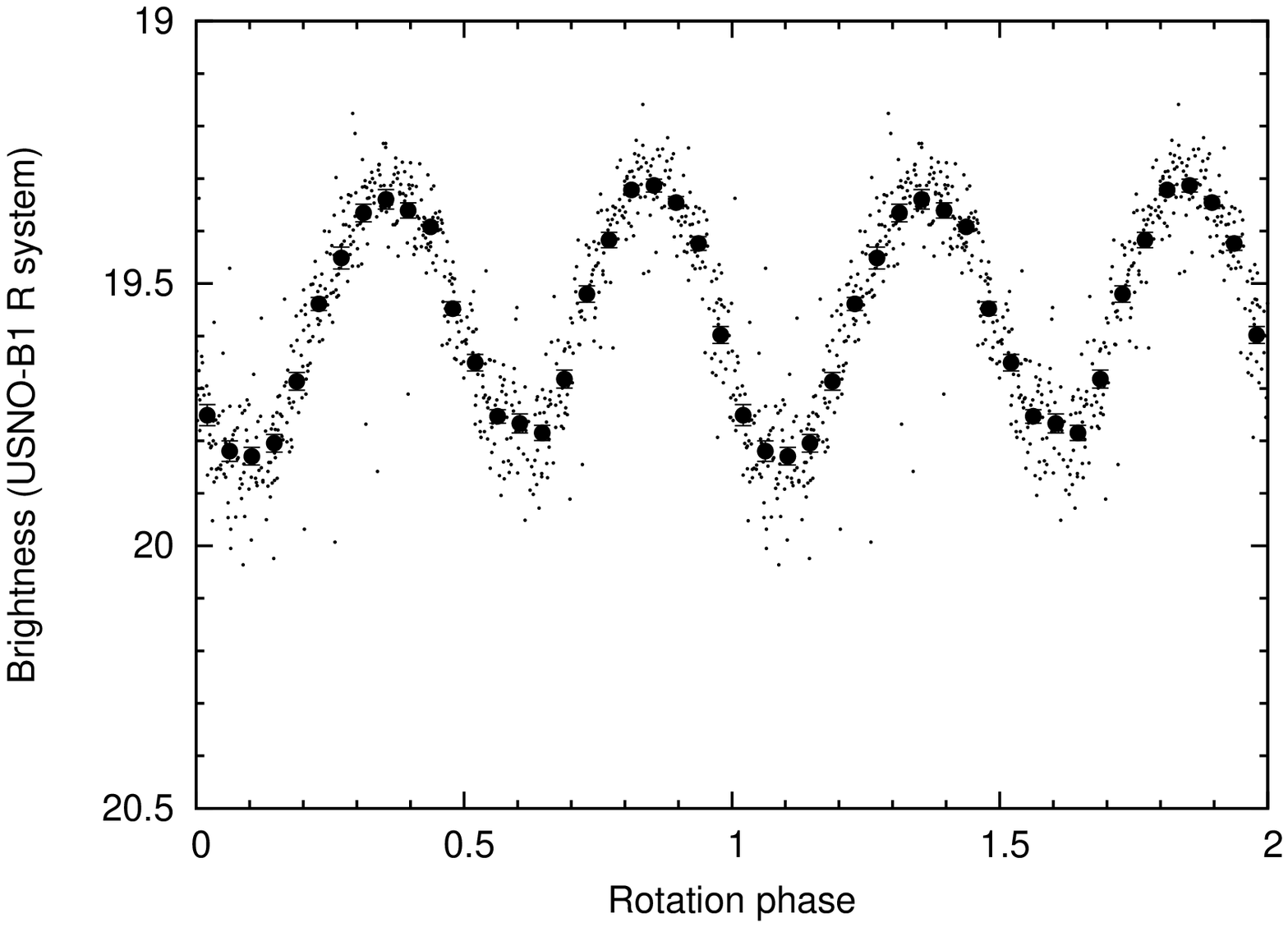}\\
58480 & 59049 & 63239\\
\includegraphics[viewport=5 40 510 355,clip,height=3.7cm]{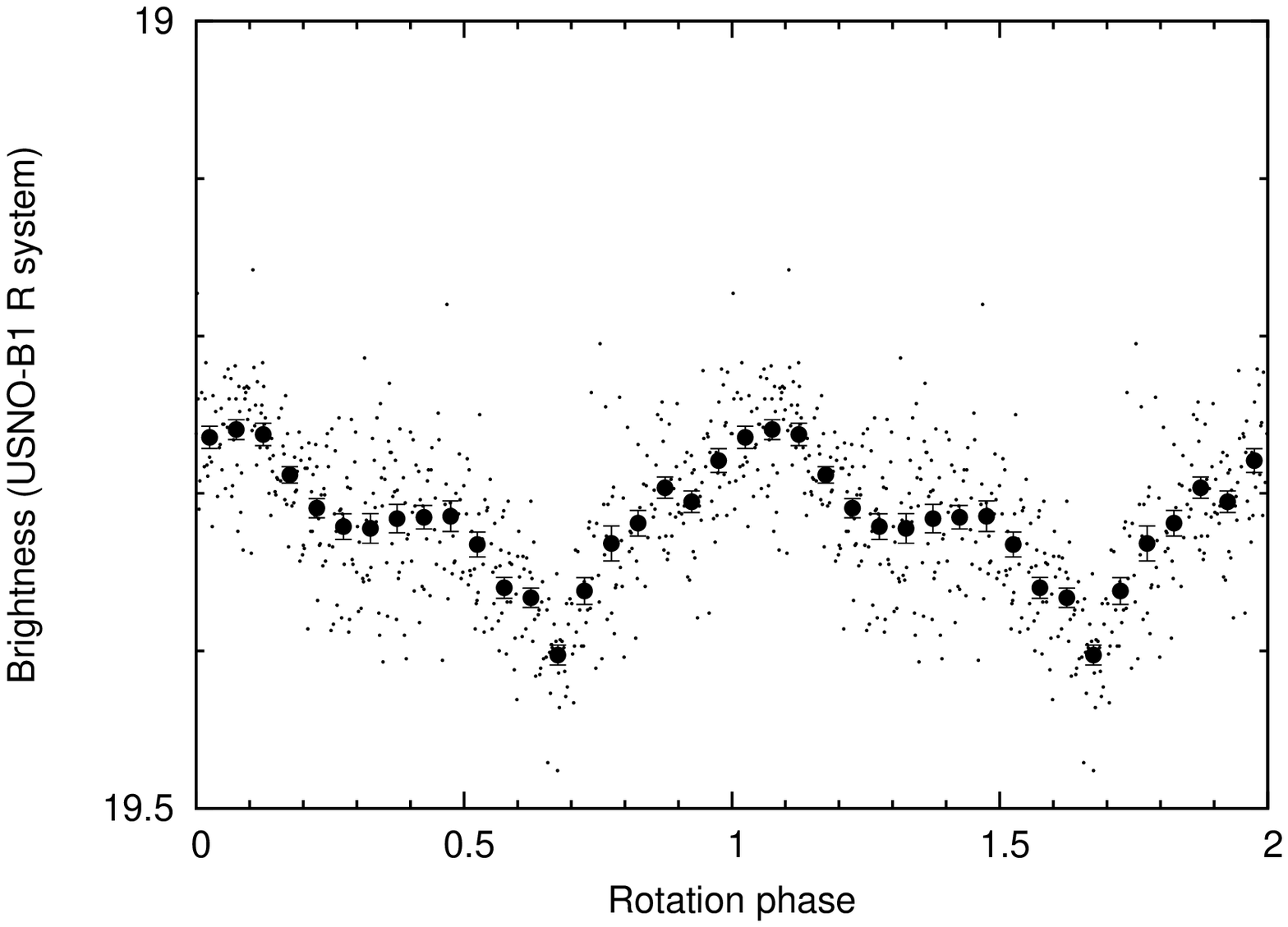}&
\includegraphics[viewport=5 40 510 355,clip,height=3.7cm]{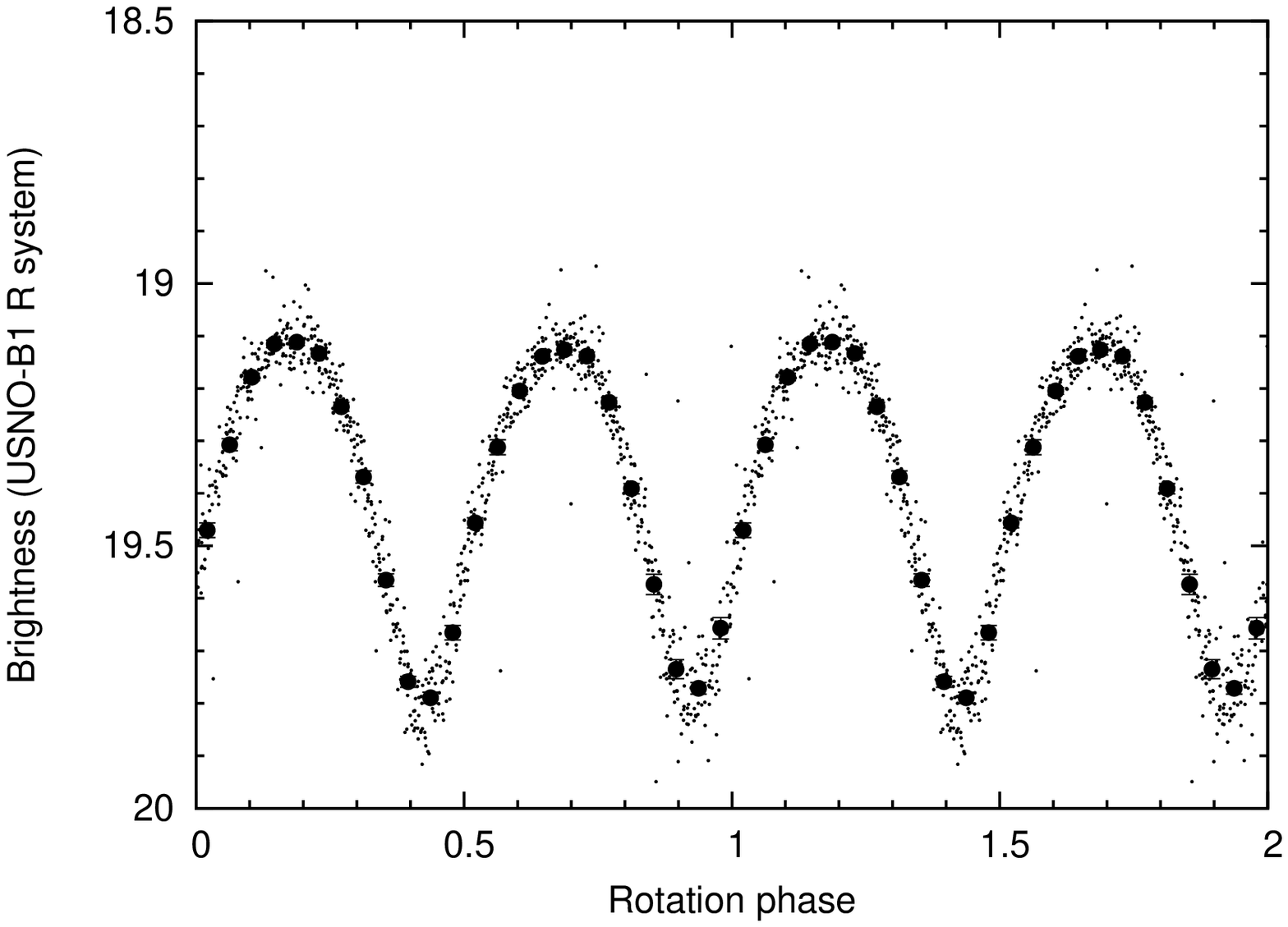}&
\includegraphics[viewport=5 40 510 355,clip,height=3.7cm]{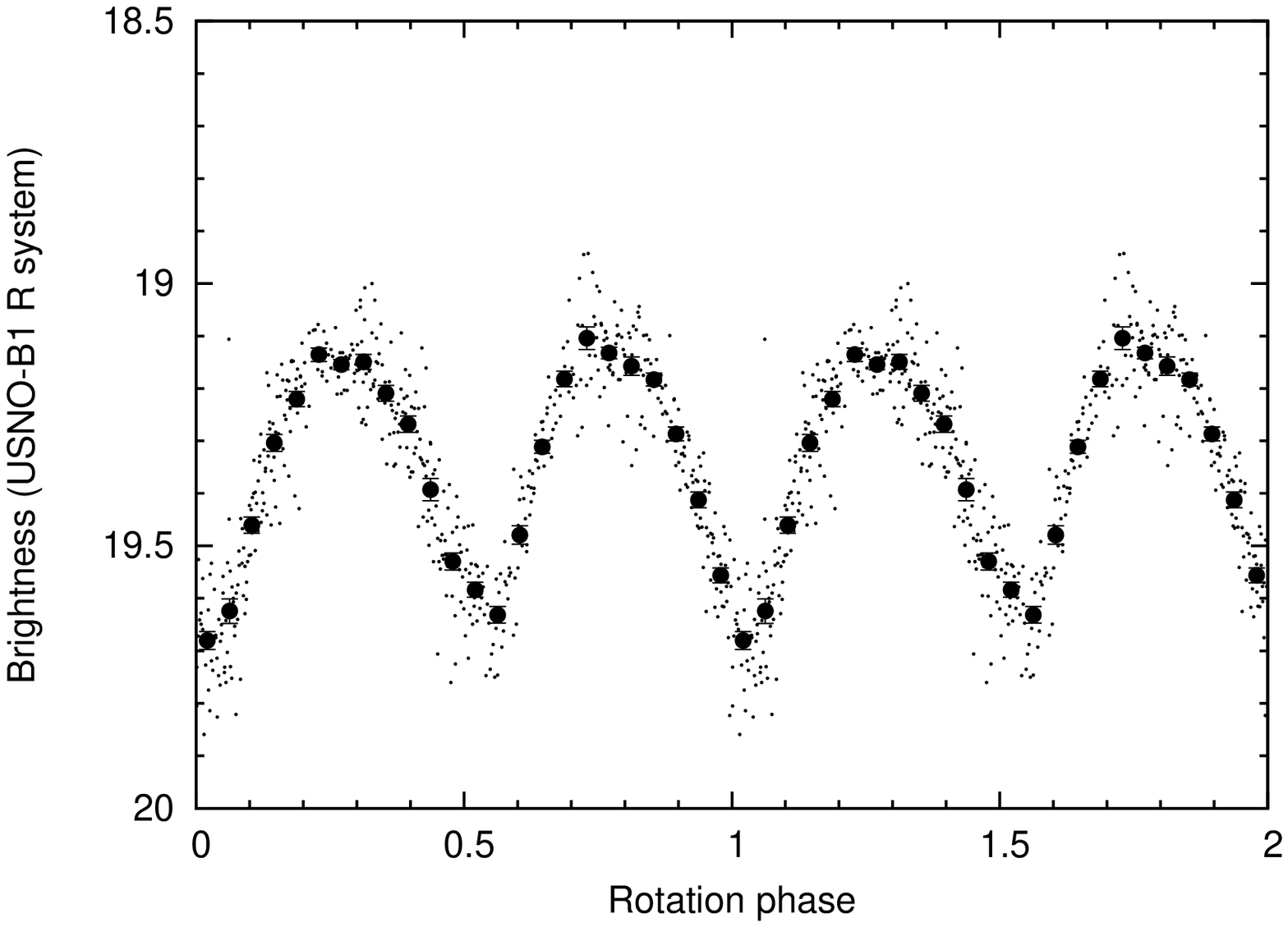}\\
\end{tabular}
\end{center}
\caption{The phased light curves of the Trojan asteroids in K2 Field 6.}
\end{figure*}

\begin{figure*}[h]
\begin{center}
\begin{tabular}{ccc}
65210 & 65223 & 65227\\
\includegraphics[viewport=5 40 510 355,clip,height=3.7cm]{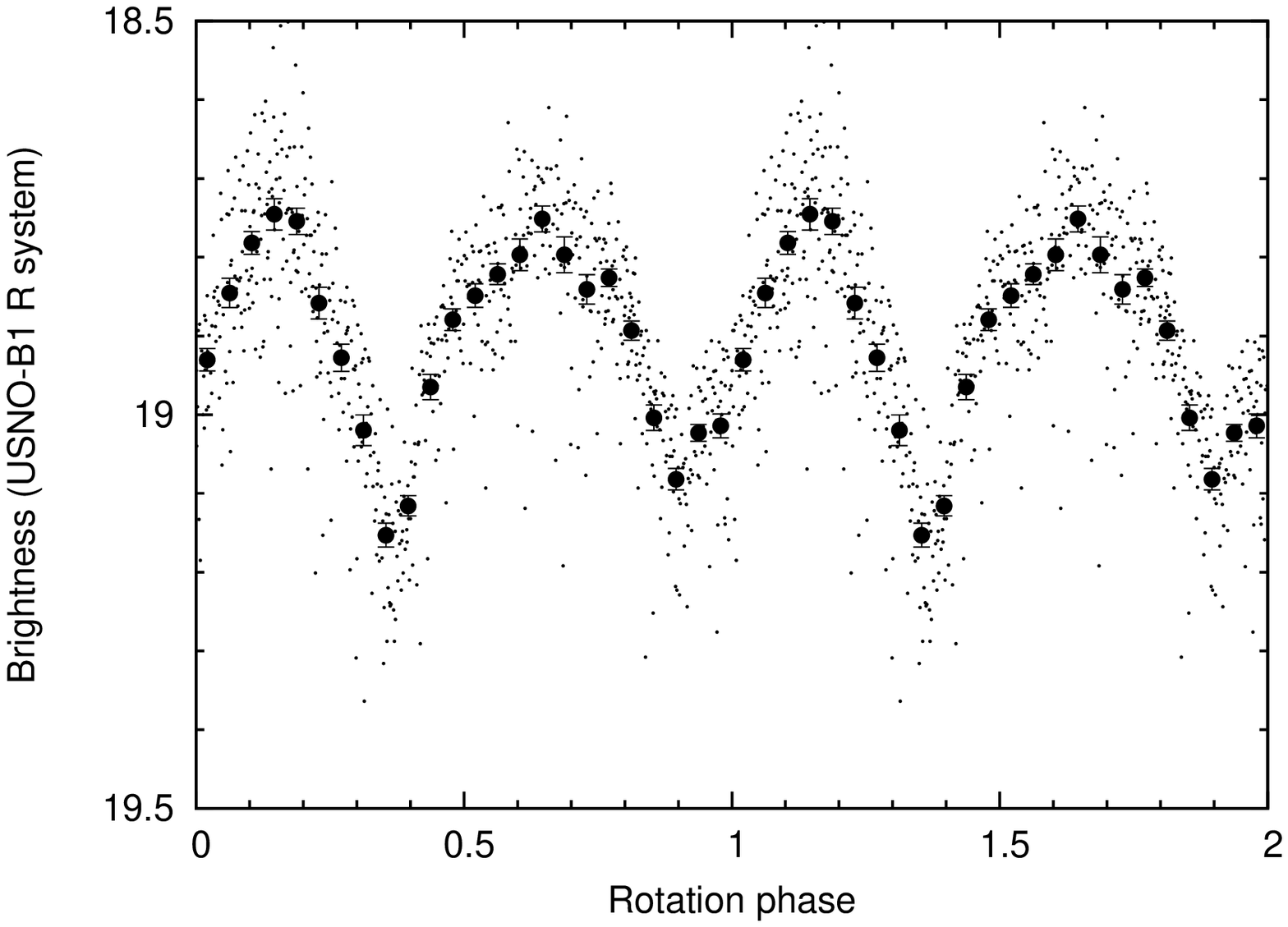}&
\includegraphics[viewport=5 40 510 355,clip,height=3.7cm]{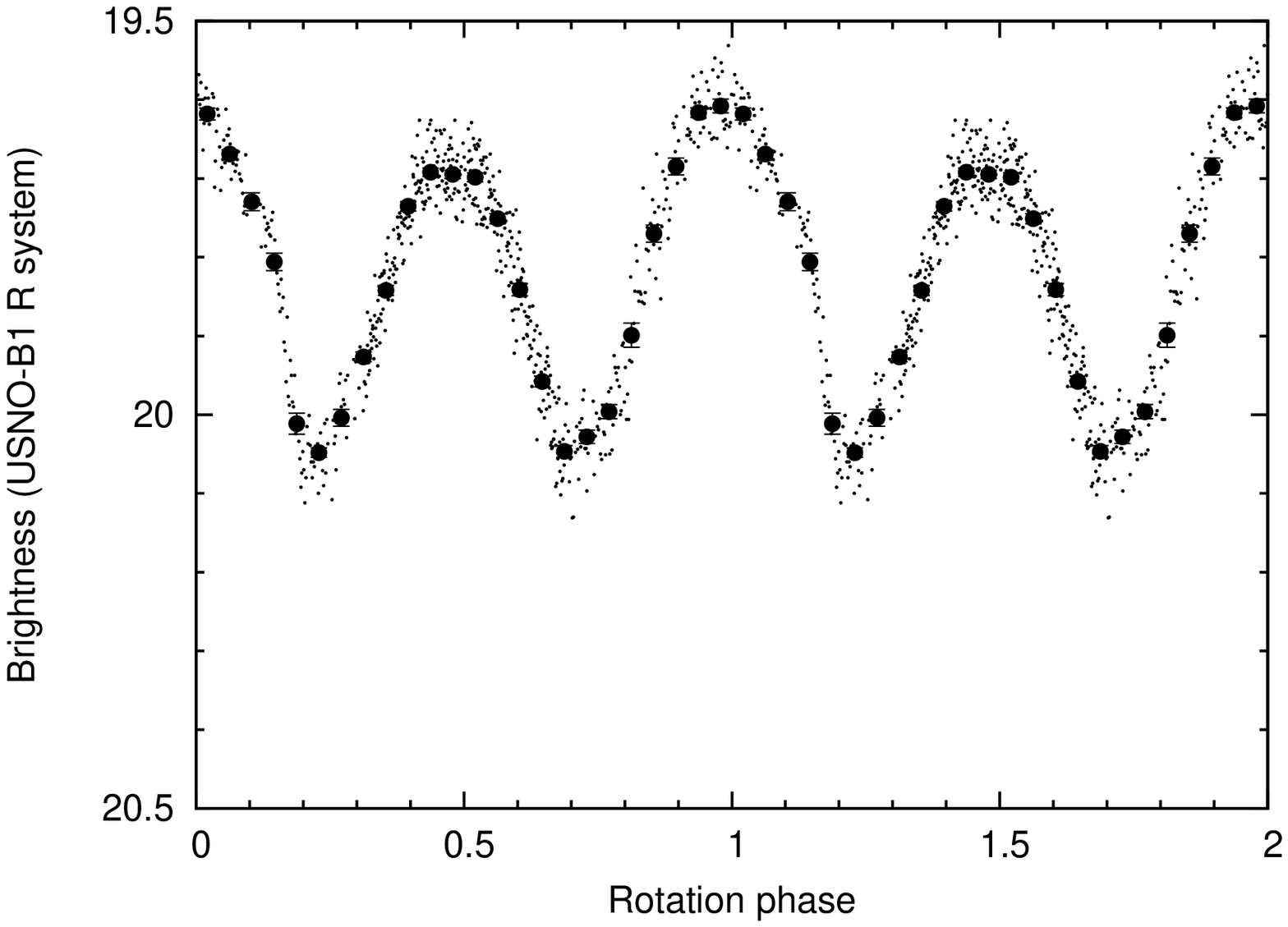}&
\includegraphics[viewport=5 40 510 355,clip,height=3.7cm]{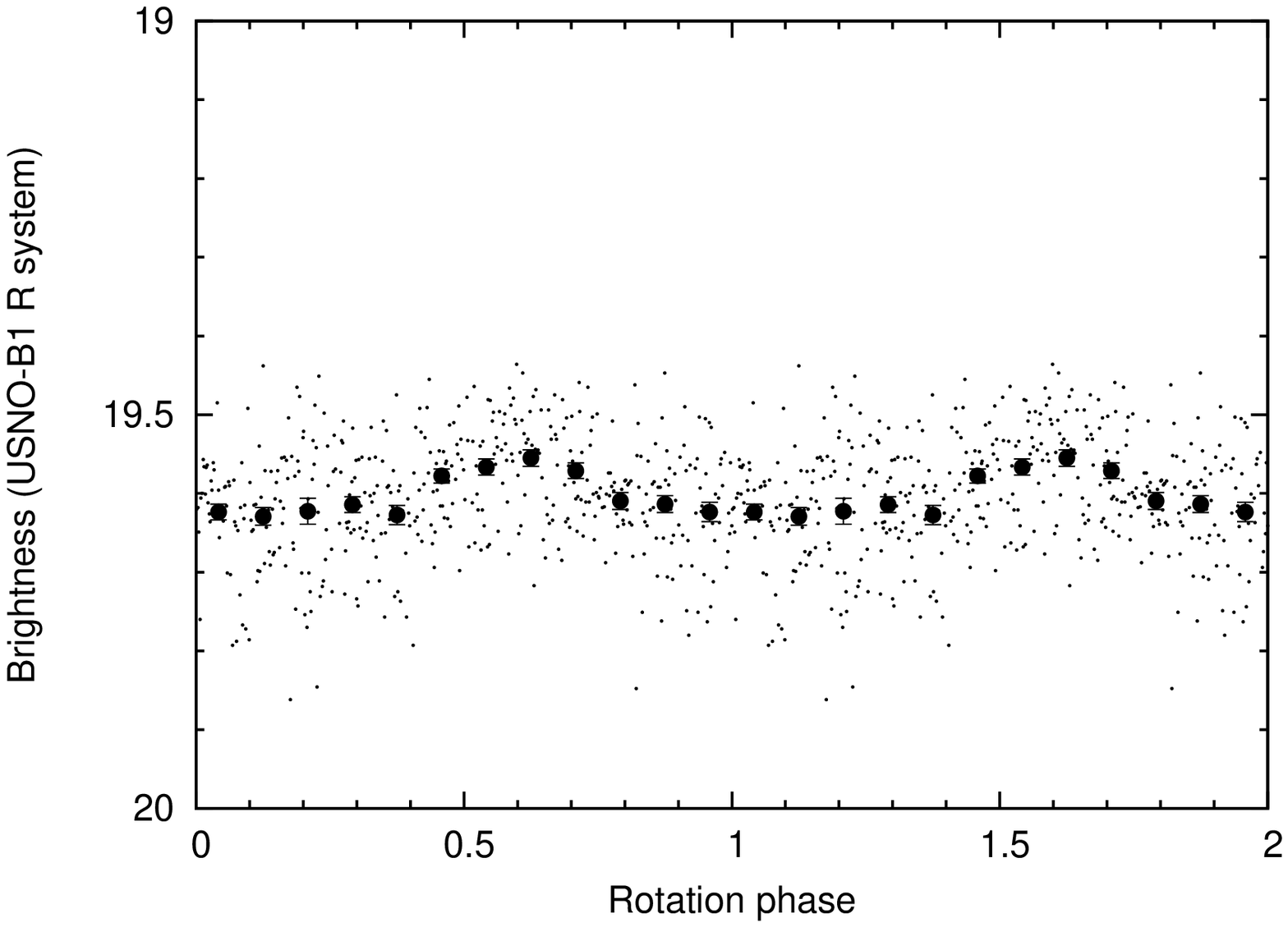}\\
65240 & 65257 & 83984\\
\includegraphics[viewport=5 40 510 355,clip,height=3.7cm]{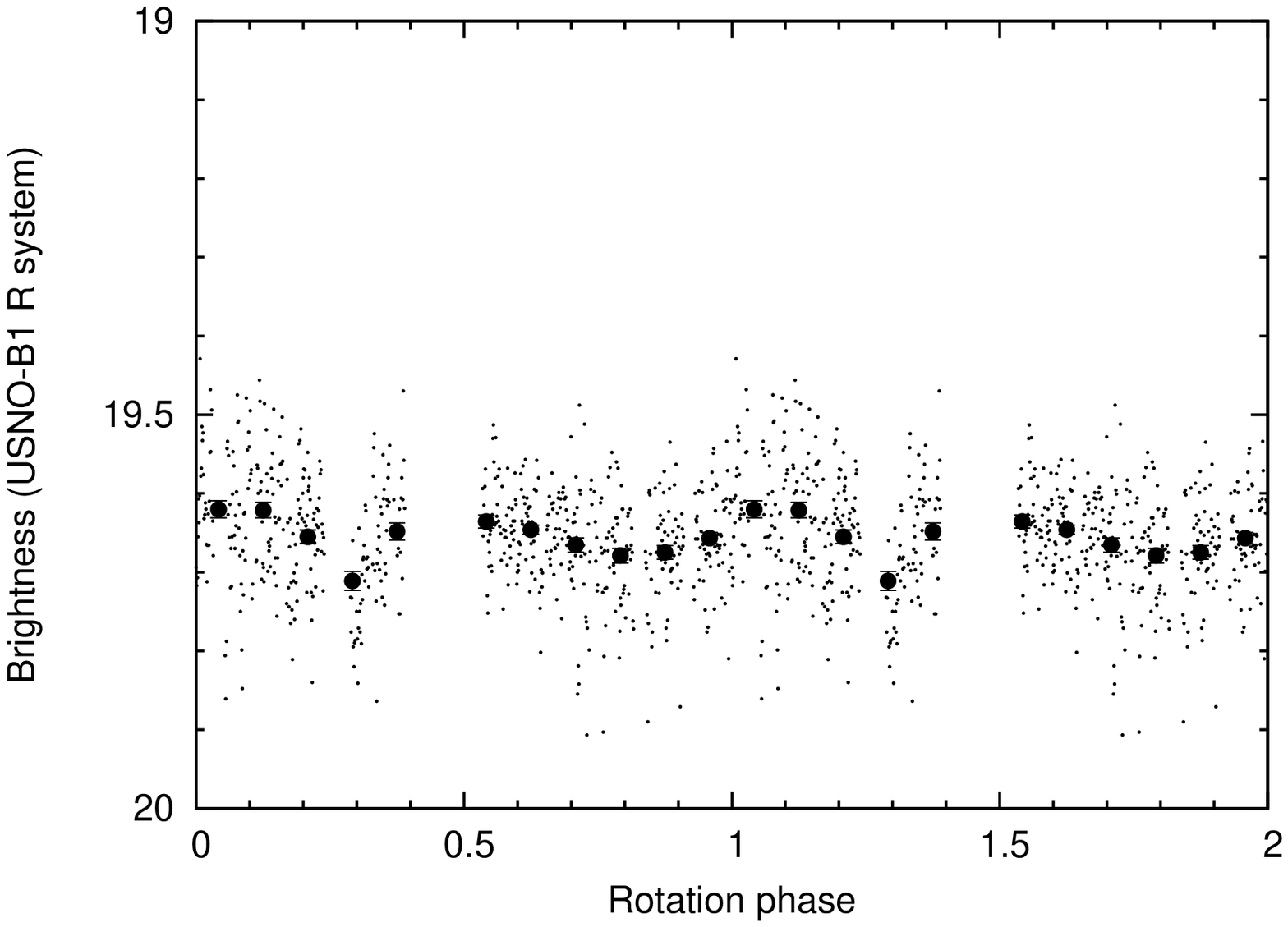}&
\includegraphics[viewport=5 40 510 355,clip,height=3.7cm]{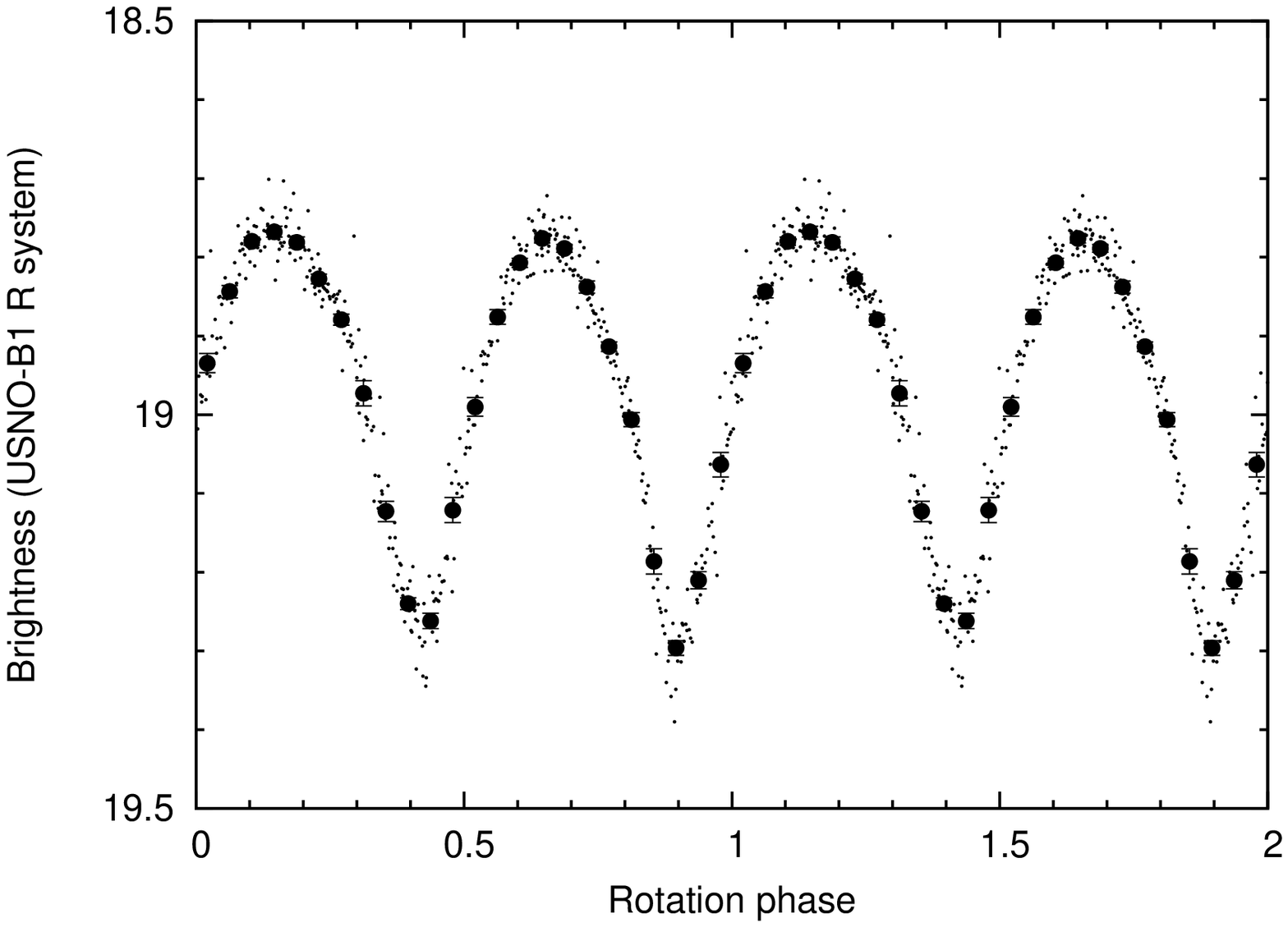}&
\includegraphics[viewport=5 40 510 355,clip,height=3.7cm]{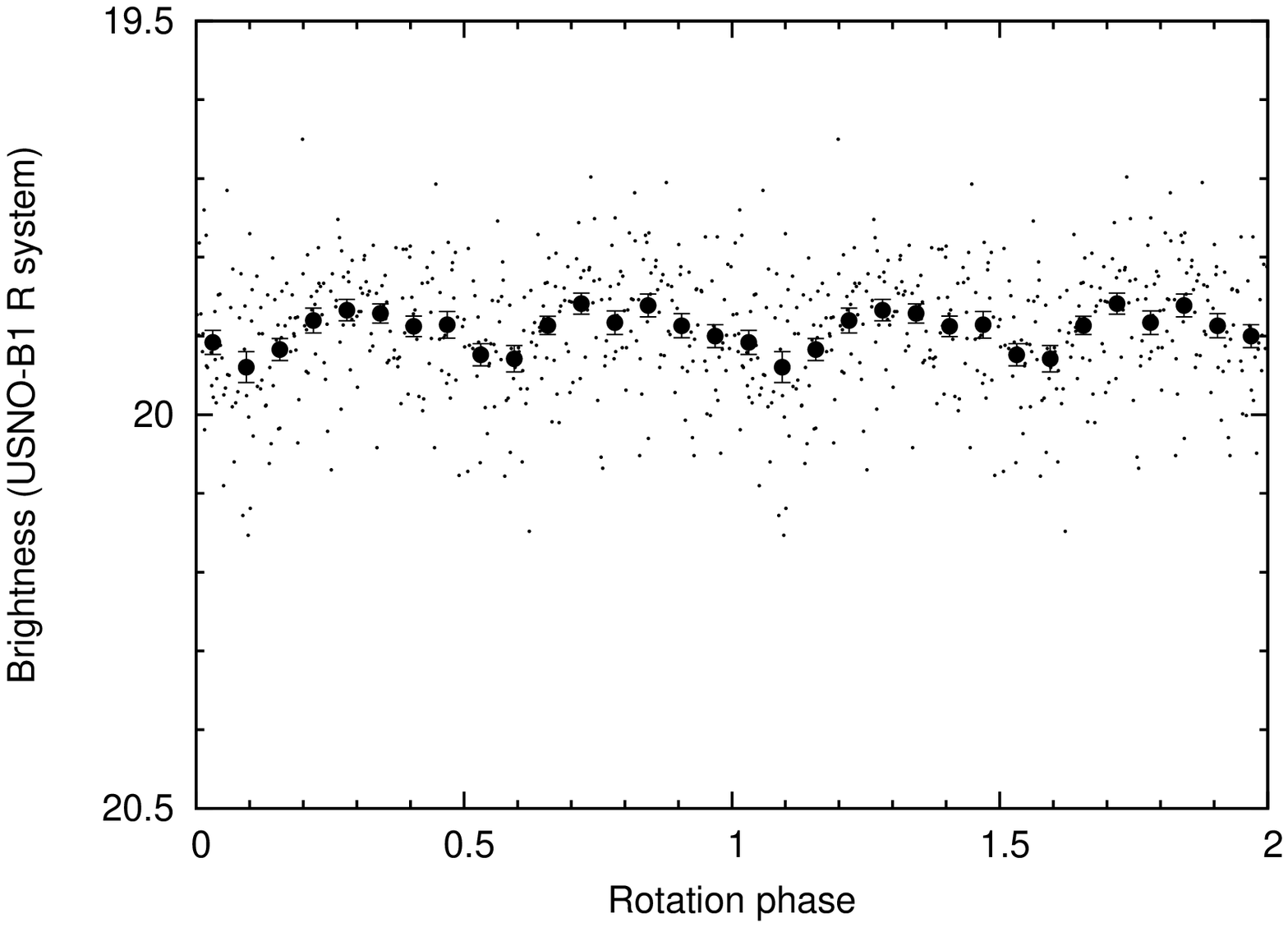}\\
88227 & 88241 & 129602\\
\includegraphics[viewport=5 40 510 355,clip,height=3.7cm]{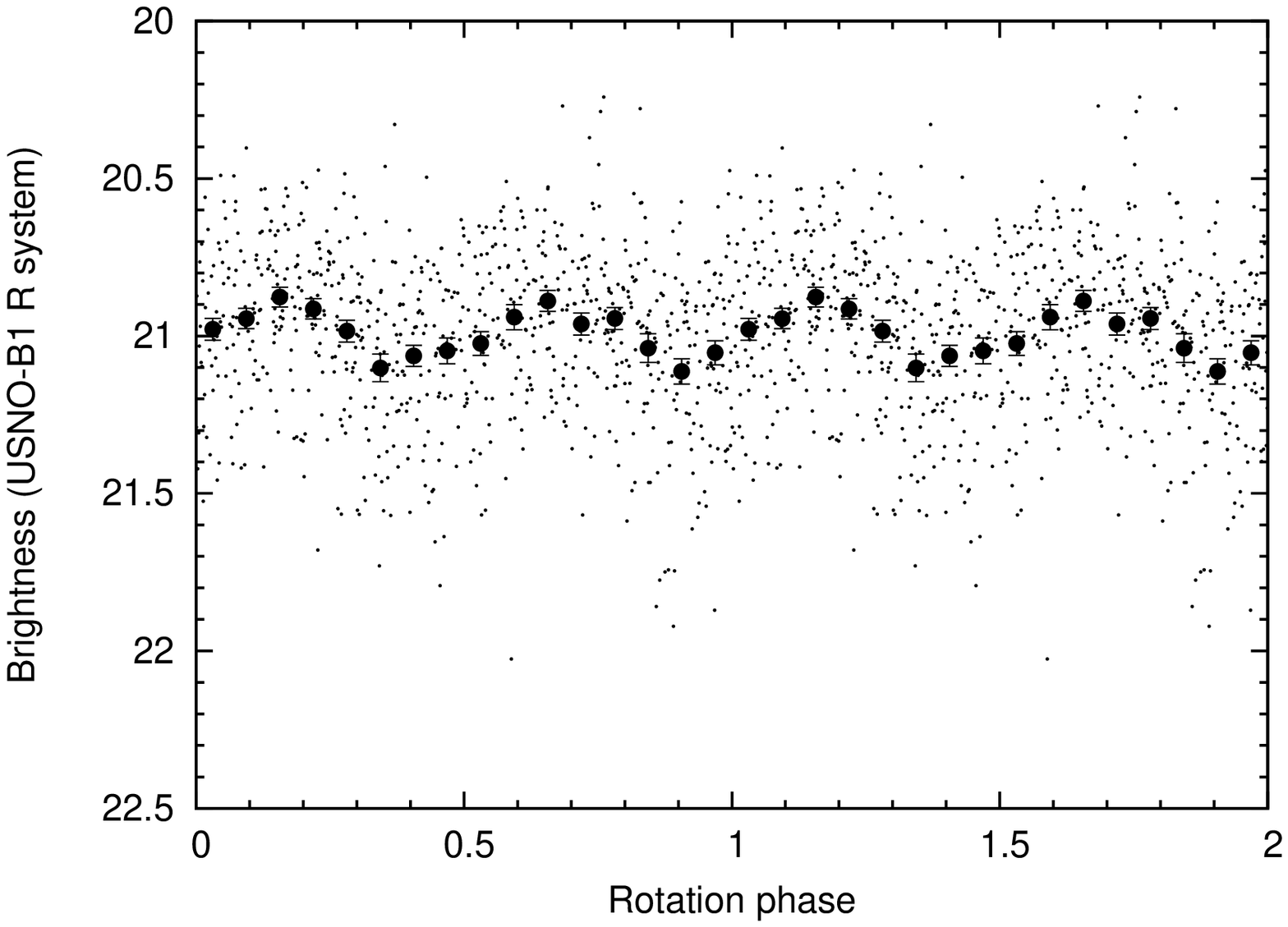}&
\includegraphics[viewport=5 40 510 355,clip,height=3.7cm]{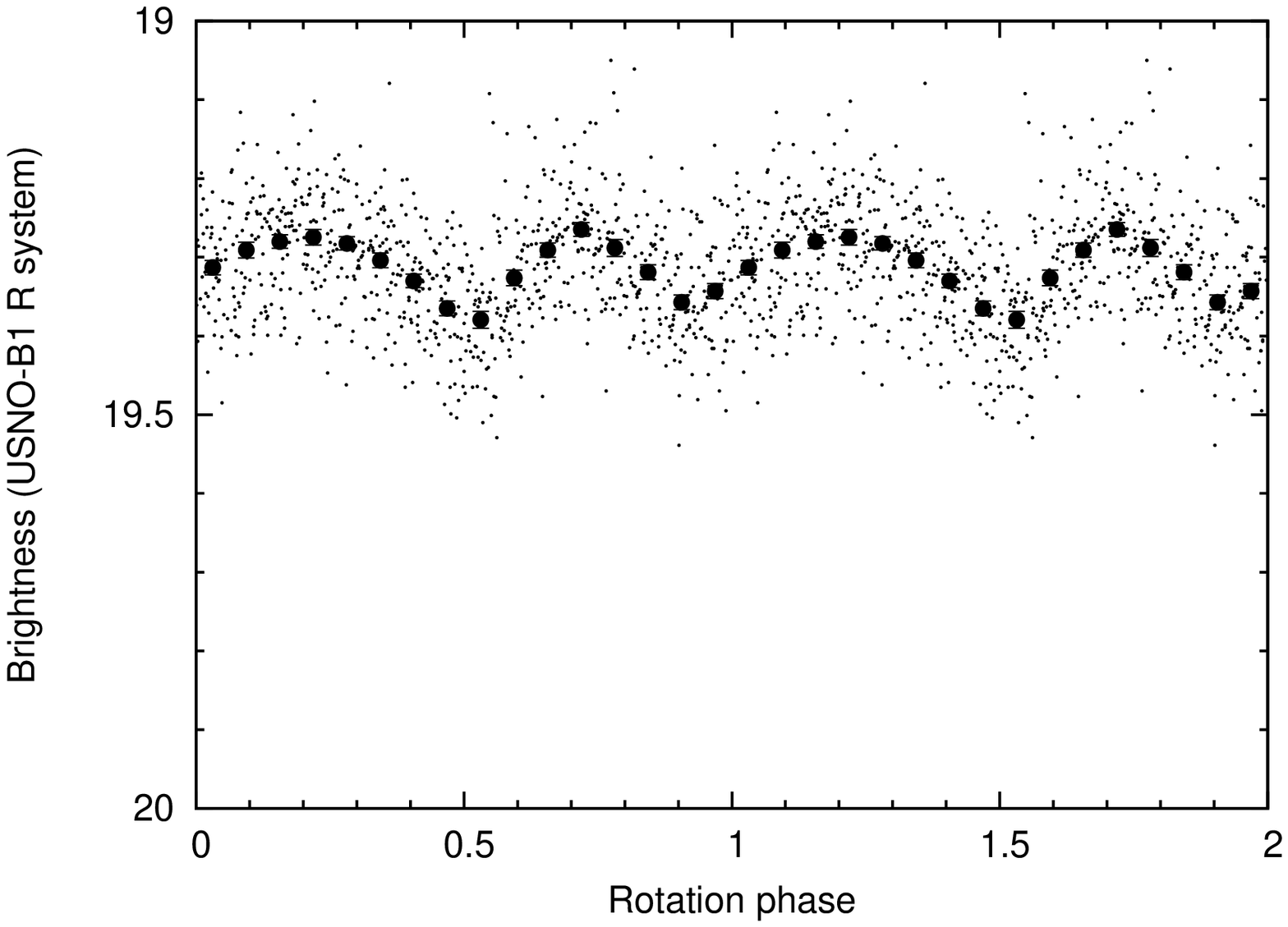}&
\includegraphics[viewport=5 40 510 355,clip,height=3.7cm]{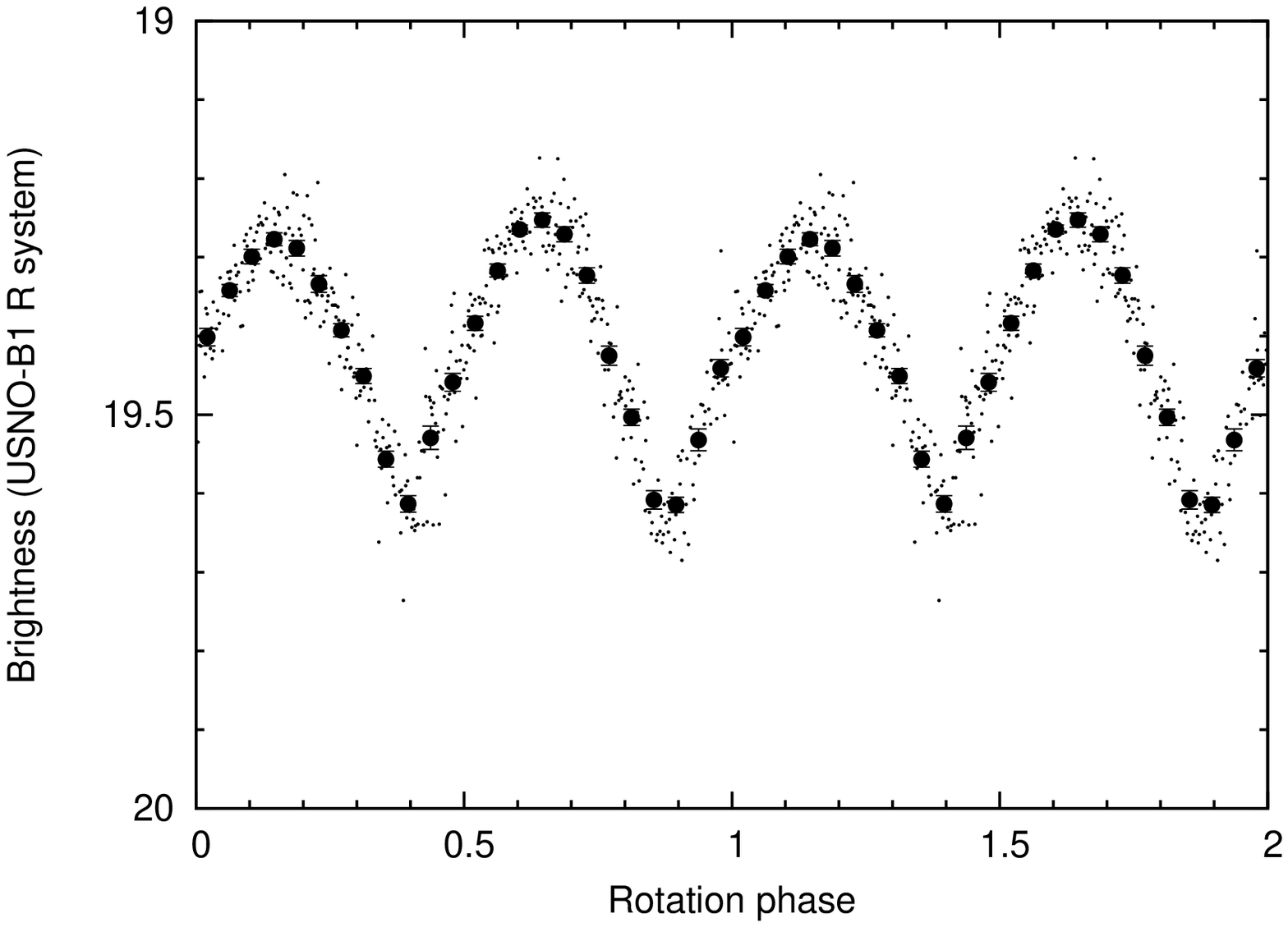}\\
228102 & 353363\\
\includegraphics[viewport=5 40 510 355,clip,height=3.7cm]{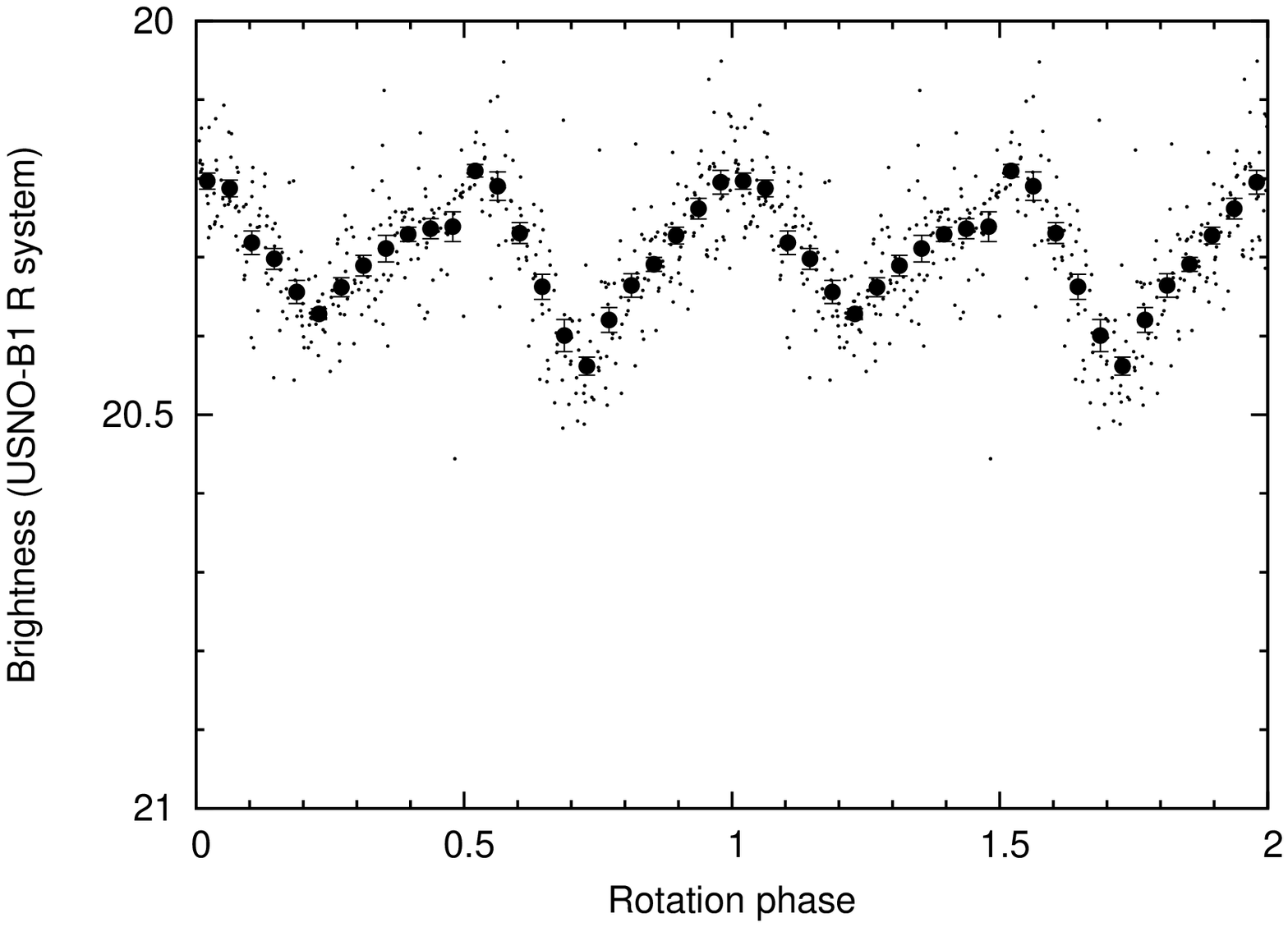}&
\includegraphics[viewport=5 40 510 355,clip,height=3.7cm]{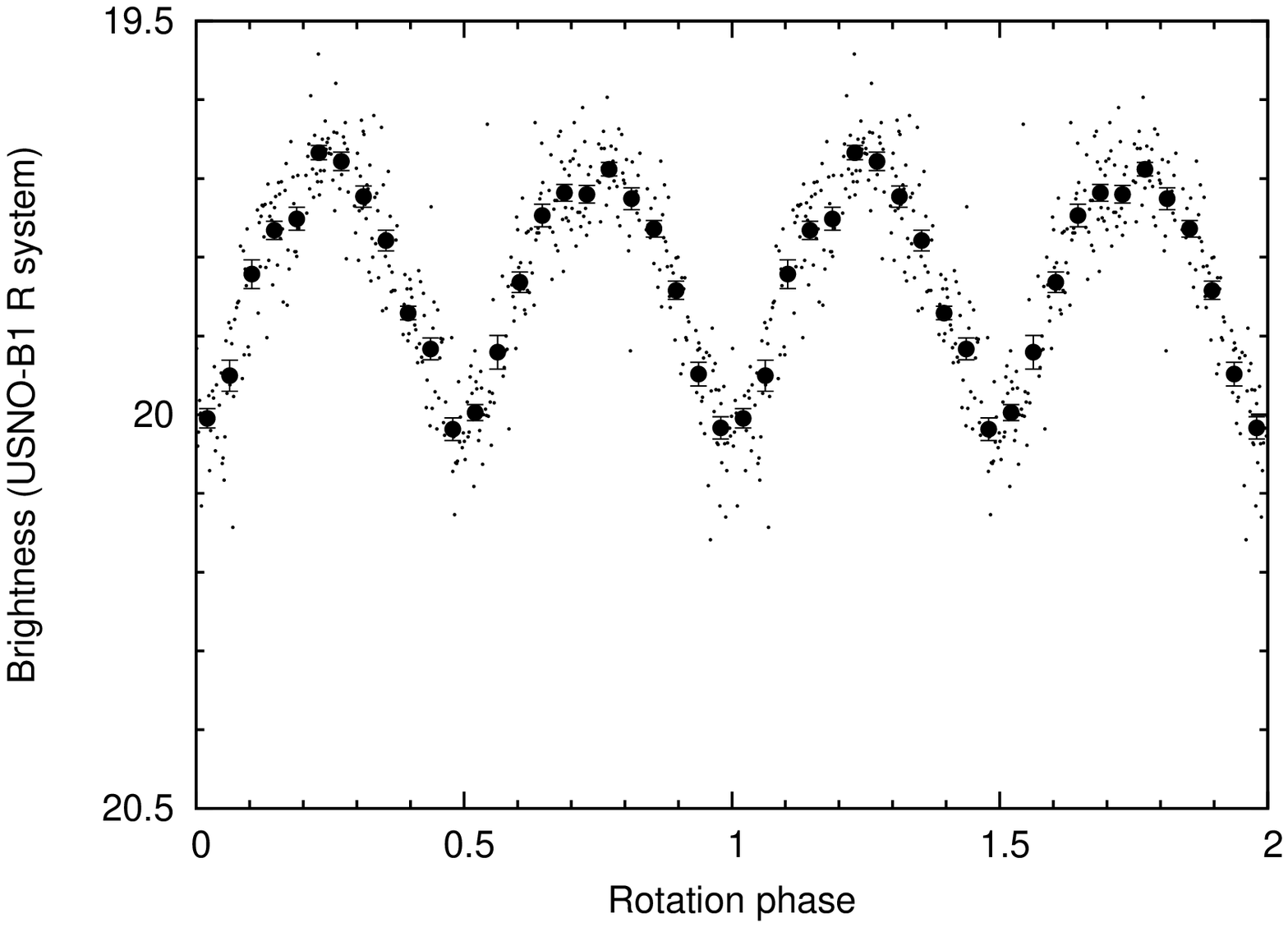}\\
\end{tabular}
\end{center}
\caption{The phased light curves of the Trojan asteroids in K2 Field 6.}
\end{figure*}

\end{appendix}


\begin{thebibliography}{99}

%
\bibitem[\protect\citeauthoryear{A'Hearn}{2011}]{AHearn2011}
A'Hearn, M.F., 2011, ARA\&A, 49, 281
%
\bibitem[\protect\citeauthoryear{Binzel \& Sauter}{2011}]{Binzel}
Binzel, R.,P. \& Sauter, L.M., 1992, Icarus, 95, 222

% 2010Sci...327..977B
% Kepler Planet-Detection Mission: Introduction and First Results
\bibitem[\protect\citeauthoryear{Borucki et al.}{2010}]{borucki2010}
Borucki, W.~J., Koch, D., Basri, G., et al. 2010, Science, 327, 977

%%
\bibitem[\protect\citeauthoryear{Brown}{2013}]{Brown2013}
Brown, M.E., 2013, ApJ, 778, L34

% 1998A&A...336.1056D
% A study of Hilda asteroids. III. Collision velocities and collision frequencies of Hilda asteroids
\bibitem[\protect\citeauthoryear{Dahlgren}{1998}]{dahlgren1998}
Dahlgren, M.
1998, A\&A, 336, 1056

% 2001A&A...366.1053D
% Updated collisional probabilities of minor body populations
\bibitem[\protect\citeauthoryear{Dell'Oro}{2001}]{delloro2001}
Dell'Oro, A.; Marzari, F.; Paolicchi, P. \& Vanzani, V.
2001, A\&A, 366, 1053

% .......................
\bibitem[\protect\citeauthoryear{Emery et al.}{2016}]{emery2016}
Emery, J. et al., 2016, The Complex History of Trojan Asteroids,
in Asteroids IV, Univ. of Arizona, Tucson.

% 2004AJ....128.1391F
% Hilda Asteroids as Possible Probes of Jovian Migration
\bibitem[\protect\citeauthoryear{Franklin et al.}{2004}]{franklin2004}
Franklin, F. A.; Lewis, N. K.; Soper, P. R. \& Holman, M. J.
2004, AJ, 128, 1391

\bibitem[\protect\citeauthoryear{French et al.}{2015}]{French}
French L.~M., Stephens R.~D., Coley D., Wasserman L.~H., Sieben J., 2015, Icar, 254, 1

\bibitem[\protect\citeauthoryear{Grav et al.}{2011}]{2011ApJ...742...40G} Grav T., et al., 2011, ApJ, 742, 40

%%%%
\bibitem[\protect\citeauthoryear{Grav et al.}{2012}]{Grav}
Grav, T., Mainzer, A.K., Bauer, J.M., et al., 2012, ApJ, 759

\bibitem[\protect\citeauthoryear{Gomes et al.}{2005}]{2005Natur.435..466G} Gomes R., Levison H.~F., Tsiganis K., Morbidelli A., 2005, Natur, 435, 466

% 2014PASP..126..398H
% The K2 Mission: Characterization and Early Results
\bibitem[\protect\citeauthoryear{Howell et al.}{2014}]{howell2014}
Howell, S.~B., Sobeck, C., Haas, M., et al.
2014, PASP, 126, 398

% 2016MNRAS.457.2908K
% Nereid from space: Rotation, size and shape analysis from Kepler/K2, Herschel and Spitzer observations
\bibitem[\protect\citeauthoryear{Kiss et al.}{2016}]{kiss2016}
Kiss, Cs., P\'al, A., Farkas-Tak\'acs, A. I.  et al.,
2016, MNRAS, 457, 2908

%%%
\bibitem[\protect\citeauthoryear{Lacerda}{2005}]{Lacerda2005}
Lacerda, P., 2005, The Shapes and Spins of Kuiper Belt Objects, PhD Thesis, Sterrewacht - Universiteit Leiden


%%%
\bibitem[\protect\citeauthoryear{Lacerda et al.}{2014}]{Lacerda2014}
Lacerda, P., Fornasier, S., Lellouch, E. et al.,  2014, ApJ. 793, L2

\bibitem[\protect\citeauthoryear{Lenz \& Breger}{2005}]{lenz2005}
Lenz, P., Breger, M., CoAst, 146, 53

\bibitem[\protect\citeauthoryear{Leone et al.}{1984}]{leone}
Leone, G., Paolicchi, P., Farinella, P., Zappal\'a{}, V. 1984, A\&{}A, 140, 265

\bibitem[\protect\citeauthoryear{Mann et al.}{2007}]{mann}
Mann, R.K., Jewitt, D., Lacerda, P., 2007, AJ, 134, 1133

\bibitem[\protect\citeauthoryear{Levison et al.}{2011}]{2011AJ....142..152L} Levison H.~F., Morbidelli A., Tsiganis K., Nesvorn{\'y} D., Gomes R., 2011, AJ, 142, 152

% 2002aste.book..289M
% Asteroids Do Have Satellites
\bibitem[\protect\citeauthoryear{Merline et al.}{2002}]{merline2002}
Merline, W. J.; Weidenschilling, S. J.; Durda, D. D.; Margot, J. L.; Pravec, P. \& Storrs, A. D.
2002, Asteroids do have satellites, in Asteroids III, W. F. Bottke Jr., A. Cellino, P. Paolicchi, and R. P. Binzel (eds), University of Arizona Press, Tucson, p.289-312

\bibitem[\protect\citeauthoryear{Molnar, Haegert, \& Hoogeboom}{2008}]{2008MPBu...35...82M}
Molnar L.~A., Haegert M., J., Hoogeboom K.~M., 2008, MPBu, 35, 82

% 2015ApJ...812....2M
% Pushing the Limits, Episode 2: K2 Observations of Extragalactic RR Lyrae Stars in the Dwarf Galaxy Leo IV
\bibitem[\protect\citeauthoryear{Moln\'ar et al.}{2015}]{molnar2015}
Moln\'ar, L.; P\'al, A.; Plachy, E.; Ripepi, V.; Moretti, M. I.; Szab\'o, R. \& Kiss, L. L.
2015, ApJ, 812, 2

%%%%
\bibitem[\protect\citeauthoryear{Mottola et al.}{2011}]{2011AJ....141..170M}
Mottola S., et al., 2011, AJ, 141, 170

%%%%
\bibitem[\protect\citeauthoryear{Morbidelli et al.}{2005}]{2005Natur.435..462M}
Morbidelli A., Levison H.~F., Tsiganis K., Gomes R., 2005, Natur, 435, 462

%%%%
\bibitem[\protect\citeauthoryear{Mueller et al.}{2010}]{Mueller2010}
Mueller, M., Marchis, F., Emery, J.P., et al., 2010, Icarus, 205, 505

\bibitem[\protect\citeauthoryear{Nesvorn{\'y}, Vokrouhlick{\'y}, \& Morbidelli}{2013}]{2013ApJ...768...45N} Nesvorn{\'y} D., Vokrouhlick{\'y} D., Morbidelli A., 2013, ApJ, 768, 45

% 2008ssbn.book..345N
% Binaries in the Kuiper Belt
\bibitem[\protect\citeauthoryear{Noll et al.}{2008}]{noll2008}
Noll, K. S., Grundy, W. M., Chiang, E. I., Margot, J.-L. \& Kern, S. D.
In The Solar System Beyond Neptune,
Eds. A.Barucci, H. Boehnhardt, D. Cruikshank, and A. Morbidelli.
Univ. of Arizona Press, Tucson, pp. 345-363.

% 2012MNRAS.421.1825P
% FITSH -- a software package for image processing
\bibitem[\protect\citeauthoryear{P\'al}{2012}]{pal2012}
P\'al, A.
2012, MNRAS, 421, 1825

% 2015ApJ...804L..45P
% Pushing the Limits: K2 Observations of the Trans-Neptunian Objects 2002 GV31 and (278361) 2007 JJ43
\bibitem[\protect\citeauthoryear{P\'al et al.}{2015a}]{pal2015}
P\'al, A.; Szab\'o, R.; Szab\'o, Gy. M.; Kiss, L. L.; Moln\'ar, L.; S\'arneczky, K. \& Kiss, Cs.
2015, ApJL, 804, 45

% 2016AJ....151..117P
% Large Size and Slow Rotation of the Trans-Neptunian Object (225088) 2007 OR10 Discovered from Herschel and K2 Observations
\bibitem[\protect\citeauthoryear{P\'al et al.}{2016}]{pal2016}
P\'al, A., Kiss, Cs.; M\"uller, Th. G., Moln\'ar, L.; Szab\'o, R.; Szab\'o, Gy. M., S\'arneczky, K. \& Kiss, L. L.
2016, AJ, 151, 117

%%%
%%%
\bibitem[\protect\citeauthoryear{Peixinho et al.}{2012}]{Peixinho2012}
Peixinho, N., Delsanti, A., Guilbert-Lepoutre, A., Gafeira, R., Lacerda, P., 2012, A\&A, 546, A86

%%%
\bibitem[\protect\citeauthoryear{Pravec \& Harris}{2000}]{Pravec+Harris}
Pravec, P. \& Harris, A.W., 2000, Icarus, 148, 12

\bibitem[\protect\citeauthoryear{Roig, Ribeiro, \& Gil-Hutton}{2008}]{2008A&A...483..911R} Roig F., Ribeiro A.~O., Gil-Hutton R., 2008, A\&A, 483, 911

% 2015ApJ...799..191S
% Binary Candidates in the Jovian Trojan and Hilda Populations from NEOWISE Light Curves
\bibitem[\protect\citeauthoryear{Sonnett et al.}{2015}]{sonnett2015}
Sonnett, S.; Mainzer, A.; Grav, T.; Masiero, J. \& Bauer, J.
2015, ApJ, 799, 191

\bibitem[\protect\citeauthoryear{Szab{\'o} et al.}{2007}]{2007MNRAS.377.1393S} Szab{\'o} G.~M., Ivezi{\'c} {\v Z}., Juri{\'c} M., Lupton R., 2007, MNRAS, 377, 1393

% 2015AJ....149..112S
% Main-belt Asteroids in the K2 Engineering Field of View
\bibitem[\protect\citeauthoryear{Szab\'o et al.}{2015}]{szabo2015}
Szab\'o, R., S\'arneczky, K., Szab\'o, Gy. M., et al. 2015, AJ, 149, 112

\bibitem[\protect\citeauthoryear{Szab\'o et al.}{2016}]{szabo2016}
Szab\'o, R., et al. 2016, A\&{}A, resubmitted

\bibitem[\protect\citeauthoryear{Tsiganis et al.}{2005}]{2005Natur.435..459T} Tsiganis K., Gomes R., Morbidelli A., Levison H.~F., 2005, Natur, 435, 459

%%%%
\bibitem[\protect\citeauthoryear{Vilenius et al.}{2014}]{Vilenius2014}
Vilenius, E., Kiss, Cs., M\"uller, Th.G., et al., 2014, A\&A, 564, A35

\bibitem[\protect\citeauthoryear{Warner et al.}{2009}]{Warner}
Warner, B.D., Harris, A.W., Pravec, P. (2009). Icarus 202, 134-146., verison as of July 19, 2016, http://www.MinorPlanet.info/lightcurvedatabase.html

\bibitem[Waszczak et al.(2015)]{2015AJ....150...75W} Waszczak, A., Chang, C.-K., Ofek, E.~O., et al.\ 2015, \aj, 150, 75

\bibitem[\protect\citeauthoryear{Wong, Brown, \& Emery}{2014}]{2014AJ....148..112W} Wong I., Brown M.~E., Emery J.~P., 2014, AJ, 148, 112

\bibitem[\protect\citeauthoryear{Wong \& Brown}{2015}]{Wong2015}
Wong, I \& Brown, M.E., 2015, AJ, 150, 147

\bibitem[\protect\citeauthoryear{Wong \& Brown}{2016}]{Wong2016}
Wong, I \& Brown, M.E., 2015, arXiv:1607.04133



\end{thebibliography}
\end{document}